\newcommand{\gev}{\, {\rm GeV}}

\newcommand{\beq}{\begin{equation}}
\newcommand{\eeq}{\end{equation}}
\newcommand{\bea}{\begin{eqnarray}}
\newcommand{\eea}{\end{eqnarray}}

\newcommand{\gsim}{\lower.7ex\hbox{$\;\stackrel{\textstyle>}{\sim}\;$}}
\newcommand{\lsim}{\lower.7ex\hbox{$\;\stackrel{\textstyle<}{\sim}\;$}}

\documentclass[a4paper,11pt]{article}
\pdfoutput=1

\usepackage{jcappub}
\usepackage[utf8]{inputenc} 
\usepackage{amsfonts,amssymb,mathrsfs,amsmath,esint,bm}
\usepackage{pdflscape} 
\usepackage[capitalise]{cleveref}
\allowdisplaybreaks

\graphicspath{{./Figures/}}
\usepackage{latexsym}
\usepackage{graphicx}
\usepackage{subfigure}
\usepackage[dvipsnames]{xcolor}
\usepackage{booktabs}
\usepackage{physics}
\usepackage{tikz}
\usetikzlibrary{decorations.pathmorphing}\usetikzlibrary{positioning,calc}
\usetikzlibrary{math}
\usepackage{cleveref}
\usepackage{comment}
\usepackage[normalem]{ulem}

\hypersetup{pdfstartview=FitV,colorlinks=true,linkcolor=blue,citecolor=red,filecolor=black,urlcolor=blue}

\def\stacksymbols #1#2#3#4{\def\theguybelow{#2}
    \def\vp{\lower#3pt}
    \def\sp{\baselineskip0pt\lineskip#4pt}
    \mathrel{\mathpalette\intermediary#1}}

\def\intermediary#1#2{\vp\vbox{\sp
     \everycr={}\tabskip0pt
     \halign{$\mathsurround0pt#1\hfil##\hfil$\crcr#2\crcr
              \theguybelow\crcr}}}

\def\trh{T_{\rm RH}}

\def\arh{a_{\rm RH}}
\def\aosc{a_{\rm osc}}

\def\be{\begin{equation}}
\def\ee{\end{equation}}
\def\bea{\begin{eqnarray}}
\def\eea{\end{eqnarray}}

\def\sp{\;\;\;,\;\;\;}

\def\a{\alpha}

\def\arh{a_{\rm RH}}
\def\aosc{a_{\rm osc}}

\def\aend{a_{\rm end}}
\def\am{a_{\rm m}}
\def\phiend{\phi_{\rm end}}
\def\rhoe{\rho_{\rm end}}
\def\He{H_{\rm end}}
\def\rhochie{\rho_\chi^{\rm end}}
\def\grh{g_{\rm RH}}
\def\rhorh{\rho_{\rm RH}}
\def\meff{m_{\chi,{\rm eff}}}
\def\bk{\boldsymbol{k}}
\def\bx{\boldsymbol{x}}

\def\lsim{\raise0.3ex\hbox{$\;<$\kern-0.75em\raise-1.1ex\hbox{$\sim\;$}}}
\def\gsim{\raise0.3ex\hbox{$\;>$\kern-0.75em\raise-1.1ex\hbox{$\sim\;$}}}

\def\inbar{\,\vrule height1.5ex width.4pt depth0pt}

\def\IC{\relax\hbox{$\inbar\kern-.3em{\rm C}$}}
\def\IQ{\relax\hbox{$\inbar\kern-.3em{\rm Q}$}}
\def\IR{\relax{\rm I\kern-.18em R}}
 \font\cmss=cmss10 \font\cmsss=cmss10 at 7pt
\def\IZ{\relax\ifmmode\mathchoice
 {\hbox{\cmss Z\kern-.4em Z}}{\hbox{\cmss Z\kern-.4em Z}}
 {\lower.9pt\hbox{\cmsss Z\kern-.4em Z}}
 {\lower1.2pt\hbox{\cmsss Z\kern-.4em Z}}\else{\cmss Z\kern-.4em Z}\fi}

\def\comment#1{}
\def\to{\rightarrow}

\def\u1x{U(1)_X}
\newcommand{\nc}{\newcommand}
\nc{\LL}{L}
\nc{\vv}{\tilde{v}}
\nc{\ccdot}{\!\cdot\!}
\nc{\gsm}{G_{SM}}
\nc{\vfive}{\mathbf{5}\oplus\mathbf{\overline{5}}}
\nc{\vten}{\mathbf{10}\oplus\mathbf{\overline{10}}}
\nc{\zhol}{Z^{\rm hol}}
\nc{\xfb}{\,{\rm fb}}

\title{Scalar Field Fluctuations and the Production of Dark Matter}

\author[a]{Marcos~A.~G.~Garcia,}
\affiliation[a]{Departamento de F\'isica Te\'orica, Instituto de F\'isica, Universidad Nacional Aut\'onoma de M\'exico, Ciudad de M\'exico C.P. 04510, Mexico}
\emailAdd{marcos.garcia@fisica.unam.mx}

\author[b]{Wenqi Ke,}
\affiliation[b]{William I.~Fine Theoretical Physics Institute, School of Physics and Astronomy, University of Minnesota, Minneapolis, MN 55455, USA}
\emailAdd{wke@umn.edu}

\author[c]{Yann Mambrini,}
\affiliation[c]{Universit\'e Paris-Saclay, CNRS/IN2P3, IJCLab, 91405 Orsay, France}
\emailAdd{mambrini@ijclab.in2p3.fr}

\author[b]{\qquad \qquad Keith A. Olive,}
\emailAdd{olive@umn.edu}

\author[d]{Sarunas Verner}
\affiliation[d]{Institute for Fundamental Theory, Physics Department, University of Florida, Gainesville, FL 32611, USA}
\emailAdd{verner.s@ufl.edu}

\abstract{One of the simplest possible candidates for dark matter is a stable scalar singlet beyond the Standard Model. If its mass is below the Hubble scale during inflation, long-wavelength modes of this scalar will be excited during inflation, and their subsequent evolution may lead to the correct relic density of dark matter. In this work, we provide a comprehensive analysis of the evolution of a spectator scalar. We examine three cases: (1) a non-interacting massive scalar, (2) a massive scalar with self-interactions of the form $\lambda_\chi \chi^p$, and (3) a massive scalar coupled to the inflaton $\phi$ through an interaction term of the form $\sigma_{n,m} \phi^n \chi^m$. In all cases, we assume minimal coupling to gravity and compare these results with the production of short-wavelength modes arising from single graviton exchange. The evolution is tracked during the reheating phase. Our findings are summarized using $(m_\chi, \trh)$ parameter planes, where $m_\chi$ is the mass of the scalar field and $\trh$ is the reheating temperature after inflation. The non-interacting scalar is highly constrained, requiring $m_\chi > 3 \times 10^{12}~\gev$ and $\trh \lesssim 7~\text{TeV}$ for an inflationary potential with a quadratic minimum.  However, when self-interactions or couplings to the inflaton are included, the viable parameter space expands considerably. In these cases, sub-GeV and even sub-eV scalar masses can yield the correct relic abundance, opening new possibilities for light dark matter candidates. In all cases, we also impose additional constraints arising from the production of isocurvature fluctuations, the prevention of a secondary inflationary phase triggered by the spectator field, and the fragmentation of scalar condensates.}

\begin{document}
\begin{flushright}
UMN--TH--4417/25, FTPI--MINN--25/02   \\
February  2025
\end{flushright}
\maketitle


\section{Introduction}
\label{sec:intro}
Scalar field fluctuations are a fundamental consequence of the quasi-de Sitter expansion during inflation~\cite{fluc}. These fluctuations, driven by quantum effects and rapid spacetime expansion, play a crucial role in the formation of primordial density perturbations that seed large-scale structure. While the inflaton field drives inflation, fluctuations in other scalar fields, known as spectator fields, can also leave significant imprints on the evolution of the universe.

An important framework for understanding these fluctuations is the stochastic approach. This approach models the evolution of fluctuations as a balance between random quantum ``kicks" and classical drift forces. Random quantum fluctuations arise due to the stochastic nature of fields at horizon crossing, while the classical drift describes the deterministic evolution of these fluctuations on superhorizon scales. This framework has been extensively applied to study inflationary fluctuations, particularly in cases where quantum noise plays a significant role in driving the dynamics~\cite{Starobinsky:1986fx, Starobinsky:1994bd}.

The cosmological implications of scalar field fluctuations are vast. Of particular significance are their effects on supersymmetry flat directions~\cite{gkm} and their role in various early-universe phenomena. These include Affleck-Dine baryogenesis~\cite{AD,Linde:1985gh,Campbell:1986qg,Enqvist:2003gh,Garcia:2013bha}, the preservation of primordial density perturbations against washout~\cite{Graziani:1988bp,Enqvist:2003gh,Enqvist:2011pt}, and their influence on reheating and thermalization processes~\cite{Allahverdi:2005mz,Olive:2006uw}. These phenomena rely on the fact that flat directions or light scalar fields remain nearly massless during inflation. However, maintaining flat directions in general supergravity models poses a challenge~\cite{Dine:1995uk}, as scalar fields typically obtain Hubble-scale masses due to inflationary dynamics. This issue can be avoided in theories with extra symmetries, such as Heisenberg symmetry~\cite{Binetruy:1987xj}, which appears in no-scale supergravity frameworks~\cite{noscale}, where flat directions can remain maintained and stable throughout inflation~\cite{GMO}.

Fluctuations in more massive scalar fields may also play an important role in cosmology by affecting the observed perturbation spectrum usually attributed to result from inflaton fluctuations. Isocurvature fluctuations in a massive spectator field or curvaton~\cite{Enqvist:2001zp,Lyth:2001nq,Moroi:2001ct}, can affect the cosmic microwave background (CMB) observables such as $n_s$ and $r$ if they decay after inflationary reheating. Initial conditions for the evolution of the curvaton (like the Affleck-Dine flat direction) are also determined by stochastic processes~\cite{Linde:2005yw,Torrado:2017qtr,Choi:2024ruu}.

This paper focuses on the impact of fluctuations of {\em stable} spectator scalar fields, which, although they do not drive inflation, can significantly contribute to the energy density of the universe and may be candidates for cold dark matter~\cite{Turner:1987vd, Peebles:1999fz, Enqvist:2014zqa, Nurmi:2015ema, Bertolami:2016ywc, Alonso-Alvarez:2018tus,ema,Markkanen:2018gcw, Tenkanen:2019aij, Choi:2019mva, Cosme:2020nac,Ling:2021zlj,Basso:2022tpd,Lebedev:2022cic,Kaneta:2023kfv}. Non-self-interacting scalars typically lead to a significant overproduction of dark matter unless the spectator mass is relatively large ($\gtrsim \mathcal{O}(10^{12})~\gev$) and the reheating temperature is relatively low ($\lesssim \mathcal{O}(100)~\gev$). Our recent work~\cite{Choi:2024bdn} has thoroughly explored enlarging the allowed parameter space when the effect of couplings between the spectator and inflaton are included, and how these scalar fluctuations can lead to the formation of scalar dark matter. In this paper, we build on these studies by examining scenarios with different inflaton potential forms near its minimum, $V(\phi) \sim \phi^k$, where $k \geq 2$, which influence the dynamics of reheating~(see~\cite{gkmo1, gkmo2} for details). Additionally, we investigate the effects of spectator scalar field $\chi$ self-interactions of the form $\lambda_{\chi} \chi^p$, where $p \geq 4$, as well as interactions between the spectator field and inflaton, $\sigma_{n,m} \phi^n \chi^m$. These terms affect the evolution of the spectator field and ultimately determine the asymptotic behavior of $\langle \chi^2 \rangle$. Our goal is to explore how these interactions give rise to distinct cosmological scenarios and outcomes, which is the central focus of this paper.

If inflation lasts long enough, scalar field fluctuations grow steadily during the quasi-de Sitter expansion, eventually reaching an asymptotic value $\langle \chi^2 \rangle$. For a massive scalar field $\chi$, these fluctuations typically saturate at $\langle \chi^2 \rangle \sim H_I^4 / m_{\chi}^2 \gg H_I^2$ , provided that the bare mass $m_{\chi}$ is much smaller than the inflationary Hubble parameter $H_I$. For sufficiently light fields, these large fluctuations can be interpreted as a nearly homogeneous background field that evolves according to classical equations of motion~\cite{longw}. However, if self-interactions become significant, they suppress the asymptotic value of $\langle \chi^2 \rangle \sim (H_I^4 / \lambda_{\chi} M_P^{4-p})^{2/p}$,  where $M_P = 1 / \sqrt{8 \pi G} \simeq 2.4 \times 10^{18}$ GeV is the reduced Planck mass. For quartic interactions ($p = 4$), this simplifies to $\langle \chi^2 \rangle \sim H_I^2 / \sqrt{\lambda_{\chi}}$, and when $\lambda_{\chi} \sim 1$, the fluctuations reduce to $\langle \chi^2 \rangle \sim H_I^2$~\cite{Starobinsky:1994bd}. Therefore, strong self-interactions reduce the energy density of the field during and after inflation, ultimately leading to a smaller contribution of $\chi$ to the present-day dark matter density. 

Although spectator fields $\chi$ do not drive inflation, they remain gravitationally coupled to the inflaton through minimal interactions, which can lead to particle production in the early universe (see Ref.~\cite{Kolb:2023ydq} for a recent review on gravitational particle production). These interactions arise via single-graviton exchange and lead to effective couplings between $\chi$ and the inflaton field. For example, in the case of an inflaton potential $V(\phi) = \frac{1}{2} m_{\phi}^2 \phi^2$, the spectator field obtains an interaction term of the form $\sigma \phi^2 \chi^2$~\cite{cmov,Clery:2022wib}, where $\sigma = (m_{\phi} / 2 M_P)^2$. However, this interaction term does not contribute to the effective mass of $\chi$, as it originates from a Lagrangian term of the form $\mathcal{L} \supset \phi^2 (\partial \chi)^2$. A similar coupling arises if $\chi$ is coupled to the Ricci curvature scalar through a term $ \mathcal{L} \supset \xi_{\chi} \chi^2 R$, where $R$ is the Ricci scalar and $\xi_{\chi}$ is a dimensionless non-minimal coupling. In such a scenario, transforming the theory to the Einstein frame introduces a term proportional to $\xi_{\chi} (m_{\phi}^2 / M_P^2) \phi^2 \chi^2$~\cite{Clery:2022wib}. Furthermore, we extend these ideas and consider more general interactions of the form $\mathcal{L} \supset \sigma_{n,m} \phi^n \chi^m$. These interactions, while typically suppressed, can still influence the evolution of $\chi$, particularly during reheating, affecting the dark matter relic abundance, and we study such possibilities in this work in some detail.

The onset of $\chi$-oscillations depends on the model parameters and may occur either before or after reheating. Once oscillations begin, the evolution of $\chi$ is determined by its self-interactions and interactions with other fields. By specifying these parameters, one can calculate the relic abundance of $\chi$, which is discussed in great detail in this paper. Notably, this production mechanism adds to and may even surpass the contribution from scalar dark matter produced directly by the inflaton condensate after inflation. We also address the constraints imposed by such dark matter production mechanisms. We provide a comprehensive study of these scenarios, particularly when the spectator scalar field serves as a stable dark matter candidate. We impose several constraints to ensure consistency with cosmological observations. First, we apply isocurvature constraints by ensuring that the isocurvature power spectrum satisfies the upper limit $\mathcal{P}_{\mathcal{S}}(k_*) < 8.3 \times 10^{-11}$, as given by \textit{Planck} observations~\cite{Planck:2018jri}. Second, we impose constraints to account for the possibility that the energy density of the spectator field grows sufficiently to dominate the total energy budget, potentially triggering a second phase of inflation. Third, we address fragmentation constraints, which arise in models where the inflaton potential takes the form $V(\phi) \sim \phi^k$ with $k \geq 4$. In this case, the effective inflaton mass may vanish, preventing inflaton decay and prematurely terminating the reheating process. We ensure that such scenarios are properly addressed to maintain viable reheating dynamics.

The structure of the paper is as follows: We first introduce the two scalar fields central to this study—the inflaton and the spectator field. In Section~\ref{sec:infevol}, we review the evolution of the inflaton for general models of inflation with varying equations of state during reheating while remaining consistent with CMB constraints. Section~\ref{sec:spectevol} provides solutions to the spectator field equation of motion, including the effects of self-interactions. In Section~\ref{sec:init}, we outline the initial conditions for both the inflaton and spectator fields, determined by the inflationary parameters. We then discuss the key cosmological constraints in Section~\ref{sec:constraints}, including those from isocurvature fluctuations, the prevention of a second period of inflation, and fragmentation during reheating. The potential role of the spectator field as a dark matter candidate is explored in the following sections: Section~\ref{Sec:spectator} examines the case of non-self-interacting spectators, Section~\ref{sec:SpectSI} focuses on self-interacting spectators, and Section~\ref{sec:specinfl} analyzes scenarios where the spectator field interacts with the inflaton. Finally, our conclusions are presented in Section~\ref{summary}.

\section{The Inflaton and Spectator}
\label{sec:inflation}

In this section, we begin by reviewing the inflationary potential used as an example, along with the evolution of the inflaton and its energy density during the reheating process. This process is characterized by an equation of state parameterized by an even integer $k$, such that $w_\phi = \frac{k-2}{k+2}$. We then proceed to study the evolution of a spectator scalar field.
\subsection{The Evolution of the Inflaton}
\label{sec:infevol}

As an example, we consider the class of inflationary models known as T-models~\cite{Kallosh:2013hoa}, which are given by the potential
\begin{equation}
    V(\phi) \; = \;\lambda M_P^{4} \left[\sqrt{6} \tanh \left(\frac{\phi}{\sqrt{6} M_P}\right)\right]^{k} \, ,
\label{Vatt}
\end{equation}
where $k$ is an even integer, and $\lambda$ is the potential scale, determined by the CMB normalization and the number of $e$-folds of inflation before horizon crossing. For $k=2$, this potential resembles the Starobinsky potential \cite{Staro}. Expanding the potential around its minimum at $\phi = 0$, we have
\begin{equation}
    \label{Eq:potmin}
    V(\phi) \; = \; \lambda M_P^4 \left(\frac{\phi}{M_{P}}\right)^k, \quad \phi \ll M_{P} \, .
\end{equation}
For $k=2$, the potential is approximately quadratic near its minimum, with the inflaton mass given by $m^2_\phi = 2 \lambda M_P^2$. Inflation in this class of models occurs at large field values and ends when the cosmological scale factor $a$ satisfies $\ddot{a} = 0$. We define the scale factor at this moment as $a(t_{\rm end}) = \aend$, and the corresponding inflaton field value $\phiend  = \phi(\aend)$ satisfies the condition $\dot \phi_{\rm end}^2 = V(\phi_{\rm end})$. An approximate expression for $\phiend$ for these models is given by \cite{gkmo2}
\beq
\phi_{\rm end} \;\simeq\;\sqrt{\frac{3}{8}}\, M_P \ln\left[ \frac{1}{2} + \frac{k}{3}\left(k+\sqrt{k^2+3}\right) \right]\,.
\eeq

The inflaton field evolves according to its equation of motion
\begin{equation}
    \label{eq:eomphi}
    \ddot{\phi} + 3H \dot{\phi} + \frac{dV(\phi)}{d\phi} \; \simeq \; 0 \, ,
\end{equation}
where $H = \frac{\dot{a}}{a}$ is the Hubble parameter. For $a>\aend$, the total inflaton energy density, $\rho_\phi = \frac12\dot \phi^2+V(\phi)$, evolves as \cite{gkmo2}
\beq
\rho_\phi(a) \; = \; \rho_{\rm end} \left(\frac{a_{\rm end}}{a} \right)^{\frac{6k}{k+2}}  \, .
\label{Eq:rhophi}
\eeq
This scaling holds until the end of reheating, at which point the radiation energy density begins to dominate. We refine this description in the next section, where the choice of initial conditions is discussed in greater detail.

We assume that reheating occurs via the decay of the inflaton, which decays into Standard Model (SM) fields as the dominant channel. We note that our analysis does not depend on the details of the decay. For the purposes of this work, we define the end of reheating when the energy density of the inflaton field falls below the radiation energy density produced by its decay. The reheating temperature is then defined by the condition
\beq
 \rhorh \; \equiv \; \rho_{\rm R}(\arh) \; = \; \rho_\phi(\arh) \; \equiv \; \alpha \trh^4\, ,
 \label{TRH}
\eeq
where $\rho_{\rm R}$ is the energy density in radiation produced by the inflaton decay, $T_{\rm RH}$ is the reheating temperature, and $\alpha = \pi^2 g_*/30$, where $g_*$ denotes the effective number of relativistic degrees of freedom at reheating, with $g_* = 427/4$ in the SM for $T_{\rm RH} > m_t$.

The inflationary slow-roll parameters are relatively insensitive to the choice of $k$. However, the inflationary scale, determined by $\lambda$, does depend on $k$ and is fixed by the normalization of the CMB spectrum, with
\beq
\lambda \; \simeq \; \frac{18 \pi^2 A_s}{6^{k/2} N_*^2} \, ,
\eeq
where $A_s(k_*) = 2.1\times 10^{-9}$ is the amplitude of the scalar perturbations measured by Planck \cite{Planck:2018jri}, and $N_*$ is the number of $e$-folds between horizon exit of the scale with wave number $k_* = 0.05$~Mpc$^{-1}$, corresponding to $\phi=\phi_*$, and the end of inflation at $\phi = \phi_{\rm end}$. The energy density at the end of inflation, $\rhoe = \rho(\phiend) = \frac32 V(\phiend)$, is fixed by $\lambda$. We will return to the issue of setting these parameters in Section \ref{sec:initinfl}.

\subsection{The Evolution of the Spectator}
\label{sec:spectevol}
In addition to the inflaton, we assume the existence of an 
additional stable scalar field, $\chi$, commonly referred to as a spectator field, because it does not affect the inflationary dynamics and it is not directly coupled to the SM.
We further assume that $\chi$ is {\it minimally} coupled to gravity, has a canonical kinetic term, and possesses a potential of the form
\beq
\label{eq:spectpot}
V(\chi) \; = \;  \frac12 m_\chi^2 \chi^2 + \lambda_\chi M_P^4 \left(\frac{\chi}{M_P}\right)^p + \sigma_{n,m} \phi^n \chi^m M_P^{4-n-m} \, ,
\eeq
which includes a bare mass term, self-interactions ($p\ge 4$), and a potential coupling to the inflaton field $\phi$. 

The spectator field evolves according to its equation of motion\footnote{We neglect the $\sigma_{n,m} \langle \chi^m \phi^{n-2} \rangle \phi$ contribution in Eq.~(\ref{eq:eomphi}) because generally it is small compared with $m_\phi^2 \phi$.} 
\begin{equation}
    \label{eq:eomS}
  \ddot{\chi}+3H \dot{\chi}+\frac{d V(\chi)}{d\chi} \; = \; \ddot{\chi} + 3H \dot{\chi} + m_{\chi, \rm eff}^2 \chi \; = \; 0 \, ,
\end{equation}
where the effective mass of the spectator field $\chi$ is given by 
\begin{equation}
    \label{eq:chieffmass}
    m_{\chi, \rm eff}^2 \; = \; m_\chi^2  + p(p-1) \lambda_\chi \langle \chi^{p-2} \rangle M_P^{4-p} +  m (m-1) \sigma_{n,m}  \langle \phi^n \rangle \langle \chi^{m-2} \rangle M_P^{4-n-m} \, .
\end{equation}
Here $\left<\cdots \right>$ represents the field expectation value at a given moment. For now, we set the coupling $\sigma_{n,m}$ to $0$, and will revisit this case in Section~\ref{sec:specinfl}.

At the end of inflation, the inflaton begins to oscillate about the minimum of its potential, dominating the energy density of the universe until reheating occurs. Similarly, the spectator field, $\chi$, begins to oscillate at $t_{\rm osc}$ corresponding to $\meff(\aend) t_{\rm osc} \simeq 1$ or 
\beq
\frac{3k}{k+2}H(\aosc) \; = \; m_{\chi, \rm eff}(\aend) \; = \; m_{\chi, \rm eff}(\aosc) \, ,
\label{Eq:aosc}
\eeq
in an inflaton-dominated universe since $t_{\rm osc} = \frac{2}{3(1+w_\phi)}H^{-1}(\aosc)$ and $w_\phi = \frac{k-2}{k+2}$. Indeed, at $\aend$, $\chi = \chi(\aend) =\chi_{\rm end}$ and remains constant up to the oscillatory phase. The subsequent evolution of $\chi$ depends on its self-interactions or couplings to other fields, such as the inflaton. 

Since we have assumed that the inflaton field dominates the energy budget during the entire reheating epoch, the evolution of $\rho_\phi$ is given by Eq.~(\ref{Eq:rhophi}) for all values of $p$, and the Hubble parameter can then be expressed as $H(a)=\He\left(\frac{ \aend}{a}\right)^{\frac{3k}{k+2}}$ for $\aend < a <\arh$. Using $\frac{d}{dt}=aH\frac{d}{da}$, Eq. \eqref{eq:eomS} can be rewritten in terms of the scale factor as:
\begin{equation}
    \He^2 \left(\frac{a}{\aend}\right) ^\frac{4-4k}{k+2}\chi ''+\frac{k+8}{k+2}\He^2 \left(\frac{a}{\aend}\right) ^{\frac{2-5k}{k+2}}\chi' +\frac{dV(\chi)}{d\chi } \; = \; 0 \, , 
    \label{eomofchi}
\end{equation}
where we denote the derivative with respect to the scale factor as $' \equiv {a_{\rm end}} \frac{d}{da}$. 
  
However, for the evolution of $\rho_\chi$, simply taking
Eq.~(\ref{Eq:rhophi}) and replacing $\phi \to \chi$ and $k \to p$ is not necessarily valid. Intuitively, for sufficiently small $p$, $m_{\chi,\rm eff}^2\propto\chi^{p-2}$ redshifts more slowly than the Hubble parameter, which scales as $H^2 \propto \phi^k$. This implies that the envelope of $\chi$, denoted by $\chi_0$, can be approximated as nearly constant over one oscillation period, with $\dot{\chi}^2 \simeq p V(\chi)$. Consequently, the total energy density of the spectator field is given by $\rho_\chi = \frac{1}{2} \dot{\chi}^2 + V(\chi) \simeq V(\chi_0)$~\cite{gkmo1}. Under these conditions, Eq.~\eqref{eomofchi} leads to the following continuity equation:
\begin{equation}
    \dot{\rho}_\chi  + 3 (1+w_\chi )H\rho_\chi=aH \rho_\chi^\prime  + \frac{6p}{p+2}H\rho_\chi \; = \; 0 \, ,
\end{equation}
where $w_\chi\equiv \frac{p-2}{p+2}$. Therefore, in the case of small $p$ where one 
neglects  $H(a>\aosc) \ll m_{\chi,\rm eff}$, we 
find $\rho_\chi(a)\propto a^{-\frac{6p}{p+2}}$, as in Eq.~(\ref{Eq:rhophi}).

For larger $p$, the situation is very different. Indeed, in this case the frequency 
$m_{\chi,\rm eff}$ redshifts faster than the expansion rate $H$ and
this rapid change of frequency affects the solution for $\rho_\chi$. This can be understood by the fact that, between two oscillations of $\chi$, the Universe has expanded such that one cannot take a coherent mean over the oscillations of $\chi$, 
because of the rapid change of the frequency $\meff$. 
Technically, we cannot write $\dot \chi^2\simeq p V(\chi)$
in Eq.~(\ref{eq:eomS}), and thus we cannot neglect 
$\frac{d}{dt}\meff$ over a timescale $H^{-1}$. 

To determine the asymptotic behavior of $\chi$, we use the ansatz $\chi(a)=c(a)a^b$ where $b<0 $ accounts for damping and $c(a)$ encodes the oscillatory behavior. Specifically, $c(a)$ is a function that satisfies the initial conditions at $\aend$, and for $a \gg \aosc$, the oscillations slow down, allowing us to neglect the higher-order derivatives: $|c''(a)| \ll |c'(a)| \ll c(a) \simeq c_0$. With this approximation, Eq.~\eqref{eomofchi} simplifies to:
\begin{equation}
  (a^b c_0)^{p-1}\lambda_\chi  M_P^{4-p}p+\frac{\left( 6+2b+b k \right)\He^2 b c _0 }{2+k}a^{b-\frac{6k}{k+2}} \; \simeq \; 0 \, ,
\end{equation}  
where
\begin{equation}
    b \; = \; -\frac{6k}{(k+2)(p-2)} \,, \quad c_0=M_P^{\frac{p-4}{p-2}}\left(36\frac{\He^2k(p-k-2)}{\lambda_\chi p (k+2)^2 (p-2)^2}\right)^{\frac{1}{p-2}} \, .
\end{equation}
For $c_0$ to be real, $p$ must satisfy $p>k+2$. In this case, the energy density scales as \begin{equation}
    \rho_\chi \; \sim \; a^{-\frac{6p k}{(k+2)(p-2)}} \, ,\label{largepregime}
\end{equation}
for large $a$. Note that this regime is characterized by small $k$ compared to $p$, so the expansion of the universe is slower, resulting in a slower redshift of $\rho_\chi$ compared to the case with the same $p$ and larger $k$. As argued earlier, the latter corresponds to the fast oscillatory regime where the energy density of $\chi$ is given by $\rho_\chi\propto a^{-\frac{6p}{p+2}}$. We then impose 
\begin{equation}
     \frac{6p k}{(k+2)(p-2)} <\frac{6p}{p+2} \, ,
\end{equation}
which implies that $p>2k+2$ in order for the behavior \eqref{largepregime} to be valid. When $p=2k+2$, \eqref{largepregime} coincides with $\rho_\chi\propto a^{-\frac{6p}{p+2}}$. An alternative way to solve the $p\leq2k+2$ cases is to rewrite   
Eq.~\eqref{eomofchi} in the form:
\begin{equation}
   \He^2 \frac{d^2}{dz^2}\chi(z)+ \frac{(k+2)^2 }{36}z^{-k-2}\frac{dV(\chi)}{d\chi} \; = \; 0 \, ,
   \label{emden}
\end{equation}
with
\begin{equation}
      z \; = \; \left(\frac{a}{\aend}\right)^{-\frac{6}{k+2}} \, .
      \label{defz}
\end{equation}
This equation is a form of the Emden-Fowler equation~\cite{emden}. For the case $p\leq 2k+2$, there is not always an exact solution, but from the previous analysis, we know that oscillations cannot be neglected for large $a$. We can use the following ansatz in Eq.~\eqref{emden}:
\begin{equation}
      \chi(z) \; = \; A z^{\alpha}\sum_{n=-\infty}^{+\infty}  \mathcal{C}_n e^{i n \omega z^{\sigma}} \, ,
      \label{fourieransatz}
\end{equation}
where $\alpha >0$, $\mathcal{C}_n$ are Fourier coefficients, and $\omega$ is the frequency of $\chi$ oscillations. We are interested in the limit $a\rightarrow \infty$ or equivalently $z\rightarrow 0$. Requiring that the imaginary part of $\chi''(z)$ vanish and matching the dominant powers of $z$, we get 
\begin{equation}
    \alpha \; = \; \frac{k+2}{p+2},\quad \sigma \; = \; \frac{p-2-2k}{p+2} \leq0 \, .
    \label{alphasigma}
\end{equation} 
This implies that the envelope of $\chi$ follows the asymptotic behavior $\chi \propto z^{(k+2)/(p+2)} \propto a^{-6/(p+2)}$, leading to $\rho_\chi \propto a^{-6p/(p+2)}$ for sufficiently large $a$ and any $p\leq 2k+2$.

As a side remark, when $p = 2$, an exact solution to Eq.~\eqref{emden} exists for any $k$ in terms of Bessel functions:
\begin{equation}
    \chi(a) \; = \; \frac{1}{k}\left(\frac{\aend}{a}\right)^{\frac{3}{k+2}}\left(\frac{6k}{k+2} {\frac{\He}{m_\chi}}\right)^{\frac{1}{k}}\left[c_1\mathcal{J}_{1/k}(\gamma)\Gamma\left(\frac{1}{k}\right)+c_2\mathcal{J}_{-1/k}(\gamma)\Gamma\left(-\frac{1}{k}\right)\right] \, .
    \label{solchibaremass}
\end{equation}
Here, $\gamma\equiv \left(\frac{a}{\aend}\right)^{\frac{3k}{k+2}}\frac{k+2}{3k} {\frac{m_\chi}{\He}}=\frac{H_{\text{osc}}}{H(a)}$. The constants $c_{1,2}$ are are determined by the initial conditions, $\chi^2(\aend) = \langle \chi^2 \rangle_{\rm end}$ from Eq.~(\ref{eq:chi2}) and $\chi^\prime(\aend) = 0$. For the case of $k = 2$, the solution simplifies to:
\begin{equation}
\chi(a) =
\sqrt{\frac{3}{2}}\frac{\He^2}{4\pi m_\chi^2 } \left(\frac{\aend}{a} \right)^\frac32  \left[ 2m_\chi \cos \left(\frac{2m_\chi ((a/\aend)^{3/2}-1)}{3\He}\right)+3\He \sin\left(\frac{2m_\chi ((a/\aend)^{3/2}-1)}{3\He}\right)\right] ,
\end{equation}
so that at late times, the energy density $\rho_\chi(a) = \frac12 {\dot \chi}^2 + V(\chi) \propto a^{-3}$. 

For $a\gg \aosc$ (i.e.,~$\gamma\gg 1$), the first-order expansion of the Bessel function gives
\beq
\mathcal{J}_{\pm 1/k}(\gamma) \; \sim \; a^{-{\frac{3k}{2(k+2)}}} \, ,
\eeq
which implies that $\chi(a) \sim a^{-3/2}$, leading to $\rho_\chi \sim a^{-3}$ for $p = 2$, independent of $k$. This is consistent with the earlier result $\rho_\chi \propto a^{-6p/(p+2)}$. This scaling behavior will be used in the next section.

For future reference, we summarize the asymptotic behavior of the spectator field energy density $\rho_\chi$ for different values of $k$ and $p$ when $a > \aosc$:
\begin{equation}
    \rho_\chi \propto \left\{\begin{aligned}  & a^{-\frac{6p}{p+2}}\,,\quad  &p\leq 2k \, ,    \\&a^{-\frac{6p k}{(k+2)(p-2)}} \, .\quad & p>2k\, .
    \end{aligned}\right.
    \label{gensol}
\end{equation}
The two expressions coincide when $p=2k+2$.

\section{Initial Conditions}
\label{sec:init}

\subsection{Inflationary Parameters}
\label{sec:initinfl}
The initial conditions for the dynamics of the scalar system are determined by inflation. For definiteness, we adopt the T-model potentials~(\ref{Vatt}). The values of the inflationary parameters are determined by the compatibility with the amplitude and tilt of the scalar and tensor power spectra, as constrained by current {\em Planck}+BICEP/{\em Keck} CMB observations~\cite{BICEP:2021xfz}. These parameters depend on the number of $e$-folds $N_*$, which is influenced not only by the inflationary scale but also by the post-inflationary expansion history, through the relation~\cite{Liddle:2003as,Martin:2010kz}
\beq
N_* \;=\; \ln\left[\frac{1}{\sqrt{3}}\left(\frac{\pi^2}{30}\right)^{1/4}\left(\frac{43}{11}\right)^{1/3}\frac{T_0}{H_0}\right] - \ln\left(\frac{k_*}{a_0H_0}\right) - \frac{1}{12}\ln\,g_{\rm RH} + \frac{1}{4}\ln\left(\frac{V_*^2}{\rho_{\rm end}M_P^4}\right) + \ln\,R_{\rm rad}\,.
\eeq
Here, $H_0=67.36\,{\rm km}\,{\rm s}^{-1}{\rm Mpc}^{-1}$~\cite{Planck:2018vyg}, $T_0=2.7255\,{\rm K}$~\cite{Fixsen:2009ug}, $a_0=1$ are the present Hubble  parameter, CMB temperature and scale factor, respectively. The reheating parameter
$R_{\rm rad}$ is defined as~\cite{Martin:2010kz}
\beq
R_{\rm rad} \;=\; \frac{a_{\rm end}}{a_{\rm rad}}\left(\frac{\rho_{\rm end}}{\rho_{\rm rad}}\right)^{1/4}\,,
\eeq
where $\rho_{\rm rad}$ and $a_{\rm rad}$ the energy density and scale factor at the beginning of the radiation-domination epoch. For practical purposes, we define this point as $|w(a_{\rm rad})-1/3|=\delta \ll 1$, where $w$ denotes the equation of state parameter of the universe.\footnote{Note that $a_{\rm rad}$ may differ from $\arh$. The equation of state at the end of reheating is defined as $w(\arh) = \frac23 \frac{k-1}{k+2}$.} 

To evaluate the contribution of reheating to $N_*$, we assume for definiteness a Yukawa-like coupling $\mathcal{L} \supset y \phi f \bar{f}$ for inflaton decay, where $f$ denotes a fermionic final state. We solve the perturbative continuity equations
\begin{align}\label{eq:reh1}
\dot{\rho}_{\phi} + 3(1+w_{\phi})H\rho_{\phi} \;&=\; -(1+w_{\phi})\Gamma_{\phi}\rho_{\phi}\,,\\ 
\label{eq:reh2}
\dot{\rho}_{R} + 4H\rho_{R} \;&=\; (1+w_{\phi})\Gamma_{\phi}\rho_{\phi}\,,
\end{align}
where $\rho_R$ is the energy density of the relativistic decay products. 
While we do not assume instantaneous reheating, we do assume instantaneous thermalization. Eq.~(\ref{eq:reh1}) is obtained as the oscillation-averaging the inflaton equation of motion~\cite{gkmo1,gkmo2}. The decay rate for a fermionic final state, $\Gamma_{\phi}$, neglecting kinematic blocking, is given by
\beq
\Gamma_{\phi} \;=\; \frac{y_{\rm eff}^2(k)}{8\pi}m_{\phi}(t)\,,
\eeq
where $y_{\rm eff}(k)\propto y$ is the effective Yukawa coupling averaged over the oscillations, and the effective inflaton mass is defined by
\beq
m_{\phi}^2(t) \;=\; V''(\phi_0(t))\,,
\eeq
with $\phi_0(t)$ representing the decaying envelope of the inflaton oscillations (see~\cite{gkmo2} for further details). Following Refs.~\cite{Garcia:2023dyf,EGNO5}, we assume a decay coupling of $y = 10^{-2}$ and $\delta=10^{-2}$, which ensures that radiation dominates over the inflaton at late times, that inflaton fragmentation is minimal (see Section~\ref{sec:fragm}), and that the reheating temperature is well above the Big Bang Nucleosynthesis (BBN) threshold. 

Table~\ref{tab:evo} summarizes the resulting inflationary parameters for T-models with $k = \{2, 4, 6,$ $8, 10\}$. The corresponding scalar spectral tilt, $n_s$, and tensor-to-scalar ratio, $r$, are also included. Generally, we find that $N_* \lesssim 56$ for $k < 4$ and $N_* \gtrsim 56$ for $k > 4$. The case $k = 4$ is special, as $R_{\rm rad} \simeq 1$ regardless of the duration of reheating.

\begin{table}[t!]
\centering
\begin{tabular}{l|l|l|l|l|l|}
\cline{2-6}
& $k=2$ & $k=4$ & $k=6$ & $k=8$ & $k=10$ \\ 
\hline
\multicolumn{1}{||l|}{$\lambda$}                                         
    &$2.1 \times 10^{-11}$&$3.3 \times 10^{-12}$ & $5.3 \times 10^{-13}$ & $8.6 \times 10^{-14}$ & $1.4 \times 10^{-14}$
    \\ \hline
\multicolumn{1}{||l|}{$\phi_{\rm end}\ [M_P]$}   
    &$0.84$ & $1.52$ & $1.98$ & $2.32$ & $2.59$
    \\ \hline
\multicolumn{1}{||l|}{$|\dot{\phi}_{\rm end}|\ [M_P^2]$}  
     &$3.7 \times 10^{-6}$ & $3.4 \times 10^{-6}$ & $3.2 \times 10^{-6}$ & $3.2 \times 10^{-6}$ & $3.1 \times 10^{-6}$
     \\ \hline
\multicolumn{1}{||l|}{$\rhoe^{1/4}\ [{\rm GeV}]$}  
&$5.2 \times 10^{15}$ & $4.9 \times 10^{15}$ & $4.8 \times 10^{15}$ & $4.8 \times 10^{15}$ & $4.8 \times 10^{15}$
\\ \hline
\multicolumn{1}{||l|}{$H_{\rm end}\ [M_P]$}  
&$2.6 \times 10^{-6}$ & $2.4 \times 10^{-6}$& $2.3 \times 10^{-6} $ & $2.2 \times 10^{-6}$ & $2.2 \times 10^{-6}$
   \\ \hline
\multicolumn{1}{||l|}{$H_I\ [M_P]$}  
& $6.5 \times 10^{-6}$ & $6.3 \times 10^{-6}$ & $6.2 \times 10^{-6}$ & $6.1 \times 10^{-6}$ & $6.1 \times 10^{-6}$
\\ \hline
\multicolumn{1}{||l|}{$\phi_*\ [M_P]$}   
    & 6.08 & 6.96 & 7.49 & 7.86 & 8.14
    \\ \hline
\multicolumn{1}{||l|}{$N_*$}  
& $54.0$ & $55.9$ & $57.5$ & $58.2$ & $58.4$
\\ \hline
\multicolumn{1}{||l|}{$n_s$}  
&$0.963$ & $0.964$ & $0.965$ & $0.966$ & $0.966$
\\ \hline
\multicolumn{1}{||l|}{$r$}  
&$0.004$ & $0.004$ & $0.004$ & $0.004$ & $0.004$
\\ \hline
\multicolumn{1}{||l|}{$y_{\rm eff}/y$}  
&$1$ & $0.712$ & $0.603$ & $0.540$ & $0.496$
\\ \hline
\multicolumn{1}{||l|}{$T_{\rm RH}\ [{\rm GeV}]$}  
&$4.3\times 10^{12}$ & $2.4\times 10^{10}$ & $6.8\times 10^{8}$ & $9.8\times 10^{8}$ & $3.7\times 10^{9}$
\\ \hline
\end{tabular}
\caption{\em {Inflationary parameters used in the numerical evolution of the fields are summarized below. For these calculations, we have fixed the Yukawa-like coupling to $y = 10^{-2}$. }}
\label{tab:evo}
\end{table}

\subsection{Initial Conditions for Reheating}
\begin{figure}[t!]
  \centering
\includegraphics[width=0.82\textwidth]{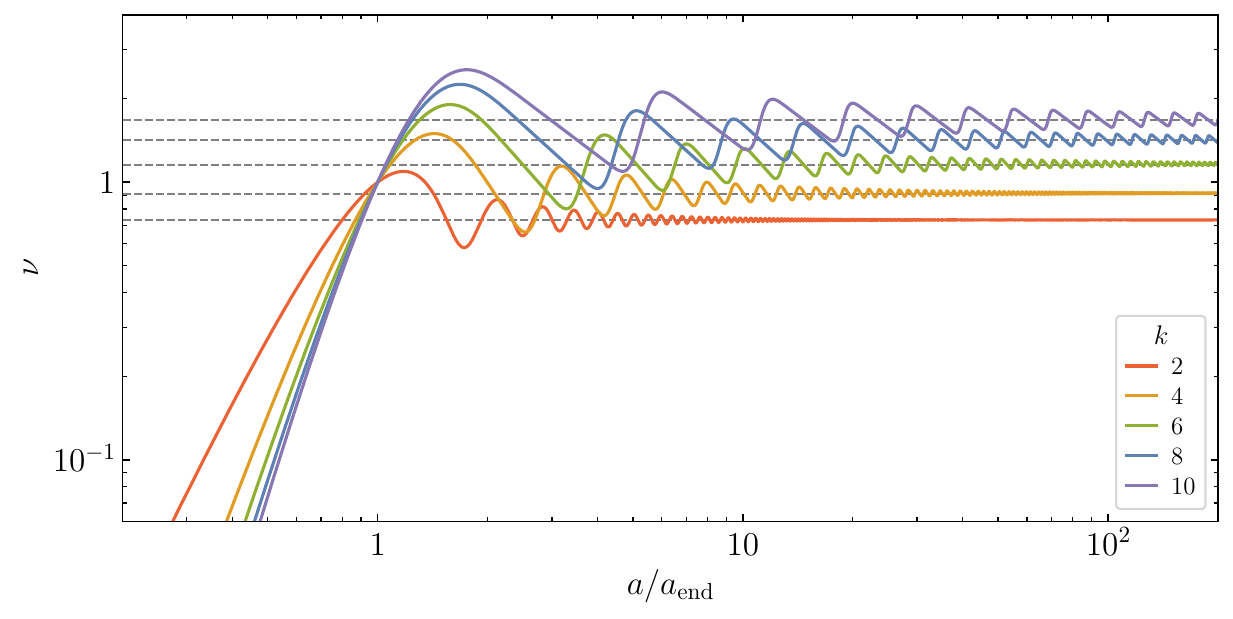}
  \caption{\em \small {Ratio of the instantaneous energy density of the inflaton relative to the power-law behavior with the initial condition $\rho_{\rm end}$ for different choices of $k$, as defined in (\ref{Eq:rhophi2}). The dashed lines represent the rescaled energy density evaluated at the first instance when $w_{\phi}=1$, marking the ``true" onset of oscillations. The parameters chosen here correspond to those listed in Table~\ref{tab:evo}. }
}
  \label{fig:rhoend}
\end{figure}

The monomial approximation (\ref{Eq:potmin}) to the full inflationary potential (\ref{Vatt}) is strictly valid only after the onset of $\phi$-oscillations around the minimum of the potential. At the end of the accelerated expansion, the inflaton is still displaced from this minimum, with the instantaneous value of the equation of state parameter given by
\beq
w_{\phi} \;=\; \frac{P_{\phi}}{\rho_{\phi}} \;=\; \frac{\dot{\phi}^2 - 2V(\phi)}{\dot{\phi}^2 + 2V(\phi)}\,,
\eeq
equal to $-1/3$, which differs from the oscillation-averaged value~\cite{Turner:1987vd, Martin:2010kz}
\beq
\langle w_{\phi}\rangle \;=\; \frac{k-2}{k+2} \,.
\eeq
As a result, the monomial approximation provides a reasonable but inexact description of the scale factor dependence of $\rho_{\phi}$ during reheating, with the initial condition $\rhoe$. The magnitude of this discrepancy is illustrated in Fig.~\ref{fig:rhoend}, where the exact energy density of $\phi$ is rescaled using the monomial approximation (\ref{Eq:potmin}) (solid curves). For $k\leq 4$, 
the approximation underestimates the asymptotic behavior of $\rho_{\phi}$, while for $k\geq 6$, it overestimates it. The dashed horizontal lines in Fig.~\ref{fig:rhoend} represent the rescaled energy density of the inflaton evaluated at the first instance when $\phi$ reaches the minimum of $V(\phi)$, corresponding to $w_{\phi}=1$. This value better captures the late-time behavior of $\rho_{\phi}$, and can be considered as the ``true'' onset of oscillations.
As a result, Eq.~(\ref{Eq:rhophi}) is modified to
\beq
\rho_\phi(a) \; = \; \nu\rho_{\rm end} \left(\frac{a_{\rm end}}{a} \right)^{\frac{6k}{k+2}}  \, ,
\label{Eq:rhophi2}
\eeq
with
\beq
\nu \;=\; \{0.73, 0.91, 1.16, 1.42, 1.68\}\qquad \text{for}\qquad k=\{2,4,6,8,10\}\,,
\eeq
assuming the nominal values in Table~\ref{tab:evo}. Equivalently, $\rho_{\phi}=\nu\rho_{\rm end}$ should be considered as the initial condition for the continuity equations (\ref{eq:reh1})-(\ref{eq:reh2}).

If the self-coupling, $\lambda_\chi$ and the coupling $\sigma_{n,m}$ between the inflaton and the spectator field in Eq.~(\ref{eq:spectpot}) are sufficiently small to prevent strong nonperturbative preheating effects,\footnote{For a detailed analysis of preheating effects in the case of large couplings $\sigma_{n,m}$ with $n = m = 2$, see Ref.~\cite{Garcia:2022vwm}.} the production of the spectator field $\chi$ is dominated by long-wavelength superhorizon modes. Assuming that the energy density of $\chi$ remains subdominant to that of the inflaton throughout inflation, the expectation value of the spectator field at the end of inflation, with $\lambda_{\chi} = 0$, can be approximated in its asymptotic limit as~\cite{fluc, Racco:2024aac, Choi:2024bdn}
\beq
\label{eq:chi2}
 \langle \chi^2 \rangle_{\rm end} \; \simeq \; \frac{3 \beta H_{\rm end}^4}{8 \pi^2 m^2_{\chi, \rm eff}(a_{\rm end})} \, ,
\eeq
where 
\begin{equation}
\label{eq:expCASEIIA}
\beta \; = \;
\begin{cases} 
1 & \text{for~~} m_{\chi, \rm eff}(a_{\rm end}) \leq \frac{3}{2} H_{\rm end} \, , \\
\exp\left[-c \pi \left(\frac{m^2_{\chi, \rm eff}}{H_{\rm end}^2}-\frac94\right) \right] & \text{for~~} m_{\chi, \rm eff}(a_{\rm end}) > \frac{3}{2} H_{\rm end} \,.
\end{cases}
\end{equation}
Here, $m_{\chi, \rm eff}^2$ in Eq.~(\ref{eq:chieffmass}) is $\chi$-dependent, and $c \sim \mathcal{O}(1)$ constant~\cite{Chung:1998bt, Markkanen:2016aes, Chung:2018ayg, Racco:2024aac, Jenks:2024fiu}. This parameter depends on the details of the inflationary model and on the transition from the quasi-de Sitter regime to the minimum of the potential. For the T-models considered here, we take $c=1.25$. It is important to note that the asymptotic limit of $\langle \chi^2 \rangle$ is generally computed in a de Sitter background and characterized by the Hubble parameter during inflation, $H_I$. However, numerical analysis indicates that for plateau-like inflationary models, the subsequent evolution of $\chi$ is more accurately described using the Hubble parameter at the end of inflation, $H_{\rm end}$. A more detailed derivation of $\langle \chi^2 \rangle_{\rm end}$ for arbitrary (even) $p$ is provided in Appendix \ref{app:sa}.

\begin{figure}[t!]
  \centering
\includegraphics[width=\textwidth]{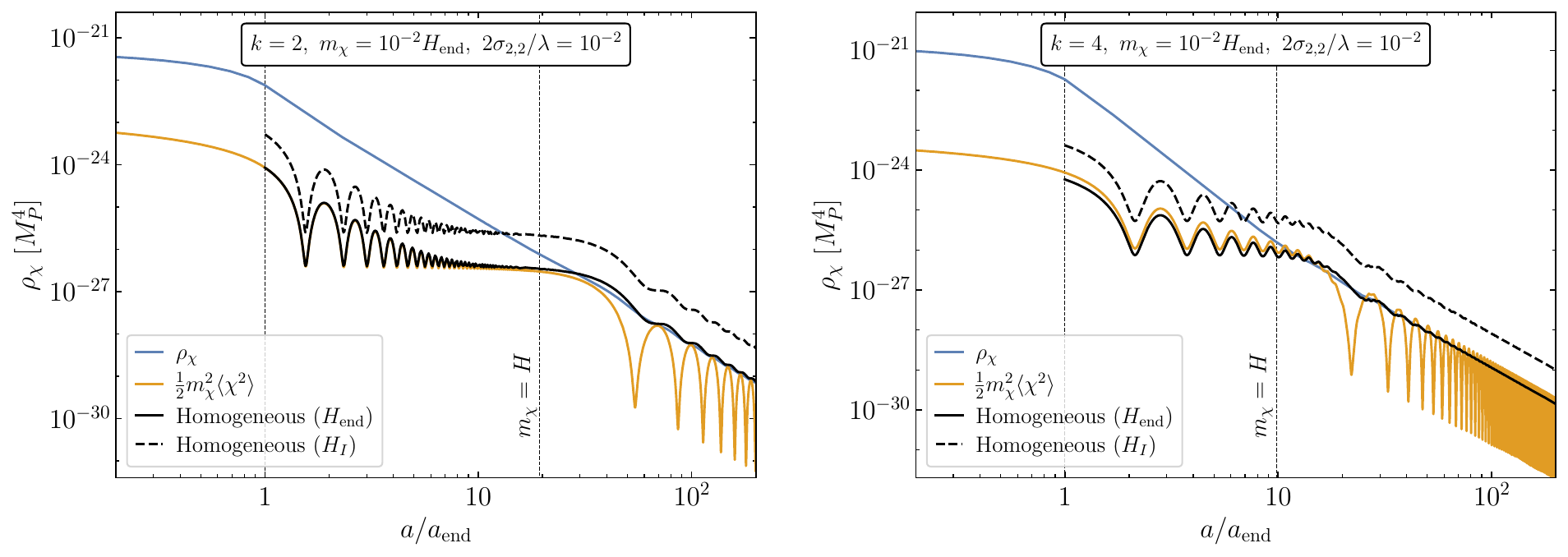}
  \caption{\em \small {Spectator energy density in three approximations. The solid blue curve represents the exact energy density given by Eq.~(\ref{eq:chirhoex}). The solid orange curve shows the potential component, calculated using the exact scalar field vev from Eq.~(\ref{eq:chirhoex}). The solid black curve illustrates the energy density assuming $\chi$ as a spatially homogeneous field, following Eq.~(\ref{eq:eomS}), with the initial condition from Eq.~(\ref{eq:chi2}) at $a = a_{\rm end}$ and $H_{\rm end}$. Finally, the dashed black curve represents the homogeneous energy density with the same initial condition (\ref{eq:chi2}), but using $H_I$ instead.}
}
  \label{fig:rhos1}
\end{figure}

Fig.~\ref{fig:rhos1} shows a comparison between the numerical solution to the spatially homogeneous equation of motion (\ref{eq:eomS}) with the initial condition given by Eq.~(\ref{eq:chi2}) and the total and potential energy densities obtained from solving the inhomogeneous equation:
\beq
\ddot{\chi} + 3H\dot{\chi} - \frac{\nabla^2\chi}{a^2} + m_{\chi,{\rm eff}}^2\chi \;=\;0\,.
\eeq
This equation is solved by imposing Bunch-Davies initial conditions deep inside the horizon for the mode functions $X_{\boldsymbol{k}}$ defined through canonical quantization, 
\beq
a(\tau)\chi(\tau,\bx) \;=\; X(\tau,\boldsymbol{x})  \;=\; \int \frac{d^3\bk}{(2\pi)^{3/2}}\,e^{-i\bk\cdot\bx} \left[ X_k(\tau)\hat{a}_{\bk} + X_k^*(\tau)\hat{a}^{\dagger}_{-\bk} \right]\,, 
\eeq
where $\tau$ denotes conformal time, related to cosmic time by $dt/d\tau=a$, and $\hat{a}^{\dagger}_{\bk}$, $\hat{a}_{\bk}$ are the creation and annihilation operators, satisfying $[\hat{a}_{\bk},\hat{a}^{\dagger}_{\bk'} ] = \delta(  \bk-\bk')$ and $[\hat{a}_{\bk},\hat{a}_{\bk'} ] = [\hat{a}^{\dagger}_{\bk},\hat{a}^{\dagger}_{\bk'} ] = 0$. The field expectation value and its energy density are computed by integrating over momentum modes:
\beq\label{eq:chirhoex}
\langle \chi^2\rangle \;=\; \frac{1}{a^2}\int \frac{d^3 \bk}{(2\pi)^3}\left(|X_{\bk}|^2 - \frac{1}{2\omega_{\bk}}\right) \,,\qquad \rho_{\chi} \;=\; \frac{1}{a^4}\int \frac{d^3\bk}{(2\pi)^3}\,\omega_{\bk}n_{\bk}\,,
\eeq
where the angular frequency of the modes and their occupation number are given by
\beq
\omega_{\bk}^2 \;=\; k^2 - \frac{a''}{a} + a^2m_{\chi,{\rm eff}}^2\,,\qquad n_{\bk} \;=\; \frac{1}{2\omega_{\bk}}\left|\omega_{\bk}X_{\bk} - iX_{\bk}'\right|^2\,,
\eeq
where primes denote differentiation with respect to $\tau$ (see e.g.~\cite{Garcia:2022vwm} for details). 

The parameters used in Fig.~\ref{fig:rhos1} are $k=2$, $m_\chi= 0.01 \He$, and $2 \sigma_{2,2} = 0.01 \lambda$. These were chosen to simplify the numerical integrations needed to produce the plot. For example, choosing smaller values of $\sigma_{2,2}$ results in the need for excessively long integration times to achieve a stable result, while for larger values of $\sigma_{2,2}$, special care must be taken to account for potential resonance effects. We see in Fig.~\ref{fig:rhos1} that $H_{\rm end}$ provides a better fit for the initial conditions of the homogeneous approximation compared to $H_I$. Additionally, this approximation closely matches the exact energy density after the field $\chi$ becomes non-relativistic and undergoes coherent oscillations. For this reason, we use $H_{\rm end}$ to determine the initial conditions for $\langle \chi^2 \rangle_{\rm end}$ rather than $H_I$, as detailed in Appendix \ref{app:sa}.

\section{Constraints}
\label{sec:constraints}

\subsection{Isocurvature Constraints}
\label{sec:isocurvature}
In this section, we address the stringent isocurvature constraints that limit the effective mass of the spectator field, $m_{\chi, \rm eff}^2$, the self-interaction term, $\lambda_{\chi}$, and the coupling between the inflaton and the spectator field, $\sigma_{n,m}$. A detailed derivation, based on the stochastic approach~\cite{Starobinsky:1986fx, Starobinsky:1994bd, Kamenshchik:2021tjh}, is provided in Appendices~\ref{app:sa} and~\ref{app:corfuncts}.

The current constraints on the isocurvature power spectrum from Planck are
\begin{equation}
    \beta_{\rm iso} \; \equiv \; \frac{\mathcal{P}_{\mathcal{S}}(k_*)}{\mathcal{P}_{\mathcal{R}}(k_*) + \mathcal{P}_{\mathcal{S}}(k_*)} < 0.038 \, ,
\end{equation}
at $95 \%$ CL, with the pivot scale $k_* = 0.05 \, \rm{Mpc}^{-1}$~\cite{Planck:2018jri}. Here $\mathcal{P}_{\mathcal{R}}$ denotes the curvature spectrum, and $\mathcal{P}_{\mathcal{S}}$ represents the isocurvature power spectrum. If the spectator field is lighter than the inflationary scale during inflation, it can generate isocurvature modes that may conflict with these constraints~\cite{Chung:2004nh, Garcia:2023awt}. This translates into an upper limit~\cite{Planck:2018jri}:
\begin{equation}
    \label{eq:isocurvatureupperlimit}
    \mathcal{P}_{\mathcal{S}}(k_*) < 8.3 \times 10^{-11} \, .
\end{equation}

The curvature power spectrum for a massive scalar field can be expressed as
\begin{equation}
   \mathcal{P}_{\mathcal{R}}(k) \; = \; \frac{H_I^2}{4\pi^2} \left(\frac{k}{a H_I} \right)^{3-2\sqrt{\frac{9}{4} - \frac{m^2_{\chi, \rm eff}}{H_I^2}}} \, ,
\end{equation}
valid for $m^2_{\chi, \rm eff}/H_I^2 \lesssim 9/4$. The isocurvature power spectrum is given by
\begin{equation}
\mathcal{P}_{\mathcal{S}}(k) \; = \; \frac{k^3}{2 \pi^2 \langle \rho_\chi \rangle^2} \int d^3 x\left\langle\delta \rho_\chi(x) \delta \rho_\chi(0)\right\rangle e^{-i \boldsymbol{k} \cdot \boldsymbol{x}} \, ,
\end{equation}
where $\rho_{\chi}$ is the spectator field energy density and $\delta \rho_{\chi}$ represents its fluctuations. For the quadratic case with $p=2$, $\rho_{\chi}(x) \propto \chi^2(x)$, and assuming Gaussianity, we approximate
\begin{equation}
\frac{\langle\delta \rho_{\chi}(x) \delta \rho_{\chi}(0)\rangle}{\langle\rho_{\chi}\rangle^2}=\frac{\left\langle\chi^2(x) \chi^2(0)\right\rangle-\left\langle\chi^2\right\rangle^2}{\left\langle\chi^2\right\rangle^2}=2 \frac{\langle\chi(x) \chi(0)\rangle^2}{\left\langle\chi^2\right\rangle^2} \, .
\end{equation}
This allows the isocurvature power spectrum to be rewritten as:
\begin{equation}
\begin{aligned}
   \mathcal{P}_{\mathcal{S}}(k) & =  \frac{k^3}{2 \pi^2} \frac{2}{\left\langle\chi^2\right\rangle^2} \int d^3 x e^{-i \boldsymbol{k} \cdot \boldsymbol{x}} \langle\chi(x) \chi(0)\rangle^2 \; = \; \frac{k^3}{2 \pi^2} \frac{2}{\left\langle\chi^2\right\rangle^2} \int \frac{d^3 q}{(2 \pi)^3} \frac{\mathcal{P}_{\mathcal{R}}(q)}{q^3} \frac{\mathcal{P}_{\mathcal{R}}(k-q)}{|\mathbf{k} - \mathbf{q}|^3} \, .
\end{aligned}   
\end{equation}
Following Refs.~\cite{Redi:2022zkt, Markkanen:2019kpv}, the integral can be evaluated analytically, yielding:
\begin{equation}
      \label{eq:isocurvspectrumfull}
      \mathcal{P}_{\mathcal{S}}(k) \; = \; \frac{H_{ \rm I}^4}{32} 
      \frac{2^{2\nu - 2}}{\pi^{3}} \frac{1}{\left\langle\chi^2\right\rangle^2} \left(\frac{k}{a_{\rm end} H_{I}} \right)^{2(n_\delta - 1)}\frac{\Gamma(2\nu)}{\Gamma(\nu)^2} \frac{\Gamma(2\nu - 3/2)}{\Gamma(2\nu)}  \frac{\Gamma\left(\frac{3}{2} - \nu\right)}{\Gamma(2 - \nu)} \, ,
\end{equation}
where $\Gamma(x)$ are the gamma functions, $n_{\delta} = 1 + 2 \frac{\Lambda_2}{H_{\rm end}}$ is the spectral tilt, $\nu = \frac{3 - (n_{\delta} -1)}{2}$, and $\Lambda_2$ are the eigenvalues obtained from the Fokker-Planck Eq.~(\ref{eqapp:fokkerplanck2}). The expectation value is given by Eq.~(\ref{inivalp}), while the eigenvalues $\Lambda_2$ are summarized in Table~\ref{tab:appeigen}. In general, these expressions are valid when the total number of $e$-folds of inflation, $N_{\rm tot}$, satisfies $N_{\rm tot} > H_I^2 / m_{\chi, \rm eff}^2$. This condition ensures that $\langle \chi^2 \rangle$ reaches a constant value due to the balance between stochastic inflationary fluctuations and the classical force driving the field toward the minimum of the potential. We remind the reader that, in the expressions for $\langle \chi^2 \rangle$, we use the Hubble parameter at the end of inflation, $H_{\rm end}$, instead of the Hubble parameter during inflation, $H_I$. As demonstrated in Ref.~\cite{Choi:2024bdn} and the previous section, fully numerical solutions for plateau-like inflationary models with a quasi-de Sitter phase exhibit better agreement when $H_{\rm end}$ is used instead of $H_I$.

For the case of a general potential $V(\chi)$, the isocurvature power spectrum can be computed directly in terms of an effective mass, $\meff^2 = \frac{d^2V}{d\chi^2}$, resulting in~\cite{Redi:2022zkt}
\begin{equation}
    \mathcal{P}_{\mathcal{S}}(k) \; \simeq \; \frac{8 m_{\chi, \rm eff}^2}{3H_{\rm end}^2} \left(\frac{k}{a_{\rm end} H_{I}} \right)^{\frac{4 m_{\chi, \rm eff}^2}{3H_{\rm end}^2}} \, , 
    \label{eq:isocurvanalytical}
\end{equation}
    where we expanded Eq.~(\ref{eq:isocurvspectrumfull}) assuming that $m_{\chi, \rm eff} \ll H_{\rm end} < H_{I}$. Using this expression, we observe that the modes that exit the horizon early during inflation will be strongly suppressed.  For the pivot mode $k_*$, which exits before the last $N_*$ $e$-folds of inflation, the isocurvature power spectrum can be approximated as:
\begin{equation}
    \mathcal{P}_{\mathcal{S}}(k_*) \; \simeq \; \frac{8 m_{\chi, \rm end}^2}{3H_{\rm end}^2} \exp\left(\frac{-4N_* m^2_{\chi, \rm eff}}{3H_{\rm end}^2} \right) \, .
\end{equation}

From this expression, we derive constraints on $m_{\chi, \rm eff}$ for different values of $N_*$. We note that there are two solutions that satisfy the isocurvature constraint. One of these solutions typically leads to a low mass, $m_\chi \ll H_{\rm end}$, and requires an extremely large total number of $e$-folds, $N_{\rm tot} > H_I^2 / m_{\chi, \rm eff}^2$.\footnote{Note that there is in principle no relation between $N_{\rm tot}$ and $N_*$.} For completeness, we list both solutions. For $N_* = 50$ and $N_* = 60$, the constraints are:
\begin{equation}
    \begin{aligned}
    \label{eq:isocorvconstrk2}
     &m_{\chi, \rm eff} (t_*) > 0.59H_{\rm end}&(50~\textrm{$e$-folds}) \,, \qquad  &m_{\chi, \rm eff} (t_*) > 0.54H_{\rm end}~&(60~\textrm{$e$-folds}) \,, \\
    \end{aligned}
\end{equation}
or the small mass solution is
\begin{equation}
         m_{\chi, \rm eff} (t_*) < 5.6 \times 10^{-6}H_{\rm end} \, .
         \label{lowmasssol}
\end{equation}

For cases dominated by the self-interaction term $\lambda_{\chi}$ in Eq.~(\ref{eq:spectpot}), the isocurvature power spectrum for $p > 2$ is given by
\begin{equation}
    \begin{aligned}
    \label{eq:isoconstrgen}
    & \mathcal{P}_{\mathcal{S}}(k_*) \; \simeq \; 1.8\sqrt{\lambda_{\chi}} \exp\left(-0.58\sqrt{\lambda_{\chi}} N_* \right) \,, \quad &(p=4)\, , \\
    & \mathcal{P}_{\mathcal{S}}(k_*) \; \simeq \; 1.0 \frac{H_{\rm end}^{2/3} \lambda_{\chi}^{1/3}}{M_P^{2/3}} \exp\left(-0.42\frac{H_{\rm end}^{2/3} \lambda_{\chi}^{1/3}}{M_P^{2/3}} N_* \right) \,, \quad &(p=6)\, , \\
    & \mathcal{P}_{\mathcal{S}}(k_*) \; \simeq \; 0.73 \frac{H_{\rm end} \lambda_{\chi}^{1/4}}{M_P} \exp\left(-0.35 \frac{H_{\rm end} \lambda_{\chi}^{1/4}}{M_P} N_* \right)\,, \quad &(p=8)\, , \\
    & \mathcal{P}_{\mathcal{S}}(k_*) \; \simeq \; 0.62 \frac{H_{\rm end}^{6/5} \lambda_{\chi}^{1/5}}{M_P^{6/5}} \exp\left(-0.32 \frac{H_{\rm end}^{6/5} \lambda_{\chi}^{1/5}}{M_P^{6/5}} N_* \right) \,, \quad &(p=10)\, . \\
    \end{aligned}
\end{equation}
In deriving these expressions, we used the expectation values $\langle \chi^2 \rangle$ from Eq.~(\ref{inivalp}), with $H_I$ replaced by $H_{\rm end}$. We also expanded the isocurvature power spectrum~(\ref{eq:isocurvspectrumfull}) in terms of $\left( \frac{H_{\rm end}}{M_P} \right)^{2-\frac{8}{p}} \lambda_{\chi}^{2/p} \ll 1$. Using the upper limit on the isocurvature power spectrum from Eq.~(\ref{eq:isocurvatureupperlimit}), we derive the following constraints on $\lambda_{\chi}$ for $N_*$ in the range of $50$–$60$ $e$-folds as well as the other solution which is mostly insensitive to $N_*$:
\begin{equation}
    \begin{aligned}
    \label{eq:isocorvconstr}
    & \lambda_{\chi} > 0.67~&(50~\textrm{$e$-folds}) \,, \qquad  &\lambda_{\chi} > 0.46~&(60~\textrm{$e$-folds}) \,, \quad &\lambda_{\chi} < 2.1 \times 10^{-21}\, ,&~(p=4)\, , \qquad \\
    & \lambda_{\chi} > \frac{1.4}{h_{\rm end}^2}~&(50~\textrm{$e$-folds}) \,, \qquad  &\lambda_{\chi} > \frac{0.77}{h_{\rm end}^2}~&(60~\textrm{$e$-folds}) \,, \quad &\lambda_{\chi} <  \frac{5.6\times 10^{-31}}{h_{\rm end}^2}\, ,&~(p=6) \,, \qquad  \\
    & \lambda_{\chi} > \frac{2.8}{h_{\rm end}^4}~&(50~\textrm{$e$-folds}) \,, \qquad  &\lambda_{\chi} > \frac{1.3}{h_{\rm end}^4}~&(60~\textrm{$e$-folds}) \,, \quad &\lambda_{\chi} <  \frac{1.6\times 10^{-40}}{h_{\rm end}^4} \, ,&~(p=8) \,, \qquad   \\
    & \lambda_{\chi} > \frac{5.9}{h_{\rm end}^6}~&(50~\textrm{$e$-folds}) \,, \qquad  & \lambda_{\chi} > \frac{2.3}{h_{\rm end}^6}~&(60~\textrm{$e$-folds}) \,, \quad &\lambda_{\chi} <  \frac{4.3\times 10^{-50}}{h_{\rm end}^6}\, ,&~(p=10) \,. \qquad 
    \end{aligned}
\end{equation}
Here, we defined the rescaled Hubble parameter as $H_{\rm end} = h_{\rm end} M_P$ for generality. In subsequent sections, we adopt the T-model of inflation with the potential given by Eq.~(\ref{Vatt}). 

If we have an interaction term $\sigma_{n,m}$ in Eq.~(\ref{eq:spectpot}), with $n = m = 2$, which dominates over the self-interaction term $\lambda_{\chi}$ and the bare mass term $m_{\chi}$, we can readily use the above constraints with $m_{\chi, \rm eff}(t_*) = \sqrt{2 \sigma_{2,2}} \phi_*$. Such scenario and its isocurvature constraints were studied in detail in~ Refs.~\cite{Garcia:2022vwm,Garcia:2023awt,Choi:2024bdn}. Values of $\phi_*$ are given in Table~\ref{tab:evo}.  From Eq.~(\ref{eq:isocorvconstrk2}), we find the constraint
\begin{equation}
    \begin{aligned}
    \label{eq:isocorvconstrsigma}
    &  \sigma_{2,2} > 0.17 \frac{H_{\rm end}^2}{\phi_*^2}&(50~\textrm{$e$-folds}) \,, \qquad  &\sigma_{2,2} >0.15 \frac{H_{\rm end}^2}{\phi_*^2}~&(60~\textrm{$e$-folds}) \,,
    \end{aligned}
\end{equation}
or from Eq.~(\ref{lowmasssol})
\begin{equation}
        \sigma_{2,2}  < 1.6 \times 10^{-11} \frac{H_{\rm end}^2}{\phi_*^2}\, .
\end{equation}
For the coupling $\frac{\sigma_{4,2}}{M_P^2}\phi^4 \chi^2$, we have \begin{equation}
    \begin{aligned}
    \label{eq:isocorvconstrsigma42}
    &  \sigma_{4,2} > 0.17 \frac{H_{\rm end}^2M_P^2}{\phi_*^4}&(50~\textrm{$e$-folds}) \,, \qquad  &\sigma_{4,2} >0.15\frac{H_{\rm end}^2M_P^2}{\phi_*^4}~&(60~\textrm{$e$-folds}) \,,
    \end{aligned}
\end{equation}
or from Eq.~(\ref{lowmasssol})
\begin{equation}
        \sigma_{4,2}  < 1.6 \times 10^{-11} \frac{H_{\rm end}^2M_P^2}{\phi_*^4}\, .
\end{equation}

In Ref.~\cite{Garcia:2023qab}, the authors demonstrated that isocurvature constraints can be avoided by introducing a nonminimal coupling term between the Ricci scalar, $R$, and the spectator scalar field in the Lagrangian, $\mathcal{L} \supset \frac{1}{2} \xi_{\chi} R \chi^2$. This coupling leads to an effective mass $m_{\chi, \rm eff}^2 \simeq -\xi_{\chi} R$, assuming that the bare mass term is negligible. During inflation, the Ricci scalar takes its de Sitter value, $R \simeq -12 H_I^2$. However, as before, we find that the analytical result aligns better with numerical solutions when $R \simeq -12 H_{\rm end}^2$ is used instead. Using the analytical approximation derived in Eq.~(\ref{eq:isocorvconstrk2}), the isocurvature constraints are satisfied for
\begin{equation}
    \begin{aligned}
    \label{eq:isocorvconstrxi}
    &  \xi_{\chi} > 0.03&(50~\textrm{$e$-folds}) \,, \qquad  &\xi_{\chi} > 0.02~&(60~\textrm{$e$-folds}) \,, 
    \end{aligned}
\end{equation}
or
\begin{equation}
        \xi_{\chi} < 2.6 \times 10^{-12} \, .
\end{equation}

\subsection{Second Period of Inflation Constraints}
\label{sec:secondinf}
In Section \ref{sec:init}, we discussed the initial conditions for the spectator condensate. After inflation, the spectator field $\chi$ evolves as a nearly homogeneous background with an energy density that remains constant until $\frac{3}{2} H \lesssim m_\chi$ (for $k = 2$). At this point, oscillations of $\chi$ begin at the scale factor $a = \aosc$. However, at $a = \aosc$, it is important that the energy density stored in $\chi$ does not dominate the total energy density of the Universe, as this would trigger a second period of inflation. If $m_\chi \ll m_\phi$, this secondary period of inflation would erase the inflationary perturbation spectrum generated by the inflaton and replace it with a new spectrum.
However, this new spectrum would have an amplitude significantly below that observed in the CMB~\cite{Graziani:1988bp,Enqvist:2003gh,Enqvist:2011pt,Choi:2024ruu}, making it inconsistent with current CMB observations and constraints.

To prevent this issue, it is necessary for oscillations to begin before the energy density of the $\chi$ condensate becomes dominant. Prior to the onset of these oscillations, the energy density of the spectator field at the end of inflation, $\rho_\chi = \rhochie$, remains constant. In contrast, the energy density of the inflaton (or radiation, if reheating has already occurred) dilutes with the expansion of the Universe. Consequently, the spectator field, despite its gravitational origin, poses a significant risk of dominating the energy density of the Universe. Furthermore, the complications arising from a second inflationary phase persist even if the $\chi$ field is unstable and decays at a later stage.

The condition that the spectator condensate remains subdominant for all $a < \aosc$ is equivalent to 
\beq
\rho(\aosc) \; >  \; \rho_\chi^{\rm end} \,,
\label{Eq:rhoosc}
\eeq
 and the start of oscillations is determined from Eq.~(\ref{Eq:aosc}). The energy density of the spectator field at the end of inflation is given by:
\begin{equation}
    \rho_{\chi}^{\rm end} \; = \; \frac{1}{2} m_{\chi, \rm eff}^2(a_{\rm end}) \langle \chi^2 \rangle_{\rm end} \, ,
\end{equation}
where,  assuming the self-interaction term $\lambda_{\chi}$ is subdominant, $\langle \chi^2 \rangle_{\rm end}$ is provided by Eq.~(\ref{eq:chi2}). Substituting these expressions along with Eq.~(\ref{Eq:aosc}) and Eq.~(\ref{Eq:rhoosc}), we obtain~\cite{Choi:2024ruu}:
\beq
\frac{3\beta }{16 \pi^2 } \frac{m_{\chi}^2(a_{\rm osc})}{m_{\chi}^2(a_{\rm end})}\He^4 \; < \; 3 H^2(\aosc) M_P^2 \; = \; \frac{(k+2)^2}{3k^2} m^2_{\chi, \rm eff}(a_{\rm osc}) M_P^2 \, ,
\eeq
or equivalently,
\beq
m_{\chi, \rm eff}(a_{\rm end}) \; > \; \frac{3 k \sqrt{\beta} \He^2}{4 (k+2) \pi M_P}
\; \simeq \; 2.0 \times 10^6 \sqrt{\beta} \left(\frac{\He}{6.4\times 10^{12}}\right)^2 \,{\rm GeV}\, ,
\label{no2inf1}
\eeq
where the latter expression is evaluated for $k=2$. When $\sigma_{n,m} = \lambda_{\chi} = 0$, the effective mass $m_{\chi, \rm eff}(a_{\rm osc})$ reduces to the bare mass term $m_{\chi}$ in these equations.

For $m_\chi \ll \He$, $\beta \simeq 1$, implying that the fluctuations of a spectator field with a mass lighter than $m_{\chi, \rm eff} < 2 \times 10^{6} \, \mathrm{GeV}$ would enter the horizon and begin oscillations too late to prevent a second period of inflation for the T-model of inflation. This condition is independent of the equation of state of the field dominating the total energy density and holds whether oscillations begin before or after the radiation-dominated reheating phase. The resulting lower bound on $\meff$ is consistent with the isocurvature constraints (\ref{eq:isocorvconstrk2}) and (\ref{lowmasssol}), leading to $m_{\chi, \rm eff} < 3.5 \times 10^{7} \, \rm{GeV}$ or $m_{\chi, \rm eff} > 3.2 \times 10^{12} \, \rm{GeV}$ for $k = 2$, with similar bounds applying to cases where $k > 2$.

When the self-interaction term dominates in the potential~(\ref{eq:spectpot}), we use Eq.~(\ref{inivalp}) to evaluate $\langle \chi^2 \rangle_{\rm end}$. To ensure that oscillations of $\chi$ begin before its energy density $\rho_\chi$ dominates the total energy density of the Universe, thereby avoiding a second period of inflation, we impose a constraint similar to Eq.~(\ref{no2inf1}) on the effective mass at $a = \aosc$:
\beq
m_{\chi,\text{eff}}(\aosc)> \frac{3k}{2k+4}\frac{\He^2 \sqrt{\beta}}{\sqrt{2} \pi M_P}
\left(\frac{\Gamma(3/p)}{\Gamma(1/p)}\right)^{\frac{p}{4}} \, .
\label{meffbound}
\eeq
Note that the lower bound on the effective mass decreases 
slightly for larger $p$ . Equivalently, for the self-coupling $\lambda_\chi$, the constraint is given by:
\beq
\lambda_\chi> 
\left(\frac{k}{k+2}\right)^p
\left(\frac{3}{p(p-1)}\frac{\Gamma(3/p)}{\Gamma(1/p)}\right)^\frac{p}{2}\, \left(\frac{3\He^4 \beta}{8 \pi^2 M_P^4}\right) \, ,
\eeq 
which gives
\begin{align}
        &&\lambda_\chi \gtrsim 7.7 \times 10^{-28}~\beta \,,  & \qquad (p=4) \, ,&& \\
        &&\lambda_\chi\gtrsim 8.8\times 10^{-31}~\beta \,, & \qquad (p=6) \, , &&\\
        &&\lambda_\chi\gtrsim 5.5\times 10^{-34}~\beta \,, & \qquad (p=8) \, ,&&
\end{align}
where we have evaluated $\He$ for $k=2$.
These values are extremely small because $\He \ll M_P$, ensuring that a second period of inflation will not occur for $p > 2$. Moreover, the lower bounds on $\lambda_\chi$ are significantly below the upper limits imposed by isocurvature constraints derived in the previous subsection. Thus, the condition is easily satisfied for all relevant cases.

\subsection{Fragmentation Constraints}
\label{sec:fragm}
For an inflaton potential (\ref{Eq:potmin}) with $k>2$, the coupling $\lambda$ not only determines the instantaneous effective mass of $\phi$, $m_{\phi}=\sqrt{V''(\phi)}$, but also governs the strength of its self-interactions. These self-interactions can lead to the excitation of spatially inhomogeneous fluctuations, $\delta\phi(t,\bx)$,  depleting the energy density of the homogeneous component $\phi(t)$~\cite{Greene:1997fu,Kaiser:1997mp,Garcia-Bellido:2008ycs,Frolov:2010sz,Amin:2011hj,Hertzberg:2014iza,Figueroa:2016wxr}. At linear order, and neglecting metric perturbations, the growth of these fluctuations is governed by the equation
\beq\label{eq:deltaphi}
\delta\ddot{\phi} + 3H\delta\dot{\phi} - \frac{\nabla^2\delta\phi}{a^2} + k(k-1)\lambda M_P^2 \left(\frac{\phi(t)}{M_P}\right)^{k-2}\delta\phi \;=\; 0\,,
\eeq
where the mode functions of $\delta\phi$ begin in the Bunch-Davies vacuum state to satisfy the commutation relations for the (quantum) field and its conjugate momentum. The oscillatory nature of the mass term in Eq.~(\ref{eq:deltaphi}) allows for the development of parametric resonance, where certain mode functions grow exponentially, pushing the system into the nonlinear regime.

Unlike the case of $k=2$, where parametric resonance is suppressed by the matter-like expansion ($w_\phi = 0$), for $k \geq 4$, the stiff equation of state causes the exponential growth of the fluctuations to accumulate during reheating, even when $\lambda$ is small. This effect enhances the instability of the inflaton field and leads to its fragmentation~\cite{Lozanov:2016hid,Lozanov:2017hjm,Garcia:2023eol,Garcia:2023dyf,Garcia:2024zir}. The homogeneous inflaton field transitions into an inhomogeneous distribution resembling a collection of inflaton quanta with nonzero momentum. This fragmentation process is accompanied by a shift in the average equation of state parameter
\beq
\langle w\rangle \;=\; \frac{k-2}{k+2}\;\longrightarrow\;\frac{1}{3}\,,
\eeq
even when $\rho_{\phi}\gg\rho_R$.

If fragmentation was complete, the effective mass of the inflaton particles would vanish, as shown by Eq. (\ref{eq:deltaphi}), preventing inflaton decay and thereby halting the reheating process. However, the growth of fluctuations is moderated by nonlinear mode-mode interactions, and the depletion of the zero mode $\phi(t)$ remains incomplete. As a result, reheating can continue in some cases, though with a modified decay rate. In such scenarios, the continuity equation (\ref{eq:reh2}) is replaced by:
\beq
\dot{\rho}_{R} + 4H\rho_{R} \;=\; \Gamma_{\delta\phi}m_{\phi}n_{\delta\phi} \;\simeq\; \Gamma_{\delta\phi}\left(\frac{m_{\phi}}{\bar{E}_{\phi}}\right)\rho_{\delta\phi}\,,
\eeq
where $\Gamma_{\delta\phi}=m_{\phi} y^2/8\pi$ is the decay rate of the free inflaton, $n_{\delta\phi}$, $\rho_{\delta\phi}$ denote the number density and energy density of these quanta, and $\bar{E}_{\phi}$ represents their mean energy. This fragmentation process results in a strong suppression of fermion production. For $k=4$, reheating can still occur after fragmentation, although with an increased duration and a substantial reduction in the reheating temperature, $T_{\rm RH}$. However, for $k\geq 6$, the effective decay rate $\Gamma_{\delta\phi}(m_{\phi}/\bar{E}_{\phi})$ redshifts faster than the Hubble rate, making reheating essentially impossible after fragmentation.

\begin{figure}[t!]
  \centering
\includegraphics[width=0.495\textwidth]{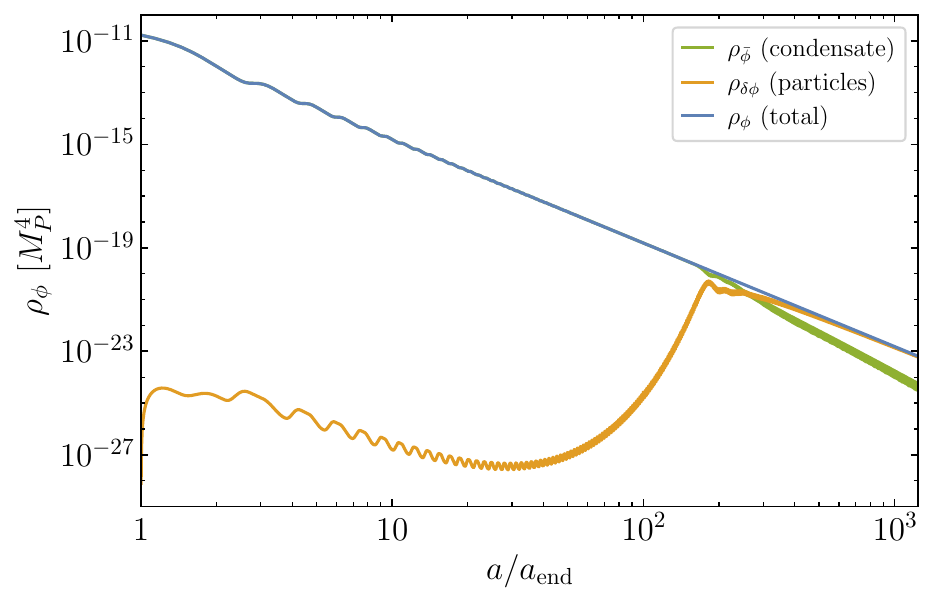}
\includegraphics[width=0.49\textwidth]{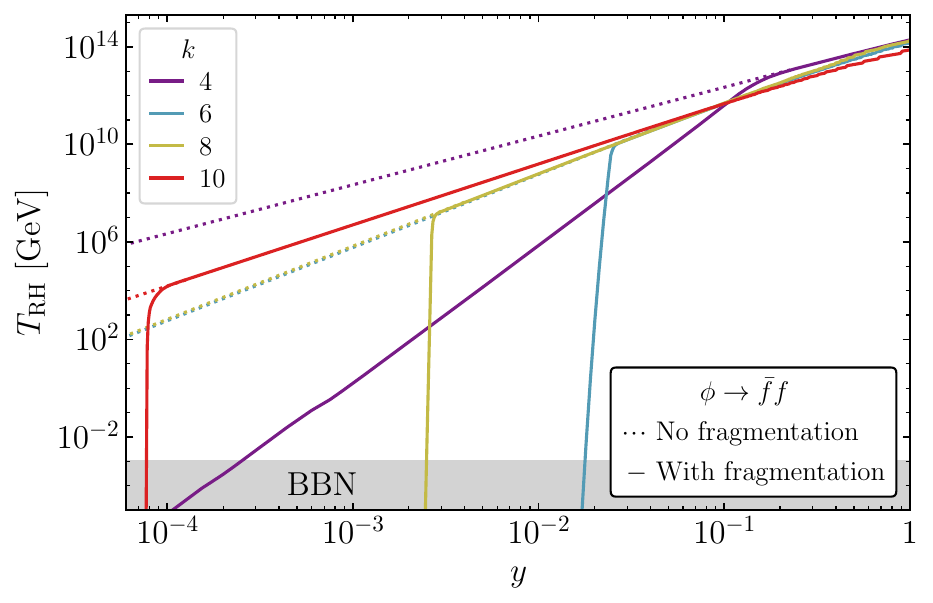}
  \caption{\em \small {Left: Evolution of the total inflaton energy density $\rho_{\phi}$ during reheating, along with the energy density of the spatially homogeneous condensate component $\rho_{\bar{\phi}}$ and the energy density of inflaton fluctuations $\rho_{\delta\phi}$, for $k = 4$, taking into account fragmentation effects. Right: Reheating temperature $T_{\rm RH}$ for inflaton decay into fermions as a function of the Yukawa coupling $y$.}
}
  \label{fig:frag}
\end{figure}

The left panel of Fig.~\ref{fig:frag} shows the evolution of the inflaton energy density during reheating for $k=4$, taking into account
the growth of inflaton fluctuations. For $a/a_{\rm end}\leq 10^2$, a spectral code using the linear approximation in Eq.~(\ref{eq:deltaphi}) tracks the energy density of the inflaton fluctuations, $\rho_{\delta\phi}$. This regime is characterized by the exponential growth of $\rho_{\delta\phi}$, driven by parametric resonance. During this phase, the energy density of the spatially homogeneous condensate component 
\beq
\rho_{\bar{\phi}} \;=\; \frac{1}{2}\bar{\dot{\phi}}^2 + V(\bar{\phi})\,,
\eeq
where $\bar{\phi}$ denotes the spatial average of $\phi$, comprises nearly the entirety of the inflaton energy density, i.e., $\rho_{\bar{\phi}}\simeq\rho_{\phi}$. For $a/a_{\rm end}>10^2$, the dynamics are tracked using the publicly available non-linear code \texttt{CosmoLattice}~\cite{Figueroa:2020rrl,Figueroa:2021yhd}. For $a/a_{\rm end}\gtrsim 180$, the backreaction of fluctuations becomes significant, and the fragmentation of the inflaton condensate into free inflaton particles is clearly observed.

The right panel of Fig.~\ref{fig:frag} depicts the effect of this fragmentation on the reheating temperature for fermionic final states. As mentioned above, for $k=4$, reheating remains possible after fragmentation, though it is significantly delayed due to the suppressed particle production rate, as observed for $y\lesssim 0.17$. For larger $k$, $T_{\rm RH}$ decreases sharply if inflaton decay is incomplete before fragmentation, with reheating becoming challenging for couplings $y \lesssim \{2.5 \times 10^{-2}, 2.7 \times 10^{-3}, 2.8 \times 10^{-5}\}$ for $k = \{6, 8, 10\}$, respectively. However, for these values of $k$, reheating temperatures above $\{1.2 \times 10^{10}~\mathrm{GeV}, 1.5 \times 10^7~\mathrm{GeV}, 1.8 \times 10^4~\mathrm{GeV}\}$ are sufficient to mitigate the effects of fragmentation~\cite{Garcia:2023dyf}. Consequently, we choose $y = 0.01$ to ensure a sufficiently high reheating temperature to avoid issues with fragmentation. Lower reheating temperatures are achievable but require tuning of $y$, as shown in the right panel of Fig.~\ref{fig:frag}. It is also worth noting that the effects of fragmentation are less severe for two-body decays into scalar particles, where the reduction in $T_{\rm RH}$ is at most an order of magnitude, allowing temperatures down to the electroweak scale. For further details, see~\cite{Garcia:2023eol, Garcia:2023dyf, Garcia:2024zir}.

As the spectator field also oscillates about its minimum, we must consider the possibility of fragmentation in this sector as well. A detailed treatment of the fragmentation of the scalar field is presented in Appendix~\ref{app:fragmentation}. In general, we follow the same procedure used for inflaton fragmentation. The self-interactions of the spectator field, characterized by the potential term $\lambda_{\chi} \chi^p$, lead to the excitation of fluctuations, $\delta \chi(t, \mathbf{x})$. However, the fragmentation of the spectator field is typically significantly suppressed. This suppression arises from the kinematic factors associated with the fragmentation process. Since the homogeneous condensate fragments into quanta $\delta \chi$ with mass $m_{\chi}$, only the higher modes of the spectator condensate contribute to fragmentation, as described by Eq.~(\ref{eq:fragmentation1}). The fragmentation rates $\Gamma_{\delta \chi}$ for the cases $p = 4, 6$, and $8$ are given by Eq.~(\ref{eq:fragp4})-(\ref{eq:fragp8}).

Thus, we expect that fragmentation of the spectator field will not significantly affect our analysis. To confirm this, we estimate the fragmentation rate and compare it to the Hubble parameter at the time when oscillations of the spectator field begin. The condition $\Gamma_{\delta \chi} < \frac{3k}{k+2}H_{\rm osc}= \meff(\aosc)$
imposes the following constraints on the coupling $\lambda_\chi$:
\begin{equation}
    \begin{aligned}
    & \lambda_\chi < 1.5 \times 10^{3} \,, \quad &(p=4)\, , \\
    & \lambda_\chi < 8.7 \times 10^{15} \,, \quad &(p=6)\, , \\
    & \lambda_\chi < 1.3 \times 10^{29} \,, \quad &(p=8)\, . \\
    \end{aligned}
\end{equation}    
Here, we have used the value for $\chi_{\rm end} = \sqrt{\langle \chi^2 \rangle}$, given by Eq.~(\ref{appeq:generalexpect6}), with $H_I$ replaced by $H_{\rm end}$ and taken for $k=2$ from Table~\ref{tab:evo}. These limits on $\lambda_{\chi}$ are significantly larger than the values considered in this work and correspond to a highly non-perturbative regime. Therefore, fragmentation of the spectator field remains negligible throughout our analysis and does not impose any additional constraints on the parameter space. This conclusion holds for all relevant values of $k$ and $p$, ensuring that the evolution of the spectator field remains predominantly governed by its homogeneous dynamics without the need to account for fragmentation effects.

\section{Non-Interacting Spectator Dark Matter
}
\label{Sec:spectator}
We begin our analysis by considering the simplest scenario of a non-interacting massive scalar field  $\chi$, with a potential given by $V(\chi)=\frac12 m_\chi^2\chi^2$. In the absence of any (non-gravitational) coupling to either SM matter or the inflaton, the spectator field is produced gravitationally. To analyze this process, we first distinguish between short-wavelength and long-wavelength modes.

\subsection{Production Through Graviton Exchange}
On small scales, the spectator field is produced through gravitational scattering of the inflaton~\cite{Garny:2015sjg,Tang:2017hvq,Chianese:2020yjo,MO,Barman:2021ugy,Haque:2021mab,cmov,Mambrini:2022uol,Haque:2023yra,hmo}. The scalar production rate is given by
\beq
\label{Eq:ratephi0}
R^{\phi^k}_0 \; = \; \frac{2 \times \rho_\phi^2}{16 \pi M_P^4} \Sigma_0^k \, ,
\eeq
where the factor of two accounts for the production of two dark matter particles per scattering, and
\begin{equation}
   \Sigma_0^k \; = \; \sum_{n = 1}^{\infty}  |{\cal P}^k_n|^2\left[1+\frac{2m^2_\chi}{E_n^2}\right]^2 
    \sqrt{1-\frac{4m_\chi^2}{E_n^2}} \, ,
    \label{Sigma0k}
\end{equation}
with $E_n = n \omega$ being the energy of the $n$-th inflaton oscillation mode, and $m_\chi$ the mass of the spectator field. The coefficients ${\cal P}^k_n$ are derived from the Fourier expansion of $V(\phi)$ in terms of $\rho_\phi$, given by
\beq
V(\phi) \; = \; V(\phi_0) \sum_{-\infty}^{+\infty}{\cal P}^k_ne^{-in\omega t}
\; = \; \rho_\phi\sum_{-\infty}^{+\infty}{\cal P}^k_ne^{-in\omega t} \,,
\label{Vexp}
\eeq
where $\phi_0$ is the envelope of $\phi(t)$. The frequency $\omega$ represents the inflaton oscillation frequency about its minimum and is given by~\cite{gkmo2}
\beq
\label{eq:angfrequency}
\omega \; = \; m_\phi \sqrt{\frac{\pi k}{2(k-1)}}
\frac{\Gamma(\frac{1}{2}+\frac{1}{k})}{\Gamma(\frac{1}{k})} \, ,
\eeq
with $m^2_\phi(\phi) = \frac{\partial^2 V(\phi)}{\partial \phi^2}$.
For $k=2$, this simplifies to $\omega = m_\phi$, making it independent of the field value. From ${\cal P}^2_2=\frac{1}{4}$, one obtains 
$\Sigma^2_0 = \frac{1}{16}$. For further details on this calculation, see \cite{gkmo2,cmov}.

The production rate (\ref{Eq:ratephi0}) appears in the Boltzmann equation for $\chi$:
\beq
   \frac{dY_{\chi}}{da} \; = \; \frac{a^{2} R^i_{\chi}(a)}{H} = \frac{\sqrt{3}M_P}{\sqrt{\rhorh}}a^2\left(\frac{a}{a_{\rm RH}}\right)^{\frac{3k}{k+2}}R^i_\chi(a) \, ,
\label{Eq:boltzmann4}
\eeq
where the comoving number density is defined as $Y_{\chi} \equiv n a^3$. In the second equality, we assume that the Hubble parameter $H$ is dominated by the inflaton energy density. The resulting density, evaluated at $\arh \gg \aend$ for scalars, is
\beq
n_0^\phi(a_{\rm RH}) \; \simeq \;
\frac{\sqrt{3}\rhorh^{3/2}}{8 \pi M_P^3}
\frac{k+2}{6k-6}
\left(\frac{\rhoe}{\rhorh}\right)^{1-\frac{1}{k}}\Sigma^k_0 \,.
\label{n0phi}
\eeq
The fraction of the present critical density can be obtained from $n(\arh)$ \cite{mybook}:
\beq
\frac{T_0^3}{\rho_c} \frac{g_0}{g_{\rm RH}} \; = \; 5.88 \times 10^6~{\rm GeV}^{-1} \, ,
\eeq
where $\rho_c^0=3M_P^2H_0^2 \simeq 8\times 10^{-47}h^2~{\rm GeV^4}$ is the present critical density, and $g_0$ and $\grh$ are the numbers of relativistic degrees of freedom at present and reheating times, respectively. Thus, the dark matter relic abundance can be expressed as
\beq
\frac{\Omega_\chi h^2}{0.12} \; = \; \frac{4.9 \times 10^7}{\rm GeV}~
\frac{\rho_\chi(\arh)}{\trh^3} \,.
\label{Omega}
\eeq
For scalars \cite{cmov}, the relic density contributions are given by:
\bea
&&
\frac{\Omega_\chi^{k=2}h^2}{0.12} \; \simeq \; 0.92 \left(\frac{\trh}{10^{10}~{\rm GeV}}\right) 
\left( \frac{m_\chi}{10^7 ~{\rm GeV}}\right) \, ,
\label{Eq:omega0k2}
\\
&&
\frac{\Omega_\chi^{k=4}h^2}{0.12} \; \simeq \; 8.3 \times 10^{4}\left( \frac{m_\chi}{10^7 ~{\rm GeV}}\right) \,,
\label{Eq:omega0k4}
\\
&&
\frac{\Omega_\chi^{k=6}h^2}{0.12} \; \simeq \; 3.6 \times 10^{6}\left(\frac{10^{10}~\rm GeV}{\trh}\right)^\frac13~\left( \frac{m_\chi}{10^7 ~{\rm GeV}}\right) \,.
\label{Eq:omega0k6}
\eea
These expressions depend on $\rhoe^{\frac{k-1}{k}}$ and use the values of $\rhoe$ from Table~\ref{tab:evo}.\footnote{The numerical coefficients assume $\trh \simeq 4.3 \times 10^{12}~{\rm GeV}$ for $k=2$. The coefficients have a small dependence on $\trh$~\cite{Liddle:2003as, Martin:2010kz, egnov}, aside from the explicit temperature dependence shown here.} This contribution to the relic density is referred to as the {\it short-wavelength} component, originating from inflaton scattering through graviton exchange. By conservation of energy, spectators with mass $m_\chi >m_\phi(a_{\rm end})$, corresponding to long-wavelength modes, cannot be produced by this channel. These long-wavelength modes are those that never exited the horizon during inflation, as their wavelengths remain smaller than the horizon size at the end of inflation. This equivalence has been demonstrated in~\cite{Kaneta:2022gug}. Note that for $k=4$, the relic abundance is independent of $\trh$, and excludes spectator masses $m_\chi \gtrsim 100$ GeV.

Of greater interest is the long-wavelength contribution to $\rho_\chi$. As discussed earlier, during inflation, $\langle \chi^2 \rangle \ne 0$, and at long wavelengths, the background condensate $\chi = \sqrt{\langle \chi^2 \rangle} $ evolves according to the equation of motion for a classical scalar field. Throughout inflation and until the condition $\frac{3k}{k+2} H(a) \simeq m_\chi$ is satisfied, the field remains frozen. Once this condition is met, the field begins to oscillate. If the scalar field is stable, the resulting energy density from these oscillations contributes to the cold dark matter density.

In the remainder of this section, we analyze the simplest case of a non-interacting massive scalar field $\chi$. In the absence of any couplings to either the SM or the inflaton, the only relevant parameters are $\trh$ and $m_\chi$, under the assumption that the inflationary parameters have been fixed by CMB measurements in the context of the chosen inflationary model. As we demonstrate, any reasonable choice of $(m_\chi,\trh)$ ensures that $\chi$-oscillations commence before the end of the reheating phase.

\subsection{Oscillations Begin During Reheating}
\label{sec:oscduringreh}
As long as Eq.~(\ref{no2inf1}) is satisfied, a second period of inflation driven by the spectator field does not occur. Instead, the spectator field begins to oscillate and eventually contributes to the dark matter density. We first consider the scenario where these oscillations begin before the end of reheating, $\aosc < \arh$. This condition corresponds to 
\bea
&&
H^2(\aosc) \; = \; { \frac{(k+2)^2}{9 k^2}} m_\chi^2 \gtrsim \frac{\rhorh}{3M_P^2}
~~\Rightarrow ~~\trh\lesssim \left( \frac{k+2}{\sqrt{3\a}k}\right)^\frac12\sqrt{m_\chi M_P} \, ,
\nonumber
\\
&&
\Rightarrow m_\chi \; \gtrsim \; 210\left(\frac{\trh}{10^{10}~\rm GeV}\right)^2~{\rm GeV}\,, \qquad (k=2) \,.
\label{Eq:mchimin}
\eea
Using $H(a) = \He (\aend/a)^\frac{3k}{k+2}$ determines the onset of oscillations
\beq
\frac{\aend}{\aosc} \; = \;  \left( \frac{(k+2) m_\chi}{3 k \He} \right)^\frac{k+2}{3k} \, .
\label{aendosc}
\eeq

Using the general inflaton potential (\ref{Eq:potmin}) and the result from the previous section (\ref{solchibaremass}), $\rho_\chi \propto a^{-3}$ for $a> \aosc$, the spectator field energy density at $\arh$ is given by
\begin{align}
\rho_\chi(\arh) & \; = \; 
\rho_\chi^{\rm end} \left( \frac{\aosc}{\arh} \right)^3 = \rho_\chi^{\rm end} \left( \frac{\aosc}{\aend} \right)^3
 \left( \frac{\aend}{\arh} \right)^3
\nonumber \\ & = 
\frac{3 \beta \He^4}{16 \pi^2}\left(\frac{3 k^2\alpha \trh^4}{(k+2)^2 M_P^2 m_\chi^2}\right)^\frac{k+2}{2k}\,,
\label{Eq:rhochiarh}
\end{align}
where we used Eqs.~(\ref{aendosc}), (\ref{Eq:rhophi}), and (\ref{TRH}). This implies the present-day energy density
\beq
\rho_\chi^0 \; = \; \frac{3 \beta\He^4}{16 \pi^2}
\left(
\frac{ 3 k^2 \alpha \trh^4}{(k+2)^2 M_P^2 m_\chi^2}
\right)^\frac{k+2}{2k}\frac{g_0T_0^3}{\grh \trh^3}\,.
\label{Eq:genericrhochi0}
\eeq

From Eq.~(\ref{Eq:genericrhochi0}), it follows that for a fixed reheating temperature, the density $\rho_\chi^0$ increases with $k$, provided that $m_\chi$ satisfies Eq.~(\ref{Eq:mchimin}) to ensure oscillations begin before reheating ends.\footnote{The term in parentheses in Eq.~(\ref{Eq:genericrhochi0}) is $(\aosc / \arh)^3$, and is always smaller than 1 if $\aosc < \arh$, as assumed here.} The dark matter abundance~(\ref{Omega}) can be expressed for different values of $k$ as~\cite{ema, Cosme:2020nac, Choi:2024bdn}:
\bea
&&
\frac{\Omega_\chi^{k=2}}{0.12} \; \simeq \; 0.71 \frac{\trh}{100~{\rm GeV}}\left(\frac{10^{12}~\rm GeV}{m_\chi}\right)^2
\left(\frac{\He}{6.4\times 10^{12}~\gev} \right)^4 \beta \,,
\label{Eq:casek2}
\\
&&
\frac{\Omega_\chi^{k=4}}{0.12} \; \simeq \; 4.8 \times 10^{12}\left(\frac{10^{12}~\rm GeV}{m_\chi}\right)^\frac32
\left(\frac{\He}{5.8\times 10^{12}~\gev} \right)^4 \beta \,,
\label{Eq:casek4}
\\
&&
\frac{\Omega_\chi^{k=6}}{0.12} \; \simeq \; 4.0 \times 10^{13}\left(\frac{10^{12}~\rm GeV}{\trh}\right)^\frac13~\left(\frac{10^{12}~\rm GeV}{m_\chi}\right)^\frac43
\left(\frac{\He}{5.5 \times 10^{12}~\gev} \right)^4 \beta \,.
\label{Eq:casek6}
\eea

\begin{figure}[ht!]
    \centering
    \includegraphics[width=\textwidth]{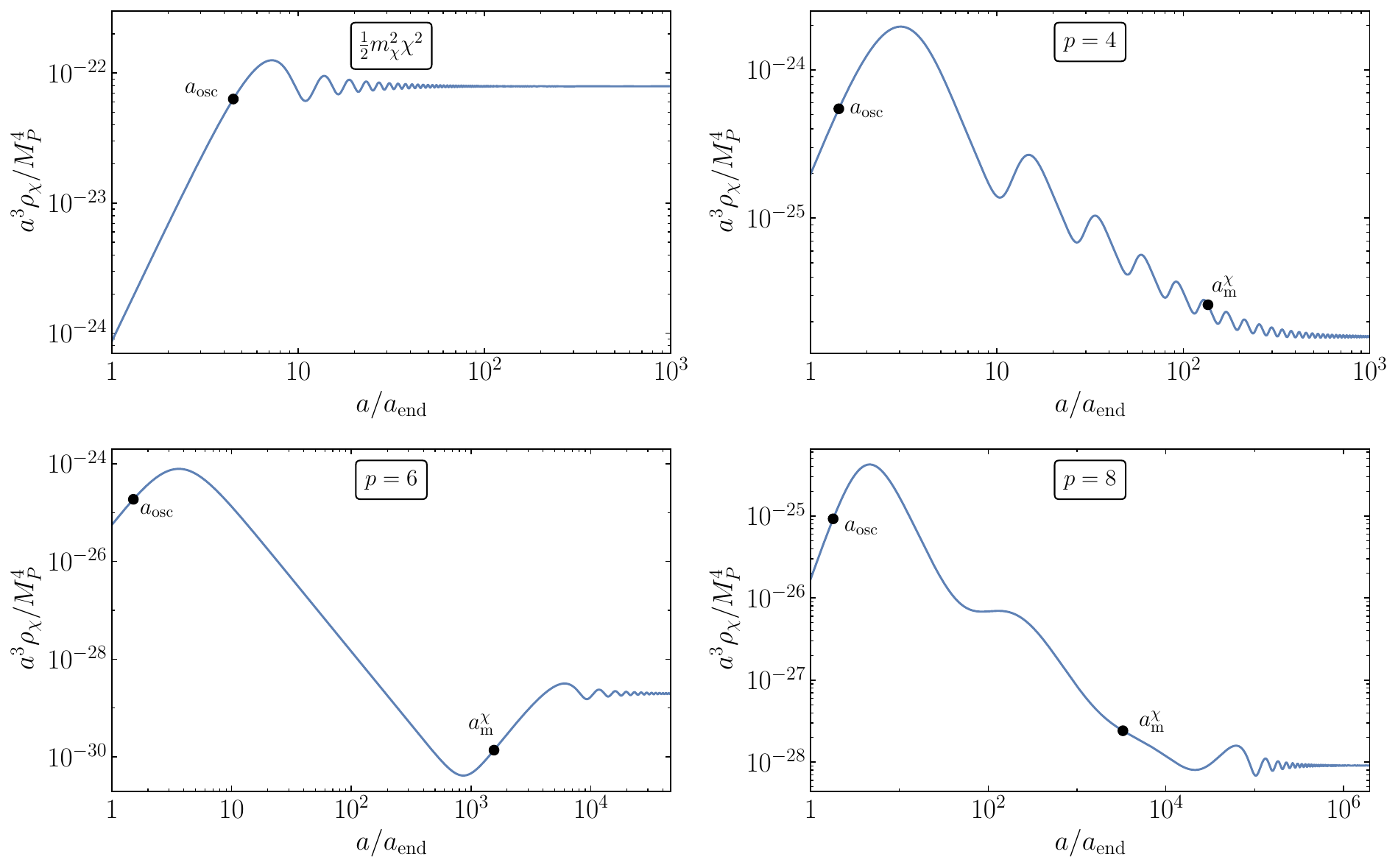}

    \caption{\it The evolution of the spectator energy density for a quadratic potential and a potential with self-interactions, $\chi^p$, with $p=4,6,8$, each including a mass term. For the quadratic potential, the parameters are $m_\chi = 10^{12}~\mathrm{GeV}$ and $\trh = 140~\mathrm{GeV}$. For $p = 4$, we set $\lambda_\chi = 1$, $m_\chi = 2.4\times 10^{10}~\mathrm{GeV}$, and $\trh = 10^{6}~\mathrm{GeV}$. For $p = 6$, the parameters are $\lambda_\chi = 10^{12}$, $m_\chi = 4\times 10^{7}~\mathrm{GeV}$, and $\trh = 10^9~\mathrm{GeV}$. For $p = 8$, we use $\lambda_\chi = 10^{24}$, $m_\chi =   10^{6}~\mathrm{GeV}$, and $\trh = 4.9\times 10^8\gev$. In all scenarios, the inflaton potential is assumed to be quadratic with $k = 2$.}
    \label{rhochinop}
\end{figure}

From the above equations, several key observations can be made. First, all expressions result in a significant overdensity for most reasonable parameter choices, except for $k = 2$, which remains viable with low reheating temperatures and large spectator masses~\cite{Choi:2024bdn}.\footnote{Minor numerical differences from \cite{Choi:2024bdn} arise due to our use of the T-model potential~\cite{Kallosh:2013hoa} instead of the Starobinsky model~\cite{Staro} employed there.} Additionally, this contribution to the density far exceeds the gravitational production described in Eq.~(\ref{Eq:omega0k2}). While the long-wavelength component can be exponentially suppressed when $\beta \ll 1$, this only occurs if $m_\chi \gtrsim \He$. In such cases, however, gravitational production becomes significant, as it can only be suppressed by phase-space effects when $m_\chi \simeq m_\phi$, requiring $\trh \lesssim 10^4$~GeV. For larger values of $k$, the long-wavelength contribution becomes increasingly dominant over the short-wavelength component.

The upper left panel of Fig.~\ref{rhochinop} shows the evolution of the comoving energy density $a^3 \rho_\chi$ (in Planck units) for $k = 2$, obtained by numerically solving Eq.~(\ref{eq:eomS}) with $m_\chi = 10^{12}$~GeV and $\trh = 140$~GeV. These values are chosen to yield the present relic density $\Omega_\chi h^2 \simeq 0.12$, as given by Eq.~(\ref{Eq:casek2}). The plot clearly illustrates that at early times ($a < \aosc$), the energy density remains constant. Once oscillations of $\chi$ begin at $a = \aosc \simeq 5 \, \aend$, the energy density scales as $\rho_\chi \propto a^{-3}$ for $a > \aosc$, in agreement with Eq.~(\ref{gensol}).

\begin{figure}[ht!]
  \centering
\includegraphics[width=0.90\textwidth]{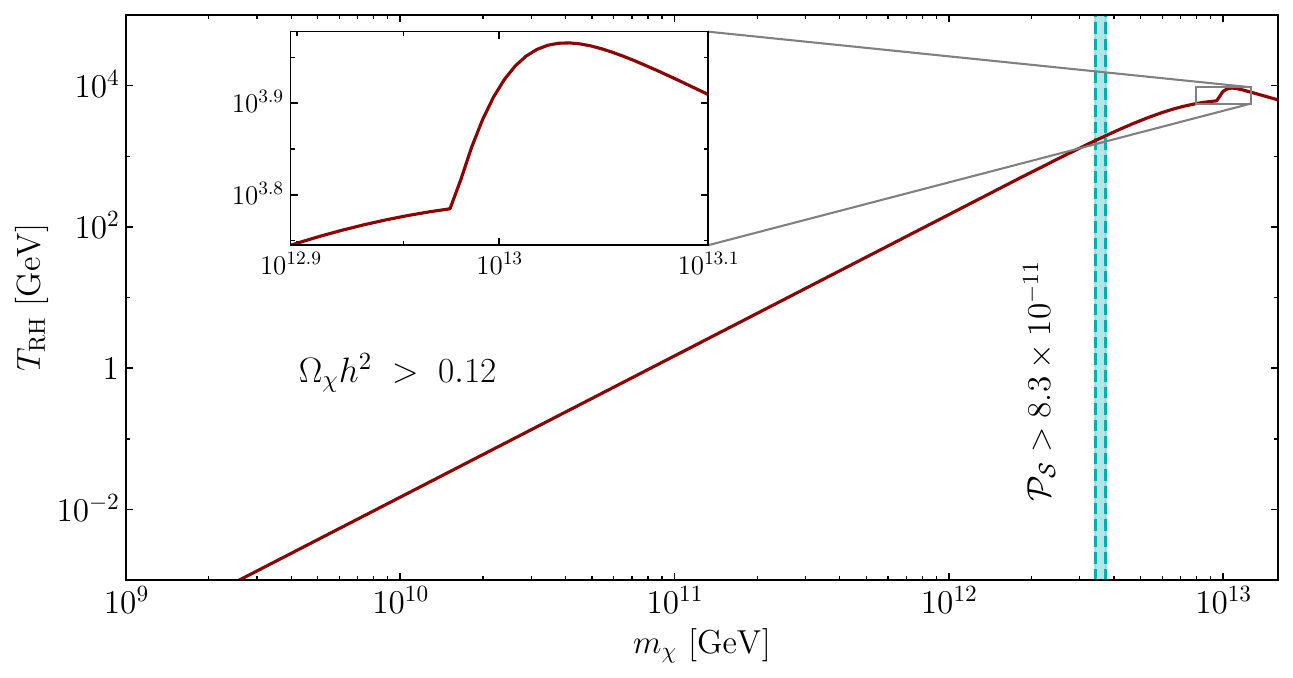}
  \caption{\em \small {The $(m_\chi,\trh)$ plane for $k=2$ is shown under the assumption that the spectator field has no self-interactions. The plot shows the total relic density, combining the short-wavelength (\ref{Eq:omega0k2}) and long-wavelength (\ref{Eq:casek2}) contributions, obtained from numerical integration of the Boltzmann equation. Points above the curve correspond to $\Omega_\chi h^2 > 0.12$. The vertical dashed lines correspond to the isocurvature constraint~(\ref{eq:isocorvconstrk2}) at 50 $e$-folds (right) and 60 $e$-folds (left).}
  }
  \label{mchitk2}
\end{figure}

The allowed region in the $(m_\chi, \trh)$ plane for $k=2$ is shown in Fig.~\ref{mchitk2}. The curve represents the points where $\Omega_\chi h^2 = 0.12$,  obtained from a numerical integration of the Boltzmann equations, which accounts for both the short-wavelength (\ref{Eq:omega0k2}) and long-wavelength (\ref{Eq:casek2}) contributions to the relic density. Points above the curve have $\Omega_\chi h^2 > 0.12$. 

The BBN constraint on $\trh \gtrsim 4$~MeV \cite{tr4} imposes a lower bound on the spectator mass, $m_\chi \gtrsim 7 \times 10^9$~GeV. However, we note that the isocurvature constraint imposes a more stringent lower bound on the spectator mass, requiring $m_{\chi} \gtrsim 3 \times 10^{12}~\mathrm{GeV}$, which is stronger than the bound derived from BBN. For masses $m_\chi \lesssim 10^{13}$~GeV, the long-wavelength contribution 
dominates. At slightly larger masses, the exponential suppression of the density becomes significant when $m_\chi \gtrsim \He$. As $m_\chi$ increases beyond this range, exponential suppression of the density becomes significant for $m_\chi \gtrsim \He$. However, at even higher masses, direct gravitational production of $\chi$ takes over, causing the upper limit on the reheating temperature to decline after reaching a maximum around $7~\mathrm{TeV}$. The entire parameter space shown in Fig.~\ref{mchitk2} satisfies the constraint in Eq.~(\ref{no2inf1}). In conclusion, this spectator dark matter model with $k = 2$ remains viable, though it requires a relatively high mass and low reheating temperature.

For $k=4$, it is not possible to satisfy the relic density constraint. Reducing the long-wavelength component with $\beta \ll 1$ requires a large spectator mass, but in this case, the gravitationally produced abundance, given by Eq.~(\ref{Eq:omega0k4})—which requires $m_\chi \lesssim 100~\mathrm{GeV}$ for any $\trh$—results in an excessive relic abundance. Consequently, it is impossible to meet the relic density constraint when both long- and short-wavelength contributions are considered. For cases with $k > 4$, the situation is even more problematic. Reducing $\beta$ does not resolve the issue, as gravitational production of $\chi$ through short-wavelength modes leads to inevitable overproduction for large $k$. In conclusion, only the case $k = 2$, with $m_\chi$ satisfying Eq.~(\ref{Eq:mchimin}) within the narrow region defined by Eq.~(\ref{Eq:casek2}) and illustrated in Fig.~\ref{mchitk2}, is viable for a quadratic potential $V(\chi)$.

One might ask how the parameter space for $m_\chi$ and $\trh$ can be expanded. A potential solution is entropy dilution through the injection of entropy from a late-decaying massive particle. However,
the preservation of the baryon asymmetry places an upper limit of order $\mathcal{O}(10^{10})$ on the amount of entropy dilution, and this assumes a
maximal baryon-to-entropy ratio (achievable e.g. in the Affleck-Dine mechanism~\cite{AD,Linde:1985gh,Campbell:1986qg,Garcia:2013bha}) is $n_b/s \sim \mathcal{O}(1)$. This constraint, combined with Eq.~(\ref{Eq:casek2}) for $k = 2$, relaxes the limit on the reheating temperature to $\trh \lesssim 10^{12}~\mathrm{GeV}$ (for $m_\chi = 10^{12}~\mathrm{GeV}$) to avoid dark matter overproduction. The allowed reheating temperature range becomes even narrower for smaller spectator masses. For $k>2$, even maximal entropy injection is insufficient to open a viable parameter space for $m_\chi$, particularly when the short-wavelength gravitational production, as described in Eqs.~(\ref{Eq:omega0k4}) and (\ref{Eq:omega0k6}), is taken into account.

Another possibility is that inflaton condensate fragmentation could affect these results, as discussed in the previous section. It has been shown in \cite{Garcia:2023eol,Garcia:2023dyf,Garcia:2024zir} that for $k>2$, the inflaton condensate can fragment due to its self-coupling. This fragmentation transforms the inflaton condensate $\phi$ into a gas of relativistic inflaton particles $\delta \phi$, effectively changing the equation of state. Specifically, the potential $V\propto \phi^k$ transitions to $V\propto \phi^4$, corresponding to $w_\phi \rightarrow 1/3$. However, this does not resolve the overdensity issue, as spectator dark matter remains overabundant even in a radiation-dominated Universe, equivalent to the case $k = 4$.

\subsection{The Quartic  Inflaton Potential with a Bare Mass: $V(\phi)\simeq \frac12 m_\phi^2 \phi^2+\lambda \phi^4$}
\label{Sec:mixedinflaton}

As shown in Eq.~(\ref{Eq:casek2}), only a quadratic potential $V(\phi)$ 
provides a viable region of parameter space at low reheating temperatures, $\trh \lesssim 7~\mathrm{TeV}$, where the spectator field can satisfy the relic abundance constraint. In contrast, a quartic potential leads to an excessive relic density. The phenomenology of reheating with a mixed potential, such as a T-model potential with $k = 4$ and a bare mass term $m_\phi$, was analyzed in \cite{Clery:2024dlk}.
It was found that although the early stages of reheating are dominated by the quartic term, the quadratic term ultimately determines the reheating temperature, significantly altering the allowed parameter space.

It is therefore useful to investigate whether the presence of a bare mass term could similarly modify the parameter space for a spectator field in the context of a quartic inflationary potential. The present-day spectator field abundance can be expressed as
\beq
\rho_\chi^0 \; = \; \rho_\chi^{\rm end}\left(\frac{\aosc}{\am}\right)^3\left(\frac{\am}{\arh}\right)^3\left(\frac{\arh}{a_0}\right)^3 \,,
\label{Eq:rhomix1}
\eeq
where $\am$ is defined as the scale factor when the quadratic term
dominates over the quartic term, or
\beq
\frac12 m_\phi^2\phi_0^2(\am) \; = \; \lambda \phi_0^4(\am)\,,~~\Rightarrow ~~\phi_0(\am) \; = \; \frac{m_\phi}{\sqrt{2 \lambda}}\,, ~~\Rightarrow ~~\rho_\phi(\am) \; = \; \frac{m_\phi^4}{2\lambda}\,,
\label{Eq:rhophiam}
\eeq
where we have assumed that $\aosc < \am < \arh$. We then find
\beq
\left(\frac{\aosc}{\am}\right)^4 \; = \;
\frac{\frac12\rho_\phi(\am)}{\rho_\phi(\aosc)} \; = \;
\frac{m_\phi^4}{3 \lambda M_P^2 m_\chi^2}\,;
~~\left(\frac{\am}{\arh}\right)^3 \; = \; \frac{\rhorh}{\frac12\rho_\phi(\am)}\,,
\label{aratiosphi}
\eeq
where we used $\rho_\phi(\aosc)=3H^2(\aosc) M_P^2=
\frac34 m_\chi^2 M_P^2 $. The coefficient of $\frac12$ in front of $\rho_\phi(\am)$ arises because, at $a=\am$, only half of the density comes from the quartic part of the potential. Substituting
into Eq.~(\ref{Eq:rhomix1}), we find
\beq
\rho_\chi^0 \; = \; \rho_\chi^{\rm end}\frac{{2^\frac14} ~\rhorh }{\rho_\phi^\frac34(\aosc) \rho_\phi^\frac14(\am)}\frac{g_0 T_0^3}{\grh \trh^3}\,.
\eeq
 Using Eq.~(\ref{Eq:rhophiam}), we obtain
\beq
\rho_\chi^0 \; = \;\frac{3 \beta \He^4}{16 \pi^2}\frac{4\lambda^{1/4}\alpha g_0}{3^\frac34\grh}
\frac{T_0^3 \trh}{(M_Pm_\chi)^\frac32m_\phi}\,,
\eeq
leading to
\begin{eqnarray}
\frac{\Omega_\chi^{\rm mix}h^2}{0.12} &\simeq&
\frac{\trh}{6 ~{\rm MeV}}
\left(\frac{\He}{5.8\times 10^{12}~{\rm GeV}}\right)^4 \lambda_\chi^{1/4} \beta
\left(\frac{10^{11}~\rm GeV}{m_\phi}\right)
\left(\frac{10^{12}~\rm GeV}{m_\chi}\right)^\frac32 \nonumber \\
& \gtrsim & \frac{\trh}{120 ~{\rm GeV}}
\left(\frac{\He}{5.8\times 10^{12}~{\rm GeV}}\right)^4 \beta
\left(\frac{10^{12}~\rm GeV}{m_\chi}\right)^2
\,,
\end{eqnarray}
where the inequality stems from $\aosc < \am$. If $\chi$ begins to oscillate during the quadratic phase for $\phi$, we
recover Eq.~(\ref{Eq:casek2}). This imposes an upper bound on $m_\phi$, as $\lambda$ is fixed by the initial conditions 
discussed in Section \ref{sec:init}, and thus produces a lower 
bound on $\Omega_\chi h^2$. While this scenario relaxes the 
constraints on $m_\chi$ and $\trh$ from the long-wavelength 
contribution relative to Eq.~(\ref{Eq:casek4}), the short-wavelength contribution—i.e., the graviton exchange term given by Eq.~(\ref{Eq:omega0k4})—remains significant. For $k = 4$, this still requires $m_\chi \lesssim 100~\mathrm{GeV}$, leaving no viable solution for dark matter in this case.\footnote{Indeed, the short-wavelength modes are always produced at the beginning of reheating.}

\subsection{Oscillations Begin After Reheating}
\label{arhltaoscp2}

The remaining possibility for a non-interacting spectator arises if $\arh < \aosc$. For this to occur, Eq.~(\ref{Eq:mchimin}) must be violated, implying that the spectator mass must be below $210 \left(\trh / 10^{10}~\mathrm{GeV}\right)^2~\mathrm{GeV}$. Even if the oscillations of $\chi$ begin before $\rho_\chi$ dominates the energy budget, it is still necessary to ensure sufficient dilution to avoid overproduction. If the spectator field fluctuations enter the oscillatory regime after the reheating phase, its present density can be written as
\bea
\rho_\chi^0 && \; = \;  \rhochie\left(\frac{\aosc}{\arh}\right)^3
\left(\frac{\arh}{a_0}\right)^3 \; = \; \rhochie\left( \frac{4\rhorh}{3m_\chi^2M_P^2}\right)^\frac34\frac{g_0T_0^3}{\grh\trh^3}
\nonumber
\\
&&
\; = \; \frac{3^\frac14 \a^\frac34g_0}{2^\frac52\pi^2\grh}\beta
\frac{\He^4T_0^3}{(m_\chi M_P)^\frac32}\,.
\eea
The present relic abundance is then given by 
\beq
\frac{\Omega_\chi h^2}{0.12}
\; \simeq \; 4.8 \times 10^{12}\left(\frac{10^{12}~\rm GeV}{m_\chi}\right)^\frac32
\left( \frac{\He}{5.8\times 10^{12}\gev} \right)^4 \beta \,.
\label{Eq:omegarh}
\eeq
Note that the relic density is identical to the case where oscillations begin before the radiation-dominated era with $k=4$, given by Eq.~(\ref{Eq:casek4}). This is not surprising because, for the 
spectator field, there is no distinction between a Universe 
dominated by a gas of SM particles or one dominated by an inflaton field redshifting as $a^{-4}$. From the perspective of the spectator, the effect of dilution is identical. The result (\ref{Eq:omegarh}) clearly demonstrates that any spectator with a mass below $M_P$ is dramatically overproduced, as it enters the oscillatory regime too late to achieve sufficient dilution. Even in an extremely fine-tuned region where 
$m_\chi\simeq \frac32 \He$, allowing the Boltzmann suppression factor in Eq.~(\ref{eq:expCASEIIA}) to play a role, the spectator is 
still overproduced via gravitational interactions unless $k = 2$ and $\trh \lesssim 7~\mathrm{TeV}$. Moreover, having $\arh < \aosc$ imposes an upper bound on $m_\chi$. Satisfying this condition together with Eq.~(\ref{Eq:omegarh}) would require an unphysically large reheating temperature. Consequently, this scenario is not viable and will not be considered further.

To summarize this scenario, fluctuations gravitationally produced during inflation are highly constrained when considering a simple quadratic potential $V(\chi) = \frac{1}{2} m_\chi^2 \chi^2$. For inflaton potentials of the form $V(\phi)\sim \phi^k$ during reheating with $k>2$, the energy density of the spectator field becomes excessive due to contributions from inflaton scattering through graviton exchange (short-wavelength modes). In the case of a quadratic inflaton potential, only a highly restricted region of parameter space survives. This region requires low reheating temperatures ($\trh \lesssim 7~\mathrm{TeV}$) and large spectator masses ($m_\chi \gtrsim 3 \times 10^{12}~\mathrm{GeV}$), as illustrated in Fig.~\ref{mchitk2} and Eq.~(\ref{Eq:casek2}). Extending this to a mixed inflaton potential, $V(\phi) = \frac{1}{2} m_\phi^2 \phi^2 + \lambda \phi^4$, does not significantly alleviate these constraints or enlarge the parameter space. Therefore, it becomes essential to examine the role of self-interactions in the spectator sector, described by a potential of the form $V(\chi) \sim \frac{1}{2} m_\chi^2 \chi^2 + \lambda_\chi \chi^p$. As we will demonstrate, these self-interactions can lead to a substantial dilution of the spectator energy density, $\rho_\chi \sim \chi^p$, before the quadratic term becomes dominant. This process can significantly reduce the relic abundance of the spectator field.

\section{Spectator Dark Matter with Self-Interactions}
\label{sec:SpectSI}

\subsection{Self-Interactions with a Bare Mass: $V(\chi) = \frac12 m_\chi^2 \chi^2 + \lambda_\chi M_P^4\left(\frac{\chi}{M_P}\right)^p$}
In the previous section, we demonstrated that a non-interacting spectator field is viable only within a highly constrained parameter space. Specifically, for a quadratic inflaton potential, the scenario requires $m_\chi \gtrsim 3 \times 10^{12}~\mathrm{GeV}$ and low reheating temperatures $\trh \lesssim 7~\mathrm{TeV}$ to avoid an overabundance of $\rho_\chi$. In all other cases, the spectator field results in an overdensity for $\rho_\chi$. On the other hand, if one considers a potential of the form
\beq
V(\chi) \; = \; \frac12 m_\chi^2 \chi^2 + \lambda_\chi M_P^4\left(\frac{\chi}{M_P}\right)^p \,,
\eeq
with $m_\chi^2< p(p-1)\lambda_\chi M_P^{4-p}\chi_{\rm end}^{p-2}$, 
the $\chi^p$ contribution to the potential dominates the evolution of $\chi$ at the beginning of its oscillatory phase. As a result, $\rho_\chi$ redshifts faster than in the pure quadratic case between $\aosc$ and $\am^\chi$, where $\am^\chi$ represents the scale factor when the potential begins to be dominated by its quadratic term. The energy density of $\chi$ at the onset of oscillations is equal to the energy density of $\chi$ at the end of inflation and is given by Eq.~(\ref{inivalp}):
\begin{equation}
    \rho_ \chi(a_\text{osc}) \; = \; \rho_\chi^{\rm end}=\frac{3\He ^4}{8\pi^2 } \beta \left(\frac{\Gamma(3/p)}{\Gamma(1/p)}\right)^{\frac{p}{2}} \, .
\end{equation}
If oscillations of $\chi$ begin during reheating, the condition $\aosc<\arh $ still implies the constraint \eqref{Eq:mchimin}, with the replacement of $m_\chi$ by the effective mass $m_{\chi, \text{eff}}(\aosc)$.

As discussed in Section \ref{sec:spectevol}, the evolution of $\chi$ depends on the relationship between $p$ and $k$. When $p\leq 2k$, the energy density of the spectator redshifts as $a^{-6/(p+2)}$. We begin by assuming this condition holds. Ignoring non-perturbative effects, the evolution of $\chi$ follows Eq.~(\ref{gensol}), assuming that self-interactions $\lambda_\chi \chi^p$ dominate the potential at the onset of oscillations. As expected, when the equation of state parameter for $\chi$, $w_\chi = \frac{p-2}{p+2}$ becomes stiffer, the spectator field energy density $\rho_\chi$ experiences larger dilution compared to the quadratic case. This stronger dilution allows for a significant relaxation of the constraints on ($m_\chi$, $\trh$), thereby opening up a much larger parameter space for viable scenarios.

As in Section \ref{Sec:mixedinflaton}, it is important to first determine the scale factor $\am^\chi$ above which the quadratic term dominates $\rho_\chi$. This should determine the duration of $\chi^p$ domination, and therefore, the amount of dilution.
Following the derivation leading to Eq.~(\ref{aratiosphi}), we define, $\am^\chi$ through
\beq
\frac12 m_\chi^2 \chi^2(\am^\chi) \; = \;  
\lambda_\chi \chi^p(\am^\chi) M_P^{4-p}
~~\Rightarrow ~~ \chi^p(\am^\chi) = \left( \frac{m_\chi^2}{2 \lambda_\chi M_P^{4-p}} \right)^\frac{p}{p-2} \, .
\label{Eq:chipamchi}
\eeq
This leads to
\beq
\frac{\am^\chi}{\aosc} \; = \; (2 \lambda_\chi)^\frac{p+2}{3p(p-2)}\left(\frac{ M_P}{m_\chi}\right)^\frac{p+2}{3(p-2)}
\left(\frac{3 \He^4 \beta}{4 \pi^2M_P^4}\right)^\frac{p+2}{6p}
\left(\frac{\Gamma (3/p)}{\Gamma (1/p)}\right)^{\frac{p+2}{12}} 
\,.
\label{amaoscpltk2}
\eeq
In this region of parameter space, the energy density at $\am^\chi$ is given by
\beq
\rho_\chi(\am^\chi) \; = \; 2 \rho_\chi^{\rm end}
\left(\frac{\aosc}{\am^\chi}\right)^\frac{6p}{p+2} \;= \; (2 \lambda_\chi)^\frac{2}{2-p}\left(\frac{m_\chi}{M_P}\right)^\frac{2p}{p-2}
M_P^4 \,,
\label{rhochiam}
\eeq
where the factor of 2 accounts for the equality of the quadratic and self-interaction terms. Note that $\rho_\chi(\am^\chi)$ is independent of $\rho_\chi^{\rm end}$. This can be understood from the fact that the energy density at $\am^\chi$ depends solely on the parameters of the potential $V(\chi)$ and not on the initial condition, as seen in Eq.~(\ref{Eq:chipamchi}).

For $p>2k$, the evolution of $\chi$, given by Eq.~(\ref{gensol}), is different, and we obtain
\beq
\frac{\am^\chi}{\aosc} \; = \; (2 \lambda_\chi)^\frac{k+2}{3kp}\left(\frac{ M_P}{m_\chi}\right)^\frac{k+2}{3k}
\left(\frac{3 \He^4 \beta}{4 \pi^2M_P^4}\right)^\frac{(k+2)(p-2)}{6kp}
\left(\frac{\Gamma (3/p)}{\Gamma (1/p)}\right)^{\frac{(k+2)(p-2)}{12k}} 
\,.
\label{amaoscpgtk2}
\eeq
Evaluating $\rho_\chi$ at $a= \am^\chi$, we find
\beq
\rho_\chi(\am^\chi)= 2 \rho_\chi^{\rm end}
\left(\frac{\aosc}{\am^\chi}\right)^\frac{6pk}{(k+2)(p-2)}=(2 \lambda_\chi)^\frac{2}{2-p}\left(\frac{m_\chi}{M_P}\right)^\frac{2p}{p-2}
M_P^4\,,
\eeq
which matches the expression in Eq.~(\ref{rhochiam}). 

To ensure an additional phase of dilution from the $\chi^p$ term of the potential, we require $\am^\chi>\aosc$. This condition imposes an upper bound on $m_\chi$, valid for both $p \le 2k$ and $p > 2k$:
\beq
m_\chi < M_P(2\lambda_\chi)^\frac{1}{p}
\left(\frac{3 \He^4 \beta}{4 \pi^2 M_P^4}\right)^\frac{p-2}{2p}
\left(\frac{\Gamma (3/p)}{\Gamma (1/p)}\right)^{\frac{p-2}{4}} \,, 
\eeq
which gives the following limits:
\begin{align}
        &&m_\chi \lesssim 2.3~\lambda_\chi^\frac14
        \beta^\frac14\times 10^{12}
        ~{\rm GeV} \,,  & \qquad (p=4) \, , \label{Eq:limitmchip4} && \\
        &&m_\chi \lesssim 1.3 ~\lambda_\chi^\frac16 \beta^\frac13\times 10^{10}~{\rm GeV} \,, & \qquad (p=6) \, , \label{Eq:limitmchip6} &&\\
        &&m_\chi \lesssim 7.6 ~\lambda_\chi^\frac18 \beta^\frac38\times 10^{8}~{\rm GeV} \,, & \qquad (p=8) \, , \label{Eq:limitmchip8} &&
\end{align}
where we have assumed the value of $\He$ corresponding to the $k=2$ case.

The amount of dilution depends on the sequence of events determining when the quadratic term in the potential begins to dominate relative to reheating and the onset of oscillations. We will therefore analyze the following three scenarios separately: $\mathbf{1}$) $\aend < \aosc < \am^\chi < \arh$, $\mathbf{2}$) $\aend < \aosc < \arh < \am^\chi$, $\mathbf{3}$) $\aend < \arh < \aosc < \am^\chi$.

\subsubsection*{1) $\aend<\aosc<\am^\chi<\arh$}

For sufficiently high $m_\chi$, the quadratic term dominates before the end of reheating. Before this occurs, the effective mass of $\chi$ is given by 
\beq
m_{\chi,\text{eff}}^2 \; = \; V''(\chi) \; = \; p (p-1)\lambda_\chi M_P^2 \left(\frac{\chi}{M_P}\right)^{p-2}\,.
\eeq
From $\aend$ to $\aosc$, $\chi$ remains nearly constant, and consequently, $m_{\chi,\rm eff}$ also stays constant. Using the expectation value for the spectator field, $\chi = \sqrt{\langle \chi^2 \rangle}$, given by Eq.~(\ref{inivalp}), we obtain
\beq
m^2_{\chi,\rm eff}(\aend) \; = \; m^2_{\chi,\rm eff}(\aosc) \; = \;
p(p-1)(\lambda_\chi)^\frac{2}{p}M_P^2
\left(\frac{3 \beta \He^4}{8 \pi^2M_P^4}\right)^\frac{p-2}{p}
\left(\frac{\Gamma(3/p)}{\Gamma(1/p)}\right)^\frac{p-2}{2} \, .
\label{Eq:mchiaosc}
\eeq
Note that the upper limits given in Eqs.~(\ref{Eq:limitmchip4})-(\ref{Eq:limitmchip8}) ensure that $m_\chi < \meff$ at the onset of oscillations. This expression simplifies for specific values of $p$:
\begin{align}
&
m^2_{\chi,\rm eff}(\aosc)\simeq 5.5\times 10^{-12} \sqrt{\lambda_\chi \beta} \left(\frac{\He}{6.4\times10^{12}~{\rm GeV}}\right)^2 M_P^2 \,,& (p=4)\, ,\\&
m^2_{\chi,\rm eff}(\aosc)\simeq 4.6 \times 10^{-16} (\lambda_\chi)^\frac13 \beta^\frac23 \left(\frac{\He}{6.4\times10^{12}~{\rm GeV}}\right)^{\frac83} M_P^2
\,,& (p=6) \, , \\&
m^2_{\chi,\rm eff}(\aosc)\simeq 2.8 \times 10^{-18} (\lambda_\chi)^\frac14 \beta^\frac34 \left(\frac{\He}{6.4\times10^{12}~{\rm GeV}}\right)^{3} M_P^2
\,,& (p=8) \, .
\end{align}
For the condition  $\rho_\phi(\am^\chi)>\rhorh$, we use 
\beq
\rho_\phi(\am^\chi)=\rho_\phi(\aosc)\left(\frac{\aosc}{\am^\chi}\right)^\frac{6k}{k+2}
\; = \; \frac{(k+2)^2}{3k^2} M_P^2~m^2_{\chi,\rm{eff}}(\aosc)
\left(\frac{\aosc}{\am^\chi}\right)^\frac{6k}{k+2}\,.
\eeq
For $p\leq 2k$, this leads to the constraint
\beq
m_\chi> M_P \left(\frac{\sqrt{3 \alpha} k}{k+2}\frac{\trh^2}{M_Pm_{\chi,\rm eff}}\right)^\frac{(k+2)(p-2)}{k(p+2)}(2\lambda_\chi)^\frac{1}{p}
\left(\frac{3 \He^4 \beta}{4 \pi^2 M_P^4}\right)^\frac{p-2}{2p}
\left(\frac{\Gamma(3/p)}{\Gamma(1/p)}\right)^\frac{p-2}{4}
\,,
\eeq
or, using Eq.~(\ref{Eq:mchiaosc}),
\beq
m_\chi > M_P
\left(
\sqrt{\frac{6 \alpha}{p(p-1)}}
\frac{k}{k+2}
\frac{\trh^2}{M_P^2}
\right)^\frac{(k+2)(p-2)}{k(p+2)}
\left[
2\lambda_\chi\left(\frac{3 \He^4 \beta}{4 \pi^2 M_P^4}\right)^\frac{p-2}{2}
\left(\frac{\Gamma(3/p)}{\Gamma(1/p)}\right)^\frac{p(p-2)}{4}
\right]^\frac{4k-2p+4}{pk(p+2)}
\,.
\label{mlimitcase1}
\eeq
For $k=2$, this reduces to
\begin{align}
    m_\chi \gtrsim  0.02 M_P (\lambda_\chi)^\frac{1}{12} \left(\frac{\He}{6.4\times 10^{12}~\gev} \right)^\frac13 \left(\frac{\trh}{M_P}\right)^\frac43 \,,& \qquad (p=4) \,.
\end{align}
For $p>2k$, the condition simplifies to
\beq
m_\chi > M_P
\left(
\sqrt{\frac{6 \alpha}{p(p-1)}}
\frac{k}{k+2}
\frac{\trh^2}{M_P^2}
\right)  \, .
\eeq
For $k = 2$, this reduces to
\begin{align}
    m_\chi \gtrsim  1.3 M_P \left(\frac{\trh}{M_P}\right)^2 \,,& \qquad (p=6) \,, \\
     m_\chi \gtrsim  0.97  M_P \left(\frac{\trh}{M_P}\right)^2 \,,& \qquad (p=8) \,.
\end{align}
For $p\leq 2k$, the present-day energy density of the spectator field is given by
\beq
\rho_\chi(a_0)= \frac12 \rho_\chi(\am^\chi)
\left(\frac{\am^\chi}{\aosc}\right)^3
\left(\frac{\aosc}{\arh}\right)^3
\left(\frac{\arh}{a_0}\right)^3\,,
\label{rhochia0plek2}
\eeq
or 
\beq
\begin{aligned}
\rho_\chi(a_0) \; = \; & \frac12 m_\chi M_P^3
(2\lambda_\chi)^{-\frac{2(k+1)}{kp}}  
\left(\frac{3 \He^4 \beta}{4 \pi^2 M_P^4}\right)^\frac{2k-p+2}{pk}
\left(\frac{\Gamma(3/p)}{\Gamma(1/p)}\right)^{\frac{(2k-p+2)}{2k}}
 \\  
 \; \times \; & \left(
\sqrt{\frac{6 \alpha}{ p(p-1)}}
\frac{k}{k+2}
\frac{\trh^2}{M_P^2}
\right)^\frac{(k+2)}{k}
\frac{g_0T_0^3}{\grh\trh^3}\,,
\label{rhochia0p}
\end{aligned}
\eeq
where the factor of $\frac12$ accounts for the fact that only half of the energy density at $\am^\chi$ redshifts as $a^{-3}$. 

For $p>2k$, we can still use Eq.~(\ref{rhochia0plek2}) while substituting for $(\am^\chi/\aosc)$, giving
\beq
\begin{aligned}
\rho_\chi(a_0)= & \frac12 M_P^4 \left( \frac{m_\chi}{M_P}\right)^{\frac{kp+2k-2p+4}{k(p-2)}}
(2\lambda_\chi)^{-\frac{2}{p-2}}  
 \left(
\sqrt{\frac{6 \alpha}{ p(p-1)}}
\frac{k}{k+2}
\frac{\trh^2}{M_P^2}
\right)^\frac{(k+2)}{k}
\frac{g_0T_0^3}{\grh\trh^3}\,.
\label{rhochia0pgtk2}
\end{aligned}
\eeq
Interestingly, whenever the condition $k/(k+2) = (p-2)/(p+2)$ holds, this expression coincides with the result found in Eq.~(\ref{rhochia0p}). This occurs for $k=2$ and $p=6$, or for $k=4$ and $p=10$, among other cases.

The evolution of the energy density of $\chi$ for $p = 4,6,8$ is shown in the upper right and lower panels of Fig.~\ref{rhochinop}. All values of $\lambda_\chi$ are chosen to satisfy the isocurvature constraints in Eq.~(\ref{eq:isocorvconstr}). In the upper 
right panel for $p=4$, we take $\lambda_\chi = 1$, $m_\chi = 2.4 \times 10^{10}~\gev$ and $\trh = 10^6~\gev$.  For this choice of parameters, oscillations begin very early with $\aosc \sim 1.4 \aend$. The energy density then redshifts as $a^{-4}$, leading to a dilution of $a^3 \rho_\chi$ as $a^{-1}$ until $a= \am^\chi \sim 136 \aend$, at which point $a^3 \rho_\chi$ becomes constant. In the lower left panel, for $p=6$, and we take $\lambda_\chi = 10^{12}$, $m_\chi =4\times  10^7~\gev$ and $\trh \sim 10^9~\gev$. Here, $\aosc \sim 1.5 \aend$, and $a^3 \rho_\chi$ initially decreases as $a^{-\frac32}$ until $\am^\chi \sim 1500 \aend$, where it becomes constant. Finally, in the lower right panel, for $p=8$, we set $\lambda_\chi = 10^{24}$, $m_\chi =  10^6~\gev$, and $\trh = 4.9 \times 10^8\gev$. In this scenario, $\aosc \sim 1.8 \aend$ and $\am^\chi \sim 3300 \aend$. All of the choices of these parameters lead to a relic abundance $\Omega_\chi h^2 \simeq 0.12$. 

We also note that the analytical approximation for $\am^\chi$ underestimates the numerically obtained value. This discrepancy arises because, at the moment when the energy density contribution from the quadratic term equals that of the self-interaction term $\chi^p$, the transition is not necessarily instantaneous. As $p$ increases, the oscillation frequency in a potential $V(\chi) \sim \chi^p$ is no longer simply given by $m_{\chi}$ but instead depends on the nonlinear structure of the potential. Consequently, the field evolution can become highly anharmonic before settling into quadratic oscillations, making the simple energy balance argument an increasingly poor approximation as it assumes an instantaneous transition.

The expression (\ref{rhochia0p}) becomes more transparent when evaluated for specific values of $p$ and $k$. For $k=2$, the relic densities are given by
\begin{align}
 &&\frac{\Omega_\chi h^2}{0.12}\simeq(\lambda_\chi)^{-\frac{3}{4}}\beta^\frac14
 \left(\frac{m_\chi}{10^{12}~\rm GeV} \right)
 \left(\frac{\trh}{47~\rm TeV}\right) \left(\frac{\He}{6.4\times 10^{12}~\gev} \right) \,,&& (p=4)\, ,
\label{Eq:k2p4}
\\&&\frac{\Omega_\chi h^2}{0.12}\simeq(\lambda_\chi)^{-\frac{1}{2}}
\left(\frac{m_\chi}{ 10^{7}~\rm GeV}\right)
\left(\frac{\trh}{8~\rm TeV}\right) \,,&& (p=6)\, ,
\label{Eq:k2p6}
\\&&\frac{\Omega_\chi h^2}{0.12}\simeq(\lambda_\chi)^{-\frac{1}{3}}
\left(\frac{m_\chi}{1~\rm GeV}\right)^\frac23
\left(\frac{\trh}{99~\rm TeV}\right) \,,&& (p=8) \, .
\label{Eq:k2p8}
\end{align}

It is interesting to compare these results with the case $\lambda_\chi=0$, obtained in Eq.~(\ref{Eq:casek2}). While the results for $\lambda_\chi=1$ and $p=4$ appear similar to the purely quadratic case (i.e., yielding the correct relic density for $m_\chi \simeq 10^{12}$ GeV with $\trh$ near or below the electroweak scale), the underlying physics is fundamentally different. For a purely quadratic potential, the relic abundance is inversely
proportional to $m_\chi^2$. In this case, a larger spectator mass results in earlier oscillations, allowing for greater dilution of $\rho_{\chi}$. In contrast, for $\lambda_\chi\neq 0$, the relic density is instead proportional to $m_\chi$ (proportional to $m_\chi^{2/3}$) for $p\leq 2k$ (for $k=2$ and $p=8$). This distinction arises because the equation of state becomes stiffer for $p>2$, leading to larger dilution. However, increasing $m_\chi$ causes the transition to quadratic domination to occur earlier, thereby reducing the duration of this dilution. Consequently, in this regime, a larger spectator mass results in a larger relic density. This behavior has important implications: one can readily reduce the relic abundance by lowering $m_\chi$, thereby opening up a large region in the ($m_\chi$, $\trh$) parameter space.

Moreover, this trend holds qualitatively for any larger value of $p$. While increasing $p$ strengthens the effect of dilution, it also shortens the duration of the $\chi^p$-dominated phase for a given $m_\chi$, leading to a longer period where the quadratic term dominates and thereby yielding a larger relic abundance, as seen in Eq.~(\ref{Eq:k2p4}). Nevertheless, even for $p = 8$, a substantial portion of the parameter space remains viable for satisfying the dark matter relic density constraint. Finally, it is important to note that lower values of $\lambda_\chi$ reduce the duration of the $\chi^p$-dominated era, thereby increasing the present-day relic abundance. 

\begin{figure}[ht!]
    \centering
    \includegraphics[width=\textwidth]{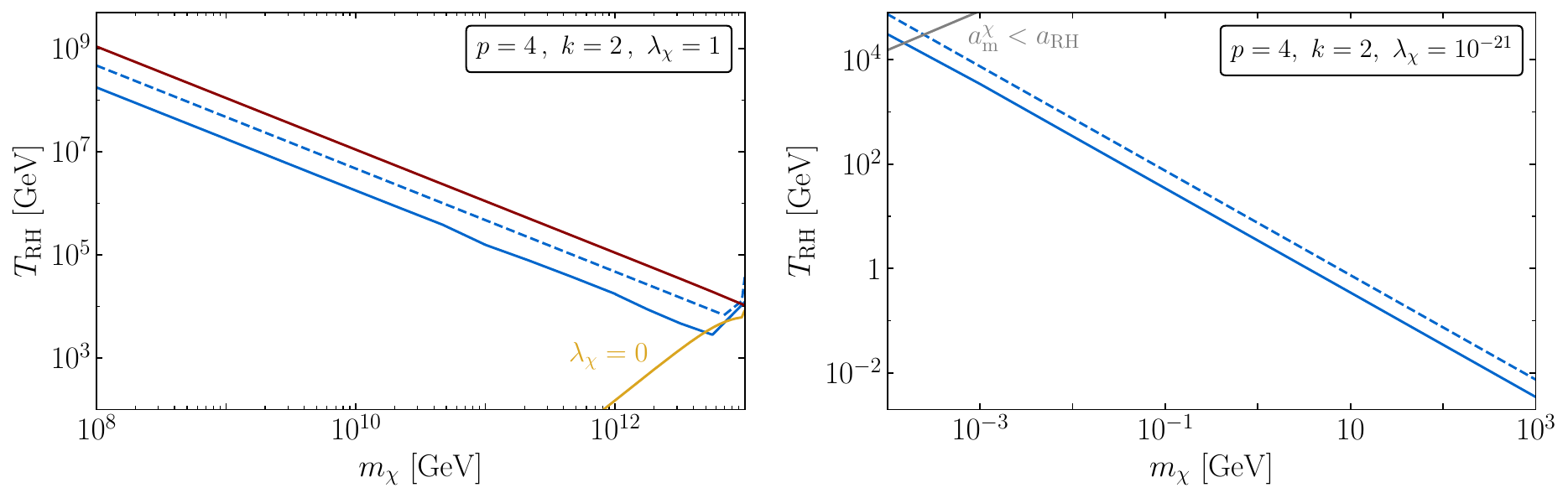}
    \includegraphics[width=\textwidth]{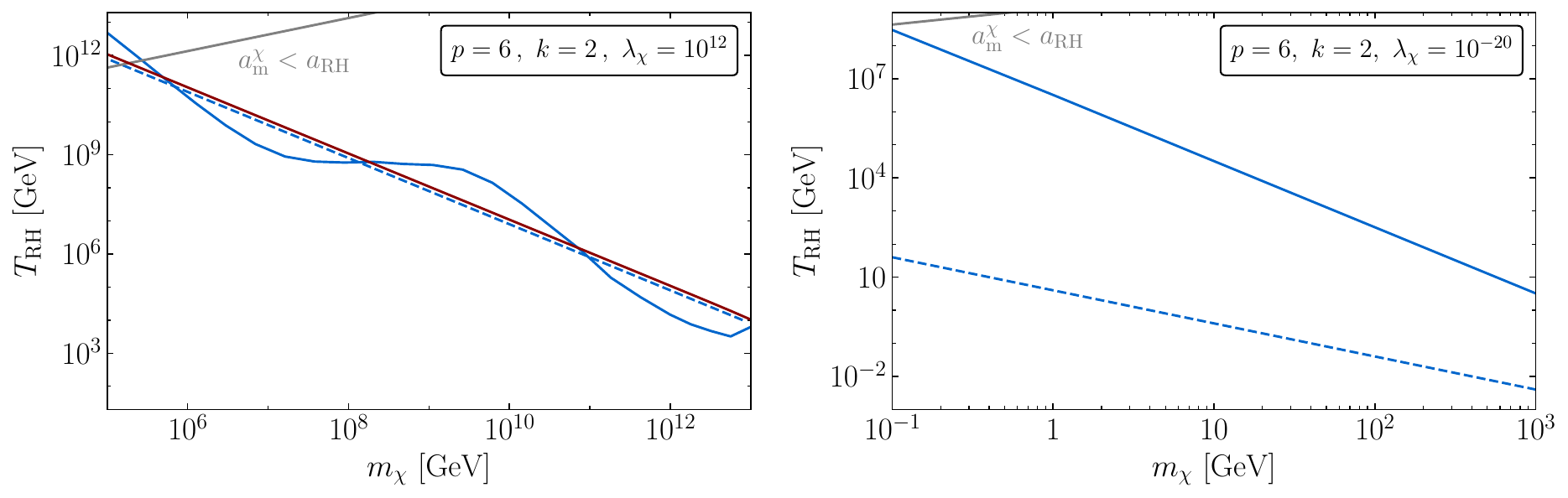}
    \includegraphics[width=\textwidth]{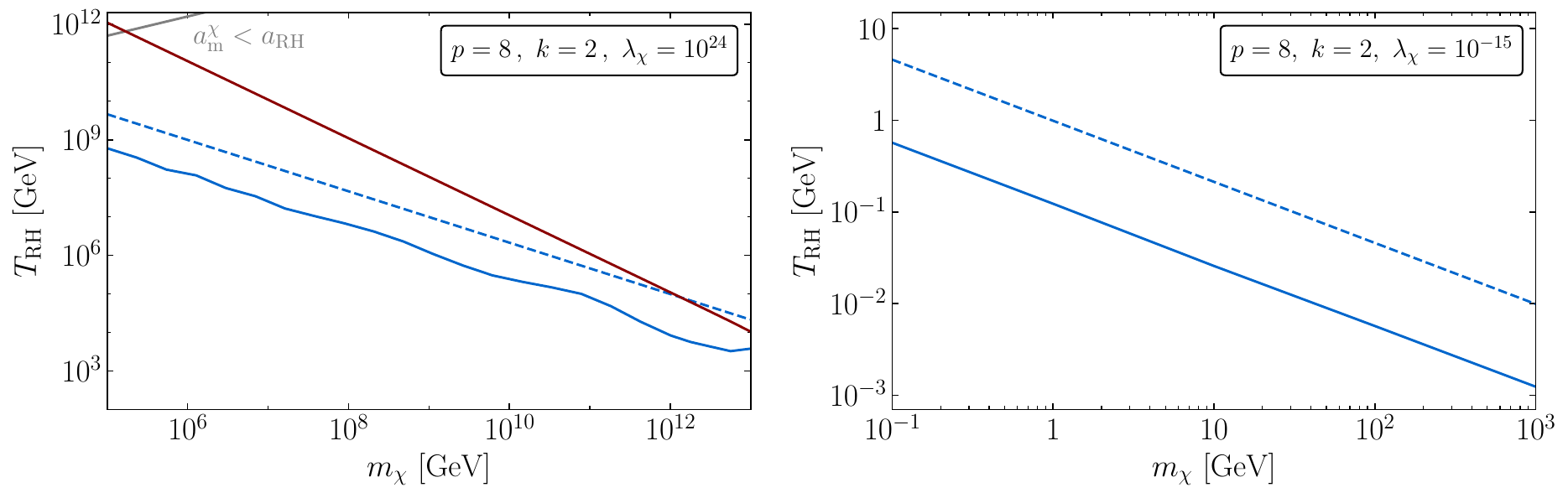}
    \caption{\it The $(m_\chi,\trh)$ plane for $k=2$  and a self-interacting spectator field with a bare mass term $\frac{1}{2}m_\chi^2\chi^2$. The blue solid lines correspond to numerical solutions obtained by integrating the equation of motion, while the dashed lines represent the analytical solutions from \eqref{Eq:k2p4}. The red solid lines denote the $\Omega_{\chi} h^2=0.12$ solutions arising solely from the short-wavelength contribution due to gravitational scattering.  The regions above the lines correspond to dark matter relic density $\Omega_{\chi} h^2>0.12$. The positively sloped gray lines indicate the transition point where $\am^\chi = \arh$. For reference, in the upper left panel, the yellow line illustrates the $(m_\chi, \trh)$ solution in the absence of self-interactions, as previously shown in Fig.~\ref{mchitk2}. 
    }
    \label{rhochip}
\end{figure}

We show in Fig.~\ref{rhochip} the parameter space allowed 
by the relic density constraint for $p=4, 6$ and $8$, each with $k=2$ and different values of $\lambda_\chi$. The blue dashed lines correspond to a full numerical integration of the equation of motion for $\chi$, whereas the solid blue lines represent our analytical results. For this choice of parameters, we observe that there is excellent agreement between the numerical and analytical results. The region above these lines leads to an overabundance of dark matter, with $\Omega_\chi h^2 > 0.12$.

In the upper left panel, we set $\lambda_\chi=1$ and $p=4$. Note that the lines of constant relic density extend well below the range shown, reaching approximately $10~\text{GeV}$. Below this threshold, obtaining the correct relic abundance would require $\trh > (\He M_P)^{1/2} \sim 2 \times 10^{15}~\text{GeV}$, which is no longer viable. Thus, the full range in masses is $\sim 10-10^{13}$~GeV and $\trh \sim 2000-2\times 10^{15}$~GeV. This result can be directly compared with the much more restricted range found in the previous section for $\lambda_\chi = 0$, as given by Eq.~(\ref{Eq:casek2}), and shown in Fig.~\ref{mchitk2}. The same result is also plotted in Fig.~\ref{rhochip} in orange, labeled as $\lambda_\chi = 0$. It is now evident that the inclusion of a quartic self-interaction term significantly enlarges the viable parameter space, compared to the purely quadratic case. The solid red line in Fig.~\ref{rhochip} represents the short-wavelength production of $\chi$, as given by Eq.~(\ref{Eq:omega0k2}). This contribution remains subdominant except for very high masses, $m_\chi \sim \He$, and small values of $\beta$. This effect is visible at the far right of the upper left panel, where the slope of the contour changes sign. This marks the transition where the bound in Eq.~(\ref{Eq:limitmchip4}) is violated, and the constraints from the previous section again become relevant.

It is also possible to find solutions with small values of $\lambda_\chi$ which satisfy the isocurvature constraint (see Eq.~\ref{eq:isoconstrgen}). 
These are shown in the right panels of Fig.~\ref{rhochip}.
In all of the cases considered, we find an acceptable relic density so long as $m_\chi \lesssim 1$~TeV, with lower masses preferred for higher reheating temperatures. Sub-GeV and sub-eV dark matter is allowed in these cases. 

For $p=4$, we see the numerical and analytical results differ by a constant factor of order 2.
This is due to (at least) two sources. As one can see from Fig.~\ref{rhochinop},  we underestimate $\am^\chi$, hence overestimating $\rho_\chi$. However using the power law dilution factor of $6p/(p+2)$ between $\aosc$ and $\am^\chi$, we are underestimating $\rho_\chi$, as the average dilution factor is smaller. The two effects conspire to the difference seen in the upper panels of Fig.~\ref{rhochip}. This discrepancy is accentuated for $p=6$ and $8$, where not only is $\am^\chi$ underestimated, the dependence on this factor is amplified. We also note that for the cases $p = 6, 8$ with $k = 2$ and small values of $\lambda_\chi$, shown in the middle right and bottom right panels of Fig.~\ref{rhochip}, the analytical and numerical results exhibit larger discrepancies. This additional mismatch arises due to the fact that, as the self-interaction coupling $\lambda_{\chi}$ decreases for $p = 6$ and $8$, the oscillation period of the spectator field becomes significantly larger. Furthermore, the field amplitude remains large over extended periods. Since our analytical approximations rely on an averaging procedure over oscillations, the error introduced by this assumption becomes larger in these cases. Additionally, as the value of $\am^\chi$ decreases with increasing $m_{\chi}$, the numerical curve oscillates. This effect induces further deviations between the analytical estimate and the exact numerical solution, leading to additional inaccuracies in the predicted evolution of $\rho_{\chi}$.

We can determine approximately the smallest $\lambda_\chi$ for which the averaging procedure is reliable, by imposing the condition that at least one oscillation is completed before $\am^\chi$. The number of oscillations depends also on $m_\chi$, because  smaller $m_\chi$ leads to larger $\am^\chi$ and hence  requires more time to complete an oscillation. To estimate the minimum value of $\lambda_\chi$, we fix $m_\chi$ to be that which allows  a large value of $\trh\sim 10^{10}\gev$ in Eq.~\eqref{Eq:k2p6}. In this case, we find that the averaging procedure is reliable for $\lambda_\chi \gtrsim 10^5 $ with $p=6$. In other words, the analytical expression is reasonably accurate for  values of $\lambda_\chi$ allowed by isocurvature constraints for large $\lambda_\chi$, whereas for small $\lambda_\chi$, we expect to have larger discrepancies. For $p>6$, the oscillations of $\rho_\chi$ get smaller and smaller,  which makes \eqref{gensol} a better estimation, so the discrepancy mainly comes from the value of $\am^\chi$, that introduces an almost constant error in the $(m_\chi,\trh)$ plane. We also observe that the discrepancy is similar (about order $10$ in $\trh$ for $p=8$) for small and large $\lambda_\chi$, as can be seen in the lower two panels of Fig.~\ref{rhochip}.

The allowed range of $(m_\chi, \trh)$ is highly sensitive to the value of $\lambda_\chi$. As illustrated in the upper right panel of Fig.~\ref{rhochip}, when taking the small coupling solution that satisfies the isocurvature constraint, the allowed parameter space shifts towards lighter spectator field masses and lower reheating temperatures, particularly when $\lambda_\chi = 10^{-21}$. For a fixed 
$m_\chi$, if $\lambda_\chi$ is smaller than the bound given in Eq.~(\ref{Eq:limitmchip4}), the resulting relic density approaches the behavior found in the non-self-interacting case discussed in the previous section. Despite the reduction in $\rho_\chi$ due to dilution between $\aosc$ and $\am^\chi$, long-wavelength production continues to dominate over the short-wavelength contribution for $\lambda_\chi < 1$. This follows from the fact that in both cases, $\Omega_\chi h^2\propto  m_\chi \, \trh$.  The two contributions are equal when $\lambda_\chi \simeq 3, 1.8\times 10^{12}$ for $p=4,6$, (for $p=8$, the value of $\lambda_\chi$ depends on $m_\chi$)  approaching non-perturbativity.

As expected, the allowed parameter space remains broad for larger values of $p$, as illustrated in the lower panels of Fig.~\ref{rhochip}. For $p=6$ and $\lambda_\chi = 10^{12}$ (noting that this is a non-renormalizable coupling suppressed by the Planck scale and does not violate perturbativity), the energy density exhibits slow oscillations. While the numerical solutions remain reliable, the analytical approximation begins to deviate. The allowed range in this case is comparable to the previous case with $p=4$ (as before, the mass range extends down to $\sim 20$~GeV, where numerical integration becomes difficult). The positively sloped gray line represents the boundary where $\am^\chi = \arh$. Below this line, the quadratic term dominates the equation of motion before reheating is complete. For $p = 4$, the corresponding line is beyond the range shown in the figure. For small values of $\lambda_\chi$ that satisfy the isocurvature constraints, the accuracy of our analytical approximations worsen. As mentioned earlier, this arises because the analytical solutions rely on averaging the energy density over many oscillations of the spectator field. However, when $\lambda_\chi$ is small, the period of oscillations becomes comparable to the timescale over which the energy density is diluted due to the enhanced redshift at higher $p$, leaving insufficient oscillations for effective averaging. For $p=8$, the slope of the $(m_\chi, \trh)$ relation changes slightly, but qualitatively, the results remain similar. Once again, the analytical solution does not match precisely with the numerical results in this regime.

For $k=4$, we obtain the following relic abundances,
\begin{align}
        &&\frac{\Omega_\chi h^2}{0.12} \simeq (\lambda_\chi)^{-\frac{5}{8}}\beta^\frac38
        \left(\frac{m_\chi}{22~\rm GeV}\right)
        \left(\frac{\He}{5.8\times 10^{12}~\gev} \right)^\frac32
        \,, && (p=4) \, , \\
        &&\frac{\Omega_\chi h^2}{0.12}\simeq(\lambda_\chi)^{-\frac{5}{12}}\beta^\frac16
       \left(\frac{m_\chi}{355~\rm keV}\right) 
       \left(\frac{\He}{5.8\times 10^{12}~\gev} \right)^\frac23
       \,, &&  (p=6) \, , \\
        &&\frac{\Omega_\chi h^2}{0.12}\simeq(\lambda_\chi)^{-\frac{5}{16}} \beta^{\frac{1}{16}}
        \left(\frac{m_\chi}{1.4~\rm keV}\right) \left(\frac{\He}{5.8\times 10^{12}~\gev} \right)^\frac14 \,, && (p=8) \,. 
          \label{Eq:k4p4}
\end{align}
In this scenario, as in all cases with $k=4$, the relic abundance is independent of the reheating temperature, $\trh$. This can be understood from the fact that, for a spectator field, the dilution of $\rho_\chi$ depends primarily on the expansion history rather than on whether the dominant energy component is the inflaton or radiation. More generally, the density $\rho_{\chi}$ tends to be higher for a given combination of ($m_\chi$, $\lambda_\chi$), provided that $\trh \lesssim \rho_{\rm end}^{1/4}$ for higher values of $k$. Furthermore, similar to the $k = 2$ case, we observe that the dark matter density increases with increasing $p$ at a significant rate. Specifically, for a fixed mass and coupling $\lambda_\chi$, the relic abundance increases by $\sim 5$ to $6$ orders of magnitude when $p$ is increased by $2$. This is due to the fact that, for larger values of $p$, the period of enhanced dilution ends sooner.

Recall that in the absence of self-interactions, there was no viable mass range for $m_\chi$ that could lead to the correct relic density for $k=4$. In contrast, when the conditions given by Eqs.~(\ref{Eq:limitmchip4})-(\ref{Eq:limitmchip8}) are satisfied, 
sub-GeV spectator masses become allowed. Importantly, this scenario avoids the usual lower limits on warm dark matter, such as the $\gtrsim$ keV constraints from structure formation or the Lyman-$\alpha$ forest, since the spectator field was never in thermal equilibrium with the Standard Model. For perturbative couplings, the short-wavelength contributions never dominate when $k=4$.

\begin{figure}[ht!]
    \centering
    \includegraphics[width=0.7\textwidth]{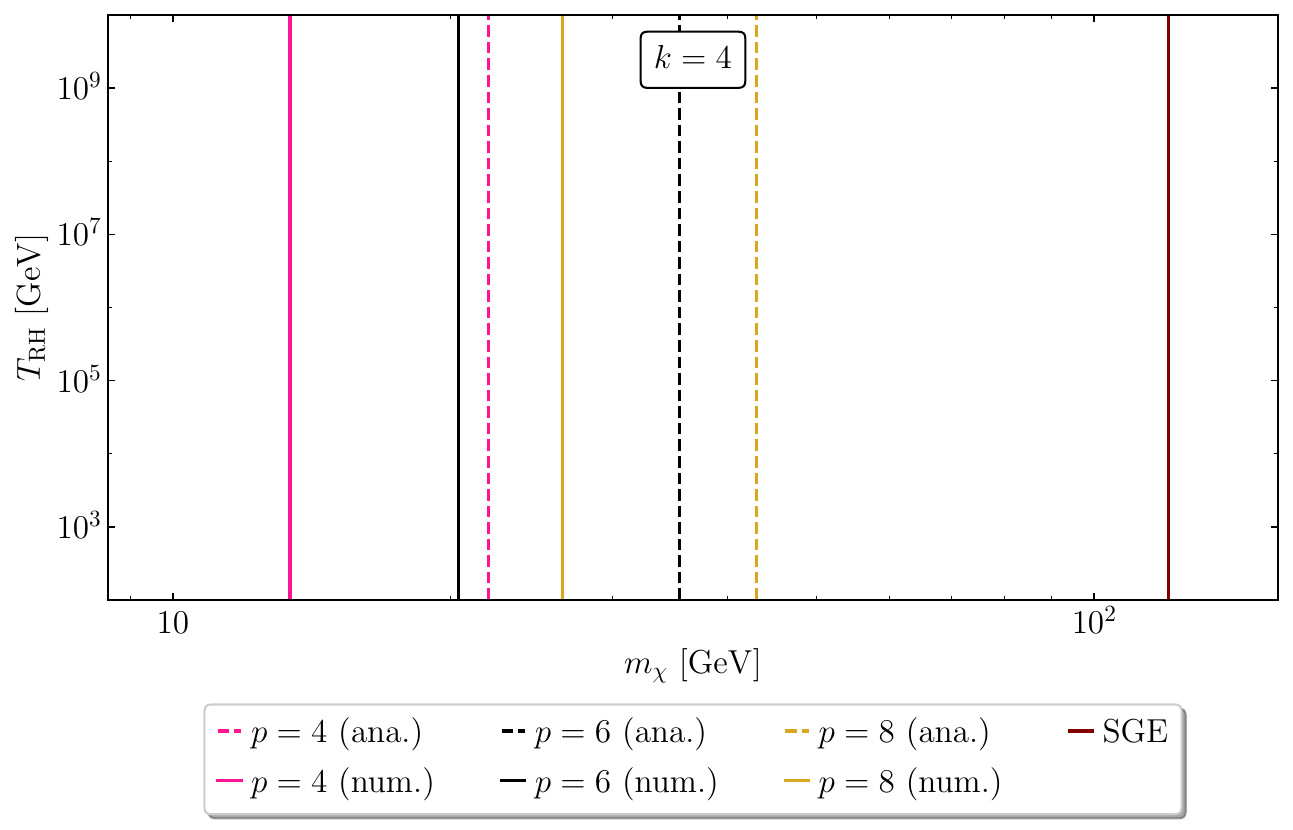}\caption{\it  
The ($m_\chi, \trh$) plane for $k=4$, with $\lambda_\chi=1$ for $p=4$, $\lambda_\chi=10^{12}$ for $p=6$, and $\lambda_\chi=10^{24}$ for $p=8$. The line colors are: pink ($p=4$), black ($p=6$), yellow ($p=8$), red (short-wavelength production from \cite{cmov} labeled SGE for single graviton exchange). Dashed: numerical solutions, solid: analytical solutions. All lines show the masses for which $\Omega_\chi h^2 = 0.12$. The density increases with $m_\chi$. }
\label{lk4}
\end{figure}

We show the ($m_\chi, \trh$) plane for $k=4$ and $p=4, 6$ and 8
taking $\lambda_\chi = 1, 10^{12}$ and $10^{24}$ respectively in Fig.~\ref{lk4}. The lines of constant $\Omega_\chi h^2$ are vertical due to the lack of dependence on $\trh$ for $k=4$. The red line (the right-most line) is the upper limit from graviton exchange ($m_\chi \lesssim 120$~GeV. The constraints from the long-    wavelength fluctuations are all stronger and the numerical results give upper limits of $\simeq 15-25$~GeV for $p=4-8$. The analytic results are all within a factor of two of the numerical results. 

For $k=6$, we obtain the following relic abundances:
\begin{align}
        &\frac{\Omega_\chi h^2}{0.12}\simeq (\lambda_\chi)^{-\frac{7}{12}} \beta^{\frac{5}{12}}
        \left(\frac{m_\chi}{0.1~\rm GeV}\right) \left(\frac{1.6\times 10^{8}~{\rm GeV}}{\trh}\right)^\frac13 \left(\frac{\He}{5.5\times 10^{12}~\gev} \right)^\frac53 \,, && (p=4) \, , \\
        &\frac{\Omega_\chi h^2}{0.12}\simeq (\lambda_\chi)^{-\frac{7}{18}}\beta^\frac29
        \left(\frac{m_\chi}{1~\rm keV}\right) \left(\frac{4.0\times 10^{6}~{\rm GeV}}{\trh}\right)^\frac13 \left(\frac{\He}{5.5\times 10^{12}~\gev} \right)^\frac89 \,, && (p=6) \, , \\
        &\frac{\Omega_\chi h^2}{0.12}\simeq (\lambda_\chi)^{-\frac{7}{24}}\beta^\frac18
        \left(\frac{m_\chi}{1~\rm eV}\right)  \left(\frac{2.0 \times 10^{4}~{\rm GeV}}{\trh}\right)^\frac13 \left(\frac{\He}{5.5\times 10^{12}~\gev} \right)^\frac12 \,, && (p=8) \, . 
        \label{Eq:k6p4}
\end{align}
In this case, even lower spectator masses are required to satisfy the relic density constraint. This follows from the fact that dilution must last longer for higher values of $k$, favoring lower $m_\chi$. However, achieving the correct relic abundance now requires relatively high reheating temperatures. The dependence on $p$ remains similar to the $k=4$ case, where larger $p$ implies a larger relic density, but now with an inverse dependence on reheating temperature when compared with the $k=2$ case.

\subsubsection*{ 2) $\aend<\aosc<\arh<\am^\chi$}
The resulting relic density in this case remains the same as found previously, with the key difference that the inequality in Eq.~(\ref{mlimitcase1}) must now be reversed, placing an upper bound on $m_\chi$. Here, the bare mass $m_\chi$ is smaller, and since Eq.~\eqref{rhochia0p} depends linearly on $m_\chi$, independent of $k$ and $p$, the relic density in this scenario is smaller and requires a higher reheating temperature. The boundary separating this region from the previously discussed scenario, where $\am^\chi < \arh$, is shown by the positively sloped gray line in Fig.~\ref{rhochip}, which is seen in the cases of $p=6$ and $p=8$.

\subsubsection*{ 3) $\aend<\arh<\aosc<\am^\chi$}
In this case, oscillations begin after reheating and the Universe is radiation-dominated. Consequently, the results derived earlier for $k=4$ remain applicable in this case. It is important to note that in scenarios \textbf{2)} and \textbf{3)}, the limit $m_\chi \rightarrow 0$ cannot be taken, as it would naively imply $\rho_\chi^0=0$. In reality, there is always a lower bound on $m_\chi$ arising from the condition $\am^\chi  < a_0$.

In summary, if the $\chi^p$ term ($p>2$) dominates before the end of reheating, the viable parameter space expands significantly compared to the purely quadratic case without self-interactions. This can be seen by comparing Eqs.~(\ref{Eq:k2p4})-(\ref{Eq:k2p8}) and (\ref{Eq:casek2}) for $k=2$. For $p>2$, dark matter masses as low as the electroweak scale can yield the correct relic abundance, assuming a suitably large value of $\trh$ and $\lambda_\chi \sim 1$. This expansion of parameter space arises from the fact that reducing $m_\chi$ extends the duration of enhanced dilution due to the non-quadratic contribution to the potential.

It is interesting to note that the relic density, $\Omega_\chi h^2$, for the long-wavelength contribution has the same dependence on $m_\chi$ (for  $p\le 2k$) and $\trh$ (for all $p$)  as the short-wavelength contribution arising from graviton exchange. The temperature dependence in both cases stems from the fact that the energy density is dominated by UV physics, i.e., scales determined at the end of inflation, and both components redshift similarly, leading to a scaling of $\trh^{\frac{4}{k} -1}$. The linear mass dependence for the short-wavelength contribution is simply due to the fact that at $\aend$, the number density does not depend on $m_\chi$,  and thus, the mass density satisfies $\Omega_\chi h^2 \propto m_\chi$. For the long-wavelength contribution, the $m_\chi$ dependence arises from two redshift factors: one from $a_{\rm osc}$ to $a_{\rm m}^\chi$ and another from $a_{\rm m}^\chi$ to the present. These effects conspire to yield a linear dependence on $m_\chi$ for $p\leq 2k$. Furthermore, the long-wavelength contribution is sensitive to the coupling, $\lambda_\chi$. For large $\lambda_\chi$, the effective mass is increased, and when $\meff(\aend)\simeq \He$, much of the power spectrum on large scales has been cut off and the two contributions become comparable. For $k=2$, we have seen that the two contributions become comparable for perturbative couplings, though for larger $k$, the short-wavelength contribution always remains sub-dominant.

\subsection{The Monomial Case: $V(\chi)= \lambda_\chi M_P^4 \left(\frac{\chi}{M_P}\right)^p$}
For completeness, we also analyze the case where $p>2$ and the spectator field follows a simple monomial potential of the form $V(\chi) = \lambda_\chi ~M_P^4 \left(\frac{\chi}{M_P}\right)^p$ which does not include a mass term for $\chi$ at the minimum of the potential. As a result, this type of model cannot serve as a viable dark matter candidate. Nevertheless, several constraints can still be placed on such scenarios, which we outline in this subsection.

As in previous cases, the spectator field must not dominate the energy density of the Universe before the onset of oscillations. Furthermore, while the energy density in $\chi$ oscillations redshifts faster than that of matter, it still contributes to the total energy density and may have observable consequences. Specifically, this contribution is constrained by BBN. For $p=4$, the $\chi$ oscillations redshift like radiation and can be directly compared to the energy density in neutrinos. For $p > 4$, the energy density contribution of $\chi$ decreases more rapidly, but it can still affect the expansion rate during BBN, providing a way to constrain the model. We examine these constraints in the following discussion.

The energy density stored in the oscillations of $\chi$ at the time of BBN, when $a = a_{\rm BBN}$ and $T_{\rm BBN} \simeq 1$~MeV, can be evaluated as:
\beq
\rho_\chi (a_{\rm BBN}) \; = \; \rho_\chi^{\rm end}
\left(\frac{\aosc}{\arh}\right)^\frac{6p}{p+2} 
\left(\frac{\arh}{a_{\rm BBN}}\right)^\frac{6p}{p+2}
 \; = \; \rho_\chi^{\rm end} \left(\frac{\rhorh}{\rho_{\rm osc}}\right)^\frac{p(k+2)}{k(p+2)}\left(\frac{g_{\rm BBN}T_{\rm BBN}^3}{\grh \trh^3}\right)^\frac{2p}{p+2}\,,
\eeq
where $g_{\rm BBN} = 43/4$ and this expression applies for $p\leq 2k$. Using the relations
\beq
\left(\frac{\aosc}{\aend}\right) \; = \; \left(\frac{9k^2 \He^2}{(k+2)^2 m_{\chi,\text{eff}}^2(\aosc)}\right)^\frac{k+2}{6k} \qquad {\rm and} \qquad \left(\frac{\aend}{\arh}\right) = \left(\frac{\alpha \trh^4}{3 \He^2 M_P^2}\right)^\frac{k+2}{6k} \, ,
\eeq
we obtain
\beq
\begin{aligned}
\rho_\chi(a_{\rm BBN})=\frac{3\He^4 \beta}{8\pi^2} \left(\frac{\Gamma   (3/p)}{\Gamma   (1/p)}\right) ^{\frac{p}{2}}\left(\frac{g_{\rm BBN}T_{\rm BBN}^3}{\grh }\right)^{\frac{2p}{p+2}}  \left(\frac{3 k^2~\alpha }{(k+2)^2M_P^2m_{\chi,\text{eff}}^2(\aosc)}\right)^{\frac{p(k+2)}{k(p+2)}}\trh^{\frac{2p(4-k)}{k(p+2)}}\,,
\end{aligned}
\eeq
where $m_{\chi,\rm eff}(\aosc)$ is given by Eq.~(\ref{Eq:mchiaosc}). 
We note that this result remains independent of $\trh$ for $k=4$.

This expression simplifies when we fix $k$ and $p$. Setting $k=2$, we obtain
\beq
\rho_\chi \; = \; \frac{M_P^4}{(\lambda_\chi)^\frac{4}{p+2}}
\left(\frac{3 \beta \He^4}{8 \pi^2 M_P^4}\right)^{\frac{6-p}{p+2}} \left(\frac{3 \alpha g_{\rm BBN} T_{\rm BBN}^3 \trh}{4 \grh p(p-1)M_P^4}
\right)^{\frac{2p}{p+2}} \left(\frac{\Gamma(3/p)}{\Gamma(1/p)} \right)^{\frac{(6-p)p}{2(p+2)}} \,.
\eeq
For $p=4$, this simplifies to
\beq
\rho_\chi(a_{\rm BBN}) \; = \; T_{\rm BBN}^4 (\lambda_\chi)^{-\frac23}
\left(\frac{3 \beta \He^4}{8 \pi^2 M_P^4}\right)^{\frac13} \left(\frac{ \alpha g_{\rm BBN} \trh}{16 \grh M_P}
\right)^{\frac43} \left(\frac{\Gamma(3/4)}{\Gamma(1/4)} \right)^{\frac23} \,.
\label{Eq:rhochiBBN}
\eeq
Unsurprisingly, smaller values of $\lambda_\chi$ correspond to smaller $\meff$, leading to delayed oscillations and, consequently, a larger energy density at BBN. This allows us to extract a lower bound on $\lambda_\chi$ to ensure agreement with BBN constraints. The energy density (\ref{Eq:rhochiBBN}) should be compared directly with the energy density of a single neutrino species at BBN, given by
\beq
\rho_\nu^0=\frac{7}{4}\frac{\pi^2}{30}T_{\rm BBN}^4 \, .
\eeq
The constraint on the effective number of neutrino species, $N_\nu < 3.18$ \cite{ysof}, can be translated into a lower bound on $\lambda$
from $\rho_\chi(a_{\rm BBN}) < 0.18 \rho_\nu$, leading to
\beq
\lambda_\chi > \left(\frac{120}{0.18\cdot 7\pi^2}\right)^\frac32 \left(\frac{3 \beta \He^4}{8 \pi^2 M_P^4}\right)^{\frac12} \left(\frac{ 43 \alpha  \trh}{16\cdot 427 M_P}
\right)^2 \left(\frac{\Gamma(3/4)}{\Gamma(1/4)} \right) \simeq 1.1 \times 10^{-25} \, ,
\label{Eq:rhochinobarek2}
\eeq
where on the right-hand side we fixed the reheating temperature to $\trh=10^{12} \gev$. This lower bound on $\lambda_\chi$ weakens for smaller values of $\trh$, implying that there is effectively little constraint on the self-coupling, and that the oscillations start early enough to avoid the BBN constraints. For $p=6$, the corresponding bound is easily derived and found to be even weaker, of order $\mathcal{O}(10^{-43})$, making it effectively negligible. 

Next, for $k=4$, as noted earlier, the dependence on $\trh$ drops out, leading to:
\beq
\begin{aligned}
\rho_\chi (a_{\rm BBN}) &= M_P^4  \left(\frac{3\He^4}{8\pi^2M_P^4}\right)^{\frac{10-p}{4+2p}}\left(\frac{\Gamma   (3/p)}{\Gamma   (1/p)}\right)^{\frac{p(10-p)}{4p+8}}\left[ \frac34  p (p-1)(\lambda_\chi)^\frac2p \right]^{-\frac{3p}{2p+4}}\left(\frac{g_{\rm BBN} T_{\rm BBN}^3\alpha^{\frac{3}{4}}}{\grh M_P^3}\right)^{\frac{2p}{p+2}} \, ,
\end{aligned}
\eeq
and for $p=4$, this simplifies to 
\beq
\rho_\chi (a_{\rm BBN}) = \left(\frac{3\He^4}{8\pi^2{M_P^4}}\right)^{\frac12} \left(\frac{\Gamma   (3/4)}{\Gamma   (1/4)}\right) \left(\frac{g_{\rm BBN}}{\grh }\right)^\frac43 \frac{T_{\rm BBN}^4\alpha}{9 \sqrt{\lambda_\chi} } \, ,
\eeq
leading to a lower limit on $\lambda_\chi$:
\beq
\lambda_\chi >  \left(\frac{120}{0.18 \cdot 7\pi^2}\right)^2 \left(\frac{3 \beta \He^4}{8 \pi^2 M_P^4}\right) \left(\frac{ 43}{427}
\right)^\frac83 \frac{\alpha^2}{81} \left(\frac{\Gamma(3/4)}{\Gamma(1/4)} \right)^2 \simeq 4.3 \times 10^{-25} \, ,
\eeq
for $\trh=10^{12} \gev$, which is similar to the bound obtained for $k=2$. Once again, for $p=6$, the constraint is significantly weaker, of order $\mathcal{O}(10^{-46})$. Note that for $p=4$ and $k=4$, the Universe is composed of two relativistic baths from the onset of oscillations down to BBN. This follows from the fact that, for $k=4$, there is no distinction between the evolution of the inflaton before reheating and the radiation-dominated epoch after reheating. Consequently, the ratio $\rho_\chi/\rho_{\rm tot}$ remains constant from $\aosc$ down to $a_{\rm BBN}$.
 
For $k=6$, we find
\beq
\rho_\chi \; = \; \frac{M_P^4}{(\lambda_\chi)^\frac{8}{3(p+2)}}
\left(\frac{3 \beta \He^4}{8 \pi^2 M_P^4}\right)^{\frac{14-p}{3(p+2)}} \left(\frac{ g_{\rm BBN} T_{\rm BBN}^3 }{ \grh \trh^\frac13 M_P^\frac83} \left( \frac{27 \alpha}{16p(p-1)} \right)^\frac23
\right)^{\frac{2p}{p+2}} \left(\frac{\Gamma(3/p)}{\Gamma(1/p)} \right)^{\frac{(14-p)p}{6(p+2)}} \,.
\eeq
Then for $p=4$, we obtain
\beq
\rho_\chi(a_{\rm BBN}) \; = \; T_{\rm BBN}^4 (\lambda_\chi)^{-\frac49}
\left(\frac{3 \beta \He^4}{8 \pi^2 M_P^4}\right)^{\frac59} \left(\frac{  g_{\rm BBN} M_P^\frac13 }{ \grh \trh^\frac13} \left(\frac{9 \alpha}{64} \right)^\frac23
\right)^{\frac43} \left(\frac{\Gamma(3/4)}{\Gamma(1/4)} \right)^{\frac{10}{9}} \,,
\eeq
leading to a lower limit on $\lambda_\chi$
\beq
\lambda_\chi >  \left(\frac{120}{0.18\cdot 7\pi^2}\right)^\frac94 \left(\frac{3 \beta \He^4}{8 \pi^2 M_P^4}\right)^{\frac54} \left(\frac{ 43 }{427}
\right)^3 \left(\frac{9\alpha}{64} \right)^2 \frac{M_P}{\trh}  \left(\frac{\Gamma(3/4)}{\Gamma(1/4)} \right)^\frac52 \simeq 6.7 \times 10^{-25} \, ,
\eeq
where in the last expression we fixed the reheating temperature to $\trh=10^{12} \gev$. 

For $p>2k$, from Eq.~(\ref{gensol}), the evolution of $\chi$ prior to reheating depends on $k$. In this case, the energy density at BBN is given by:
\beq
\rho_\chi (a_{\rm BBN}) \; = \; \rho_\chi^{\rm end}
\left(\frac{\aosc}{\arh}\right)^\frac{6pk}{(k+2)(p-2)} 
\left(\frac{\arh}{a_{\rm BBN}}\right)^\frac{4p}{(p-2)}
=\rho_\chi^{\rm end} \left(\frac{\rhorh}{\rho_{\rm osc}}\right)^\frac{p}{(p-2)}\left(\frac{g_{\rm BBN}T_{\rm BBN}^3}{\grh \trh^3}\right)^\frac{4p}{3(p-2)}\,.
\eeq
This leads to 
\beq
\rho_\chi(a_{\rm BBN}) \; = \; (\lambda_\chi)^{-\frac{2}{p-2}}\left( \frac{3k^2 \alpha}{(k+2)^2 p (p-1)}\right)^\frac{p}{(p-2)}
 \left(\frac{g_{\rm BBN}T_{\rm BBN}^3}{\grh M_P^3}\right)^\frac{4p}{3(p-2)} M_P^4 \, .
 \eeq
Finally, the case where oscillations begin after reheating is identical to the previous case with $k=4$.

In summary, for a pure monomial potential $V(\chi)\sim \lambda_\chi \chi^p$, the bounds on the parameter space $(\lambda_\chi, \trh)$ from BBN constraints are very weak, with $\lambda_\chi \gtrsim 10^{-25}$ for $\trh \gtrsim 10^{12}$ GeV for $p=4$ and $k=2$, as shown in Eq.~(\ref{Eq:rhochinobarek2}). The allowed parameter space increases for smaller $\trh$ when $k<6$ and/or larger values of $p$. For $k=6$, where the dependence on $\trh$ differs, we obtain a conservative bound,
$\lambda_{\chi}\gtrsim 6.7\times 10^{-25}\left(\frac{10^{12}~\gev}{\trh}\right)$.

\section{Spectator Dark Matter Interacting with the Inflaton}
\label{sec:specinfl}
As demonstrated in the previous subsection, self-interactions of the spectator field significantly expand the allowed parameter space in the $(m_\chi, \trh)$ plane, permitting spectator masses to range from sub-GeV (even sub-eV) scales up to nearly the inflaton mass, of order $10^{13}$~GeV. It was previously shown in Ref. \cite{Choi:2024bdn} that interactions between the inflaton and the spectator field can similarly open up the parameter space for couplings of the form $\sigma \phi^2 \chi^2$, with $\sigma \sim m_\phi^2/M_P^2 \sim 10^{-10}$, assuming $k=2$ for the inflaton potential. The introduction of such an interaction generates an effective mass, $m_{\chi,\rm eff}^2 \sim \sigma \phi^2$, which can induce sufficient redshifting of $\rho_\chi$, provided that the bare mass satisfies $m_\chi \ll m_{\chi,\rm eff}(a_{\rm end})$.\footnote{A bare mass is necessary to ensure that $\chi$ remains a viable dark matter candidate.} This scenario closely parallels the case with a self-interaction term, where for $p = 4$, $\lambda_\chi \chi^2 \leftrightarrow \sigma \phi^2$ , or $\sigma \leftrightarrow \lambda_\chi$. 

In this section, we generalize these results to include a coupling of the form $\sigma_{n,m}\phi^n \chi^m$ and extend our analysis to higher values of $k$. To isolate the effects of the inflaton coupling, we set the self-interaction term to zero, $\lambda_\chi=0$. However, this formalism naturally accommodates self-interactions by identifying the case $n=0$, $m=p$.

When the interaction term dominates in $dV/d\chi$, the equation of motion for the spectator field can be expressed as:
\begin{equation}
    \chi''+\frac{k+8}{k+2}\frac{1}{a}\chi'+m\sigma_{n,m}\frac{\phiend^n M_P^{4-n-m}}{\He^2}a^{\frac{4k-4-6n}{k+2}}\chi^{m-1} \; = \; 0 \, .
    \label{interacteom}
\end{equation} 
Here, we have taken the expectation value $\langle \phi^n \rangle = \phiend^n (\aend/a)^{6n/(k+2)}$, and assumed that the contribution from the bare mass term, $m_\chi^2$, is negligible.

The simplest case, with $k=n=m=2$, was previously studied in \cite{Choi:2024bdn}. Before generalizing the solutions to arbitrary values of $k, n,$ and $m$, we briefly review those results. In this case, Eq.~(\ref{interacteom}) simplifies to
\begin{equation}
   \chi'' + \frac52  \frac{\chi'}{a} + 4 \tilde{\sigma} \frac{\chi}{a^{2}} \; = \; 0 \, , \qquad \tilde{\sigma} \; = \; 2 \sigma_{2,2} \frac{ M_P^2}{m_\phi^2} \, ,
   \label{Eq:chidiff}
\end{equation}
where we used the relation $\lambda \phiend^k = 2\He^2 M_P^{k-2}$, which reduces to  $4H_{\rm end}^2 M_P^2 = m_{\phi}^2 \phi_{\rm end}^2$ for $k=2$.

To solve this equation, we assume a general ansatz of the form $\chi(a) = \chi_{\rm end} a^{-A+i B}$. Substituting this into Eq.~(\ref{Eq:chidiff}), we find that the general solution is given by
\begin{equation}
\begin{aligned}
&&
\chi(a)=
\chi_{\rm end}\left(\frac{\aend}{a}\right)^\frac34
\left[
\cos\left(B\log\left[\frac{a}{\aend}\right]\right) +\frac{3}{4B}\sin\left(B\log\left[\frac{a}{\aend}\right]\right)
\right]
\,,
\label{eq:chi(a)}
\end{aligned}
\end{equation}
with $A = \frac{3}{4}$ and $B = 2 \sqrt{ {\tilde \sigma}-\frac{9}{64}}$. 
The constants of integration have been set by assuming $\chi({\aend}) = \chi_{\rm end}$ and $\chi^\prime(\aend) = 0$. A real value for $B$ then requires ${\tilde \sigma} \ge 9/64$. For smaller ${\tilde \sigma}$, the solution is no longer oscillatory and instead takes the form
\beq
\chi(a) \; = \;  \chi_{\rm end} \times \frac{1}{2c} \left(\frac{\aend}{a}\right)^\frac{c+3}{4} \left( c-3 + \left(\frac{a}{\aend }\right)^\frac{c}{2} (c+3) \right) \, ,
\eeq
where $c= 8 \sqrt{\frac{9}{64} - {\tilde \sigma}}$. These solutions break down when the bare mass term for $\chi$ can no longer be neglected in Eq.~(\ref{Eq:chidiff}). This occurs when ${\tilde \sigma}$ is very small, or at large $a$, when $\phi^2$ becomes negligible. In this regime, Eq.~(\ref{Eq:chidiff}) is modified by replacing the third term with $4 {\tilde \sigma}/a^2 \to m_\chi^2 a/H_{\rm end}^2 \aend^3$, and the spectator field transitions to a matter-like behavior, with $\rho_\chi \propto a^{-3}$ and $\chi \propto a^{-3/2}$.

When $\sigma_{2,2} \phiend^2 \ll m_\chi^2$, the resulting density follows the results in Section~\ref{Sec:spectator} for non-interacting fields. The densities of $\chi$ at $\arh$ and $a_0$ are given by Eqs.~(\ref{Eq:rhochiarh}) and (\ref{Eq:genericrhochi0}), respectively. However, for larger ${\tilde \sigma}$, the effective mass due to the interaction with the inflaton becomes significant and evolves during the reheating process as $\meff^2 \propto \phi^2(a) \propto a^{-3}$, leading to the scaling relation:
\beq
 \rho_\chi(a) \; = \;  \frac{1}{2} m^2_{\chi, \rm eff} \chi^2(a) \; \simeq \;  \frac{3H_{\rm end}^4 \beta}{16\pi^2}  \left(\frac{\phi(a)}{\phi_{\rm end}} \right)^2 \left(\frac{a_{\rm end}}{a}\right)^{\frac32} \, .
 \label{s22sol}
\eeq
In this case, the spectator field energy density scales as $\rho_\chi(a) \propto a^{-\frac92}$, implying that $\rho_\chi$ dilutes more rapidly than $\rho_\phi$, thereby preventing the spectator field from dominating the total energy budget before reheating is completed.

As in the case of spectator self-interactions, the parameters $\sigma$, $m_\chi$, and $\trh$ determine the relative order of the scale factors $\aosc, \am^\chi$, and $\arh$:
\begin{align}
 & \qquad{\tilde  \sigma}  < \frac{9}{32} \, , &\am^\chi < a_{\rm osc} \, , \label{achiltaosc} \\
    & \qquad {\tilde  \sigma} < \frac34 \frac{m_\chi^2 M_P^2}{\alpha \trh^4} \, , &\am^\chi < a_{\rm RH} \, , \label{acltarh}\\
   & \qquad {\tilde  \sigma} < \frac{9}{16}  \left( 1- \frac{4 M_P^2 m_\chi^2}{3 \alpha \trh^4}\right)\, , &a_{\rm RH} < a_{\rm osc} \, .
   \label{arhltaosc}
\end{align}
Each of these cases leads to distinct evolution for $\rho_\chi$, as discussed in detail in \cite{Choi:2024bdn}.

For ${\tilde \sigma} < \frac{9}{32}$, the relic density is given by Eq.~(\ref{Eq:genericrhochi0}), with the replacement $m_\chi^2 \to 2\sigma_{2,2} \phiend^2$, leading to
\begin{equation}
    \frac{\Omega_\chi h^2}{0.12} \; \simeq \tilde \sigma^{-1}\frac{\trh}{25~{\rm TeV}} \, .
    \label{Eq:omegaIAbis}
\end{equation}

For ${\tilde \sigma} > 9/32$, the spectator field begins to oscillate while dominated by the effective mass $\propto \phi$, until $a=\am^\chi$,
where the oscillations transition to being dominated by the bare mass. The energy density must then be evolved from $a_{\rm end}$ as follows:
\begin{equation}
\begin{aligned}
 \rho_\chi(a_0) \; = \; & \frac{3\beta}{16\pi^2} H_{\rm end}^4 \left( \frac{a_{\rm end}}{a_{\rm osc}} \right)^3 \left( \frac{a_{\rm osc}}{\am^{\chi}} \right)^{3+2A}   \left( \frac{\am^{\chi}}{a_{\rm RH}} \right)^3 \left( \frac{a_{\rm RH}}{a_0} \right)^3  \\ \; = \; & \frac{3 \beta}{16\pi^2} H_{\rm end}^4 \left( \frac{a_{\rm osc}}{a_{\rm end}} \right)^{2A} \left( \frac{a_{\rm end}}{\am^\chi} \right)^{2A}  \left( \frac{a_{\rm end}}{a_{\rm RH}} \right)^3  \left( \frac{a_{\rm RH}}{a_0} \right)^3 \, ,
 \label{rhochiIIA}
\end{aligned}
\end{equation}
where $A$ is defined from Eq.~(\ref{eq:chi(a)}). For the limited range $9/32 < {\tilde \sigma} < 9/16$, the condition $\aend < a_{\rm osc}$ holds. However, for larger ${\tilde \sigma} > 9/16$,  oscillations begin immediately at the end of inflation, i.e., $a_{\rm osc} = \aend$. In this case, the exponential suppression in the initial value of $\langle \chi^2 \rangle$, alluded to earlier,
must be incorporated. The resulting density fraction is then given by:
\begin{equation}
    \Omega_{\chi} h^2 \; = \; 6.5 \times 10^5~\beta~\frac{H_{\rm end} m_{\chi} T_{\rm RH}}{\mathrm{GeV} M_P^2 \sqrt{\tilde \sigma} } \, , \label{eq:caseIIAOmegachi}
\end{equation}
or equivalently,
\beq
\frac{\Omega_\chi h^2}{0.12} \; \simeq \; \frac{\beta}{\sqrt{\tilde \sigma}}
\left(\frac{\trh}{10^{10}~{\rm GeV}}\right)
\left(\frac{m_\chi}{1.7\times 10^7~{\rm GeV}}\right) \,.
\label{eq:caseIIAOmegachi2}
\eeq

\begin{figure}[t!]
  \centering
\includegraphics[width=0.74\textwidth]{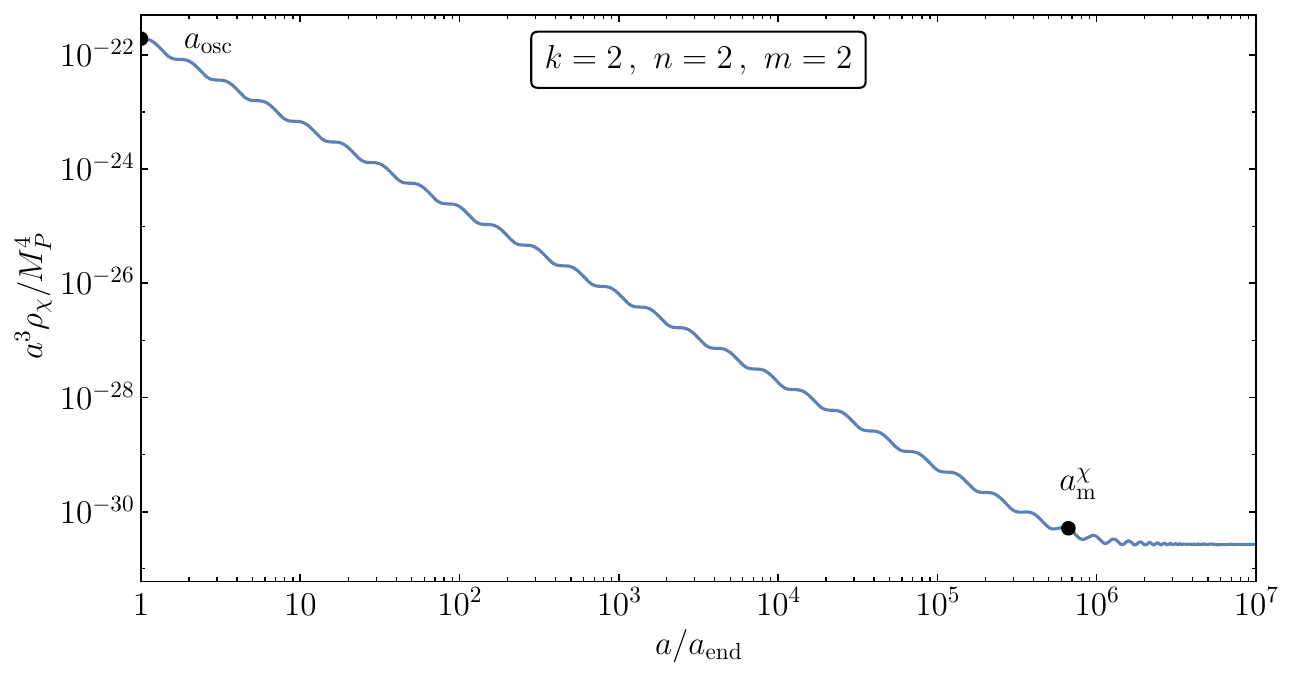}
  \caption{\em \small {
  The evolution of the comoving energy density of the spectator field, $\rho_\chi a^3$, is shown for the case $k = m = n = 2$ with an interaction term of the form $\sigma \phi^2 \chi^2$. The chosen parameters are $\sigma = 1.6 \times 10^{-10}$ and $m_\chi = 5 \times 10^4 \, \mathrm{GeV}$. 
}}
  \label{fig:rhochi2AB}
\end{figure}

The numerical solution to the equation of motion for $\chi$ in this case 
with ${\tilde \sigma} = 1.7$ is shown in Fig.~\ref{fig:rhochi2AB}. For this choice of $\tilde \sigma$, the 
onset of oscillations coincides with the end of inflation, $a_{\rm osc} = a_{\rm end}$, while the 
transition to quadratic domination occurs at $\am^\chi \simeq 6.1 \times 10^5 \, \aend$, and 
reheating completes at $a_{\rm RH} \simeq 1.3 \times 10^6 \,\aend$. In 
contrast to the case shown in the upper left panel of   Fig.~\ref{rhochinop}, where oscillations begin after a period of frozen evolution, here $\chi$ begins oscillating immediately at the end of inflation, leading to a rapid dilution of its energy density as $\rho_\chi \propto a^{-9/2}$. Once the quadratic term dominates at $a > \am^\chi$, the density redshifts $\rho_\chi \propto a^{-3}$.

\begin{figure}[t!]
  \centering
\includegraphics[width=0.74\textwidth]{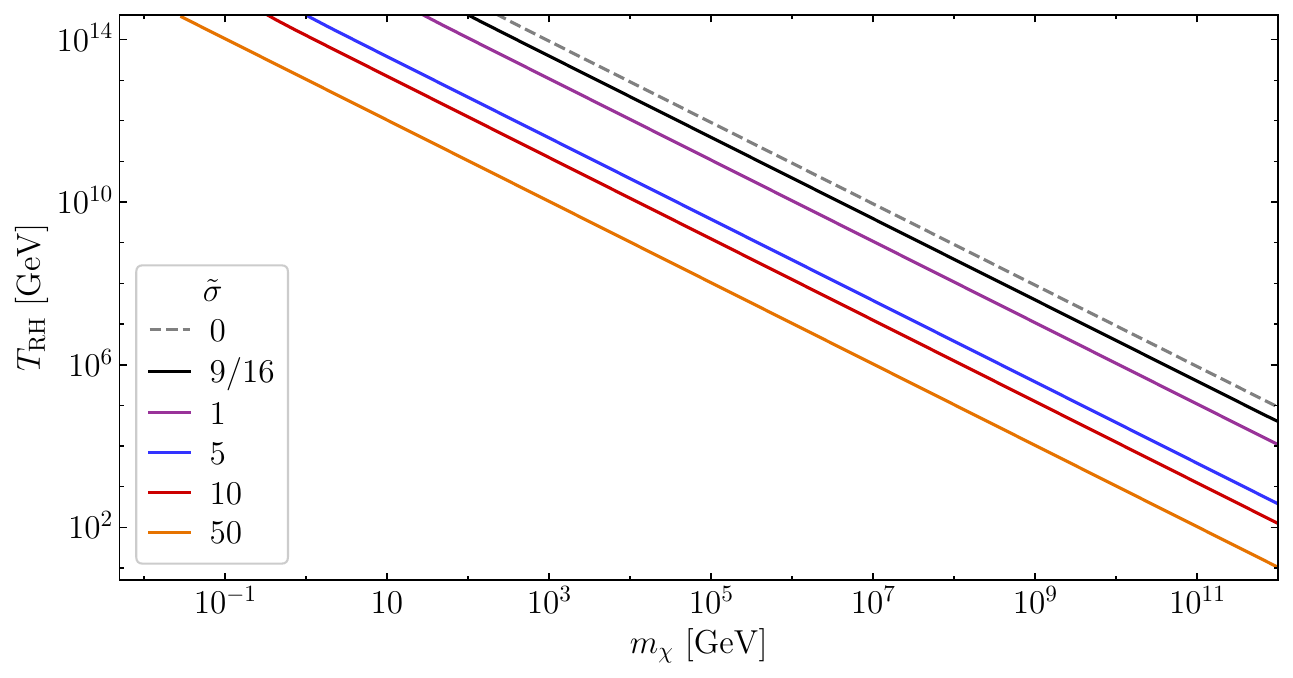}
  \caption{\em \small {The reheating temperature, $T_{\rm RH}$, as a function of the bare mass, $m_\chi$, satisfying $\Omega_{\chi}h^{2}=0.12$ for $\tilde{\sigma} = \frac{9}{16}$(black), 1 (purple),  $5 $(blue), $10$(red), $50$(orange), $0$(gray dashed). The latter neglects the contribution from large-scale fluctuations and corresponds to the result in \cite{cmov}. For $m_{\chi} < m_{\rm \chi, min}$ (corresponding to the endpoint of the lines), no value of $T_{\rm RH}$ satisfies $\Omega_{\chi} h^{2} = 0.12$. 
}
}
  \label{fig:test2}
\end{figure}

In Fig.~\ref{fig:test2}, we show the allowed parameter space for ${\tilde \sigma} \ge 9/16$ in the $(\trh,m_\chi)$ plane. The curves correspond to $\Omega_\chi h^2 = 0.12$ for fixed values of $\tilde \sigma$ as indicated.  For ${\tilde \sigma} \ge 9/16$, the relic density is saturated by the contribution from inflaton scattering, as $\beta \ll 1$. The dotted line in Fig.~\ref{fig:test2} shows the result from Ref.~\cite{cmov} due to inflaton scattering through single graviton exchange (i.e., with $\tilde \sigma = 0$) while neglecting the dominant contribution from large-scale fluctuations. In this case, there is a direct relation between $m_\chi$ and $T_{\rm RH}$, as given in Eq.~(\ref{Eq:omega0k2}). For further details, see \cite{Choi:2024bdn}.

If we generalize the previous results by allowing $k > 2$ while keeping $n=m=2$, we obtain solutions similar to Eq.~(\ref{s22sol}). The energy density of the spectator field can then be expressed as
\begin{equation}
    \rho_\chi(a) \; = \;  \frac{1}{2} m^2_{\chi, \rm eff} \chi^2(a) \; \simeq \;  \frac{3\He^4 \beta}{16\pi^2}  \left(\frac{\phi(a)}{\phi_{\rm end}} \right)^2 \left(\frac{a_{\rm end}}{a}\right)^{\frac{6}{k+2}} \, .
\end{equation}
Since the inflaton field scales as $\phi(a)^2 \propto a^{-\frac{12}{k+2}}$, the overall scaling of the spectator field energy density is given by $\rho_\chi(a) \propto a^{-\frac{18}{k+2}}$. The dependence of the field amplitude and energy density on the scale factor $a$ for different values of $k$ is summarized in Table~\ref{tab:fieldscaling}.

\begin{table}[ht!]
\begin{center}
\begin{tabular}{l|l|l|l|l||}
\cline{2-5}
&General&$k=2$ & $k=4$ & $k=6$  \\ 
\hline
\multicolumn{1}{||l|}{$\chi$}                                           & $\propto a^{-\frac{3}{2+k}}$  & $\propto a^{-\frac{3}{4}}$&$\propto a^{-\frac{1}{2}}$&$\propto a^{-\frac{3}{8}}$  
    \\ \hline
\multicolumn{1}{||l|}{$\phi$}                                           & $\propto a^{-\frac{6}{2+k}}$ &$\propto a^{-\frac{3}{2}}$& $\propto a^{-1}$& $\propto a^{- \frac{3}{4}}$            
    \\ \hline
\multicolumn{1}{||l|}{$\rho_{\phi}$}  
     & $\propto a^{-\frac{6k}{2+k}}$ &$\propto a^{-3}$& $\propto a^{-4}$& $\propto a^{-\frac{9}{2}}$  
\\ \hline
\multicolumn{1}{||l|}{$\rho_\chi$}  
   & $\propto a^{-\frac{18}{2+k}}$ &$\propto a^{-{\frac{9}{2}}}$& $\propto a^{-3}$& $\propto a^{-\frac{9}{4}}$  \\ 
    \hline
\end{tabular}
\caption{\em Field amplitude $\phi(a)$ and energy density $\rho_\chi(a)$ dependence on the scale factor $a$ for different values of $k$.}
\label{tab:fieldscaling}
\end{center}
\end{table}

More generally, we can rearrange Eq.~(\ref{interacteom}) into an Emden-Fowler equation:
\begin{equation}
    \frac{d^2}{dz^2}\chi+A z^{n-k-2}\chi^{m-1} \; = \; 0 \, ,
\end{equation} 
where $z$ defined in Eq.~\eqref{defz}, with $z=a^{-\frac{6}{k+2}}$, and $A=\frac{(k+2)^2 m \sigma_{n,m}M_P^{4-n-m}\phiend^n}{36\He^2}$. This equation can be identified with Eq.~\eqref{emden}, up to a constant factor in the second term, with the substitutions $p \rightarrow m$ and $k \rightarrow k - n$ (while keeping $z$ unchanged). The asymptotic behavior of $\chi$ follows the same discussion as long as $k - n \geq 2$. When $k=n$ and $m=2$, there exists an analytical solution, equivalent to the one discussed in \cite{Choi:2024bdn} for $k = n = m = 2$. However, if $m > 2$, assuming slow oscillations and using the power-law ansatz $\chi(z) \simeq c_0 z^b$, one finds $b=0$, which is inconsistent. To correctly account for the effects of oscillations, we instead use the ansatz~\eqref{fourieransatz}, which gives the solution $\chi\propto z^{\frac{2}{m+2}}\propto a^{-\frac{12}{(n+2)(m+2)}}$ for $z\rightarrow 0$. 

\begin{figure}[t!]
  \centering
\includegraphics[width=0.74\textwidth]{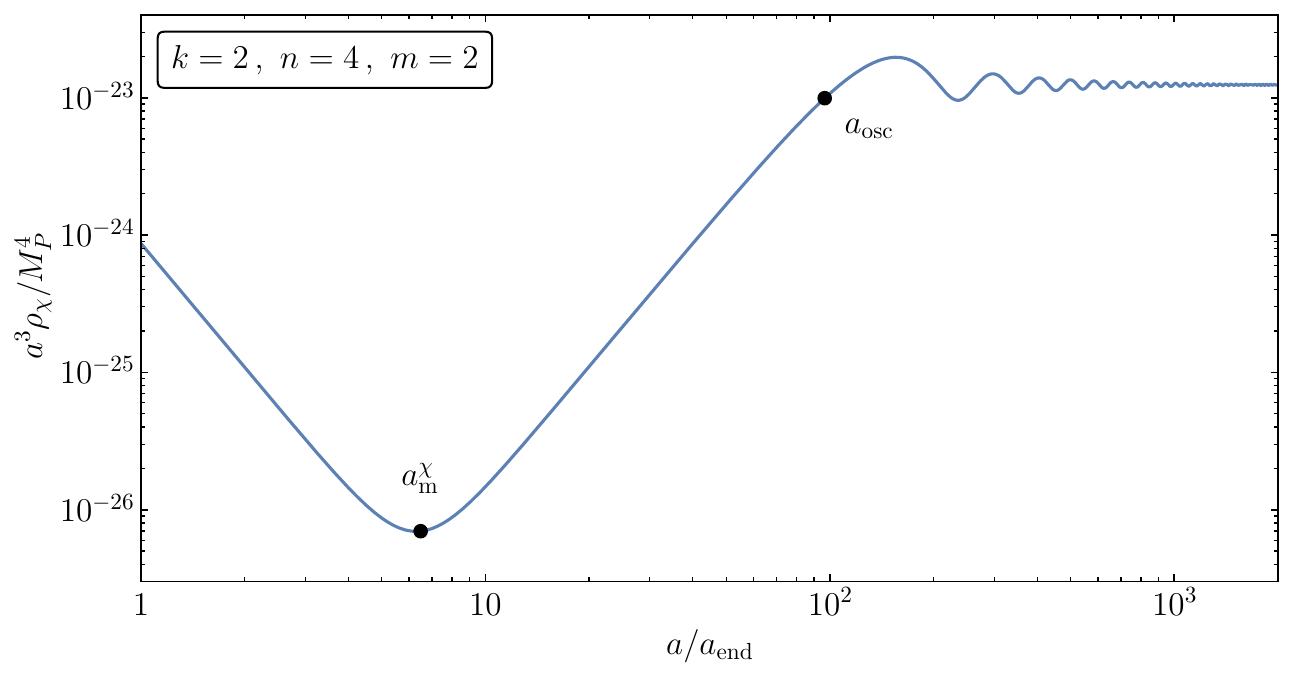}
  \caption{\em \small {The evolution of the $\chi$ energy density in the presence of an interaction term $\frac{  \sigma _{4,2}}{M_P^2}\phi^4 \chi^2$. We take $ {\sigma}_{4,2}= 10^{-12}$ and a bare mass $m_\chi=10^{10}\gev$. Initially, the redshift of $\rho_\chi$ is governed by the interaction term, leading to a scaling behavior of $\rho_\chi \propto a^{-6}$. As the Universe evolves, the bare mass term eventually dominates, causing the energy density to remain frozen until $H \sim m_\chi$. Thereafter, $\rho_\chi$ redshifts as $a^{-3}$, consistent with the expected analytical predictions.}
}
  \label{k2n4m2}
\end{figure}

When $k\leq n-2$ and $m=2$, there exists an exact solution in terms of Bessel functions. Expanding this solution to first order for large $a$, one finds $\chi \sim a^0$ so $\rho_\chi \sim a^{-\frac{6n}{k+2}}$, implying that $\rho_\chi \sim a^{-\frac{6n}{k+2}}$. Since this decreases as $a^{-6}$ or faster, the interaction term $\phi^n\chi^m$ rapidly becomes subdominant in the energy density compared to the bare mass term. This behavior is illustrated in Fig.~\ref{k2n4m2}. For $k\leq n-2$ and $m>2$, the ansatz \eqref{fourieransatz} gives $\chi\propto a^{\frac{6(n-k)}{(m-2)(k+2)}} $, leading to $\rho_\chi\propto a^{\frac{12n  -6km}{(k+2)(m-2)}}$. The exponent in the expression for $\rho_\chi$ can be positive, zero, or negative, leading to different physical implications. If the exponent is positive or zero, $\rho_\chi$ will quickly dominate over the inflaton energy density, violating the assumption that Eq.~\eqref{interacteom} remains valid. If the exponent is negative, the competition between the redshift of the interaction term and the bare mass term must be examined to determine which eventually dominates.
The energy density of $\chi$, given by $ \rho_\chi\simeq \sigma_{n,m}\phi^n\chi^m$, then evolves as:
\begin{equation}
    \rho_\chi \propto\left\{
    \begin{aligned}
      a^{-\frac{6m(2+k-n)}{(k+2)(m+2)}-\frac{6n}{k+2}}  ,&&   k\geq n+2\text{ and }m\leq2( k-n) \, ,\\a^{-\frac{6(k-n)m}{(m-2)(k+2)}-\frac{6n}{k+2}},&& k\geq n+2\text{ and }m>2( k-n ) \, ,\\ a^{-\frac{6(1-\sqrt{1-4A})}{n+2}-\frac{6n}{n+2}},&& k=n
    \text{ and }  m=2, \text{ }A\leq\frac{1}{4}\, ,\\ a^{-\frac{6}{n+2}-\frac{6n}{n+2}},&& k=n
    \text{ and }  m=2, \text{ }A>\frac{1}{4} \, ,\\     a^{-\frac{12m}{(n+2)(m+2)}-\frac{6n}{n+2}}  ,&  & k=n
    \text{ and }  m>2 \, ,\\  a^{-\frac{6n}{k+2}},& & k\leq n-2 \text{ and }m=2 \, ,\\  a^{\frac{6nm   -6km}{(k+2)(m-2)} -\frac{6n}{k+2}},& & k\leq n-2 \text{ and }m>2 \, .\end{aligned}\right.
    \label{lotsofposs}
\end{equation}
Note the presence of $\propto a^{-\frac{6n}{k+2}}$ in every expression, which is simply the redshift effect induced by $\phi^n$. As a result, for $n > k$, the effective mass of $\chi$ redshifts faster than the Hubble parameter, $H \propto a^{-\frac{6k}{k+2}}$. This prevents $\chi$ from entering the oscillatory regime until the bare mass term eventually dominates the potential. The validity of this result holds as long as the inflaton energy density remains the dominant contribution to the total energy budget.

To better understand the effect of the coupling of $\chi$ to the inflaton, consider $m=2$. The effective mass 
of the spectator field evolves as $\meff\sim \phi^{\frac{n}{2}}\sim a^{-\frac{3n}{k+2}}$ for $a>\aend$. This can be contrasted with the effective mass generated from self-interactions, $\meff \sim a^{-3\frac{p-2}{p+2}}$ for $p \le 2k$ for $a > \aosc$.
We then easily understand that, for $k=2$ for example,
values of $n\ge2$ will cause $\meff$ to redshift towards the bare mass contribution, $m_\chi$,
faster than the case of a self-interaction of the type $\lambda_\chi \chi^4$ ($p=4$). This means there is a shorter period of time before the bare mass dominates
leading to a  
greater density of dark matter. We also remark that, for $n>k$,
$\meff$ redshifts {\it faster} than $H$, which means $\chi$
will never enter in the oscillatory regime before the bare mass term dominates. Nevertheless, the energy density $\rho_\chi$ decreases since the effective mass decreases due to the coupling with the inflaton.

From Eq.~(\ref{lotsofposs}), we see that there are many possible evolutions for $\rho_\chi$ depending on the relative values of $k,n$, and $m$. To highlight the effect of the interactions between the inflaton and spectator fields, we conclude this section with a specific example: $k=2$, $n=4$, and $m=2$.

Assuming that the energy density of $\chi$ is initially dominated by the interaction term $\frac{ \sigma_{4,2}}{M_P^2}\phi^4 \chi^2$, the evolution of $\rho_\chi$ is characterized by three distinct periods. First, the redshift is driven by the interaction term, and the energy density decreases as $\rho_\chi \propto a^{-6}$ according to \eqref{lotsofposs}. This behavior is illustrated in Fig.~\ref{k2n4m2}, where for $a/\aend \lesssim 6$, the quantity $a^3 \rho_\chi$ decreases as $a^{-3}$. Next, when the bare mass term becomes comparable to the interaction term at $\am^\chi $, and if the Hubble parameter is still larger than the bare mass, the energy density becomes effectively frozen until the long-wavelength modes enter the horizon. In the figure, this is seen as $a^3 \rho_\chi \propto a^3$ for the interval $\am^\chi \lesssim a \lesssim \aosc$. Finally, once the oscillations of $\chi$ begin at $a = \aosc$, the energy density redshifts as matter, i.e., $\rho_\chi \propto a^{-3}$. This behavior is seen in Fig.~\ref{k2n4m2} for $a > \aosc \sim 100 \, \aend$.

\begin{figure}[t!]
  \centering
\includegraphics[width=0.7\textwidth]{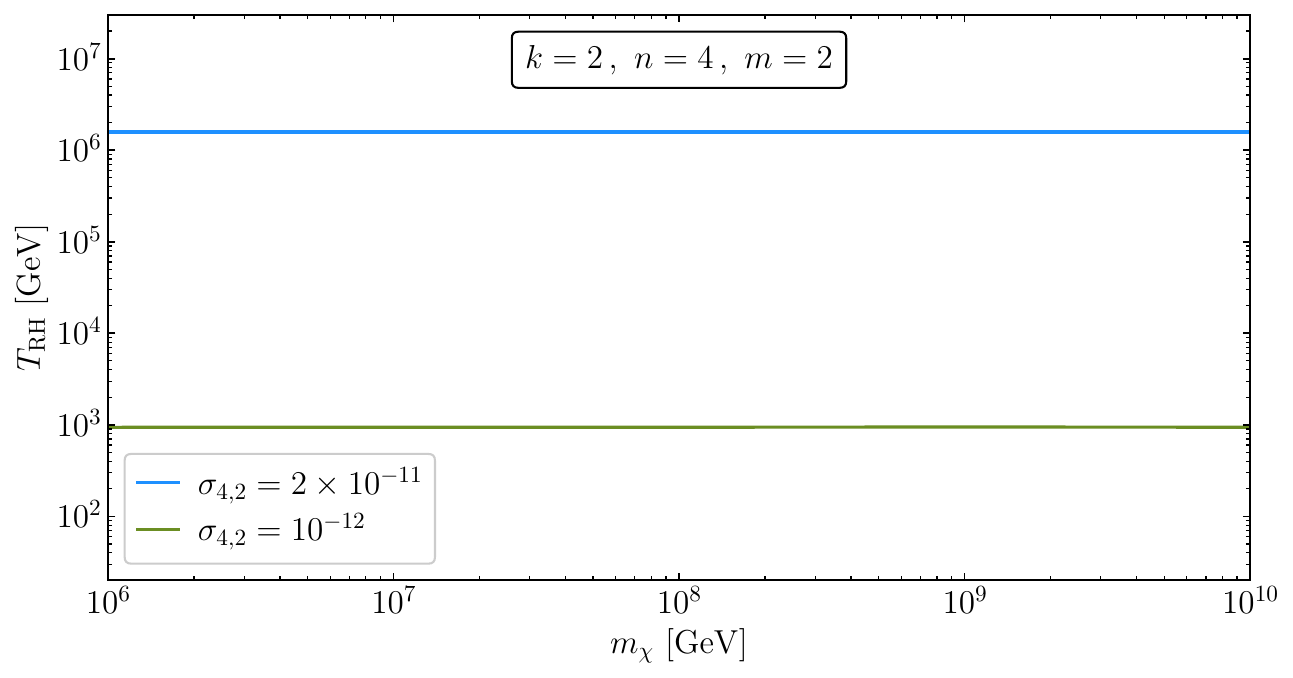}
  \caption{\em \small {The $(m_\chi,\trh)$ solutions satisfying $\Omega_\chi h^2=0.12$ (accounting only for long-wavelength production) in the presence of an interaction term $\frac{\sigma_{4,2}}{M_P^2} \phi^4 \chi^2$.}
}
  \label{k2n4m2mT}
\end{figure}

In Fig.~\ref{k2n4m2mT}, we show the $(m_\chi,\trh)$ parameter space satisfying $\Omega_\chi h^2=0.12$ for $n=4$ and $k=m=2$, considering two values of the coupling $\sigma_{4,2}$. Even for a coupling as small as $\sigma_{4,2} = 10^{-12}$ (which remains consistent with isocurvature constraints), the parameter space significantly broadens, with the solution independent of $m_\chi$. This independence arises from the fact that $\chi$ remains constant until $a = \am^\chi$, where its energy density is set by the interaction term $\frac{1}{2}\rho_\chi(\am^\chi)=\frac{ \sigma _{4,2}}{M_P^2}\phi^4 \chi(\am^\chi )^2=\frac{1}{2}m_\chi ^2\chi(\am)^2=\frac{1}{2}m_\chi ^2\chi_{\rm end}^2$. Since the energy density remains frozen until $a_{\rm osc}$, with $H(\aosc)=\frac{2}{3}m_\chi$, we find
\begin{equation}
    \left(\frac{\am^\chi}{\aend}\right)^3 \; = \; \left(\frac{2 \sigma _{4,2}\phiend^4}{M_P^2 m_\chi^2}\right)^{\frac{1}{2}} \, , 
\end{equation} 
and
\begin{equation}
\left(   \frac{ \aosc}{\aend}\right)^3 \; = \; \left(\frac{3\He}{2m_\chi }\right)^2 \, .
\end{equation}
Requiring that $\am^\chi <\aosc$ imposes an upper bound
\begin{equation}
    \sigma _{4,2}<\frac{81}{32}\left(\frac{\He^2M_P}{\phiend^2 m_\chi}\right)^2\simeq \frac{1.5\times 10^{15}\gev^2}{m_\chi^2} \, .
    \label{sigtilde}
\end{equation} 
Therefore, the present-day energy density is given by
\begin{equation}\begin{aligned}\rho_\chi^0&=\rho_\chi(\aosc)\left(\frac{\aosc}{\aend}\right)^3 \left(\frac{\aend}{\arh}\right)^3\left(\frac{\arh}{a_0}\right)^3\\&=\frac{1}{2}m_\chi^2 \chi_{\rm end}^2 \left(\frac{3\He}{2m_\chi }\right)^2\left(\frac{\aend}{\arh}\right)^3\left(\frac{\arh}{a_0}\right)^3\\&=\frac{1}{2}  \chi_{\rm end}^2 \left(\frac{3\He}{2  }\right)^2\left(\frac{\aend}{\arh}\right)^3\left(\frac{\arh}{a_0}\right)^3 \, .
\end{aligned}
\end{equation}
If the radiation energy density dominates over $\rho_\chi$ at $\arh$,\footnote{This requirement corresponds to $\rho_\phi(\aosc)>\rho_\chi(\aosc)$, which imposes $\sigma_{4,2}\gtrsim 6.5 \times 10^{-25}$.} then the last two factors become independent of $m_\chi$, leading to a relic density that does not depend on $m_\chi$.

On the other hand, for large $ \sigma _{4,2}$ violating \eqref{sigtilde}, the oscillations start 
without a frozen period. In this case, we can evaluate the effective mass of $\chi$ at the end of inflation:
\begin{equation}
    \meff^2\simeq 2\frac{ \sigma _{4,2}}
{M_P^2}\phiend^4>\frac{8.3\times 10^{51}\gev^4}{m_\chi^2} \, .
\end{equation}
For $m_\chi \lesssim 10^{12} \gev$, this is significantly larger than $\He^2$, leading to an exponential suppression of the relic density through the $\beta$ factor.


\section{Conclusion}
\label{summary}

In this work, we have studied in detail the phenomenology of a spectator field $\chi$, produced during inflation.
To concentrate on the evolution of $\chi$, we have chosen the  T-model attractors \cite{Kallosh:2013hoa} as a viable phenomenological example for inflation which easily allows for different
equations of state, parameterized by $k$, and therefore different cosmic expansion rates during the period of reheating. For these models, the inflaton potential can be approximated by $V(\phi) \sim\lambda \phi^k$, near the minimum of the potential. We further assumed a general form for the spectator potential,
$V(\chi)=\frac{1}{2}m_\chi^2 \chi^2 + \lambda_\chi \chi^p + \sigma_{n,m}\phi^n\chi^m$
and studied in turn the effects of turning on or off parameters in this potential. 

The set of initial conditions we employed were discussed in Section \ref{sec:init}. For each value of $k$, we use the the CMB observables (primarily $A_s$ and $n_s$)
to fix the inflationary parameters such as $H_I$, $\rhoe$, and $\He$. During inflation, fluctuations in the spectator field tend to grow as $\langle \chi^2 \rangle \propto H_I^3 t$ up to a maximum value determined by $H_I$ and the effective mass of the spectator $\meff$. The details of the determination of $\chi^2_{\rm end}$ are given in Appendix \ref{app:sa}. Once the initial conditions are fixed, the inflaton and spectator fields evolve according to their equations of motion as described in Section \ref{sec:inflation}. We also applied several constraints on the spectator parameters coming from 1) the production of isocurvature fluctuations, 2) the avoidance of a second period of inflation, and 3) the fragmentation of the inflaton condensate. 
These were described in Section \ref{sec:constraints}. 
The most severe of these is the production of isocurvature fluctuations which sets a lower limit on the effective mass of the spectator to be larger than roughly $\He/2$ or significantly lighter ($\lesssim 5 \times 10^{-6}\He$). These limit apply to either the bare mass or the self-coupling responsible for the bare mass. While not a trivial constraint, it is relatively easy to ensure that a second period of inflation is not triggered by the spectator, as this would have disastrous consequences on the spectrum of density fluctuations. Finally, fragmentation can generally be controlled by a suitably 
large reheating temperature \cite{Garcia:2023dyf}.

The simplest case, $\lambda_\chi=\sigma_{n,m}=0$, is that of a stable non-interacting (other than gravity) scalar field with only a bare mass term. In this case, 
the spectator fields tend to be overproduced leading to an unacceptably large relic density. After inflation, $\langle \chi^2 \rangle$ is frozen until the expansion rate (dominated by the inflaton) drops sufficiently so that $H \simeq m_\chi$, at which point the spectator begins to oscillate and its energy density redshifts as $a^{-3}$. For $k=2$, the energy density, $\rho_\chi$ is sufficiently diluted so that $\Omega_\chi h^2 = 0.12$ can be found for $m_\chi \gtrsim 3 \times 10^{12}~\gev$. When $m_\chi \sim \He$ large scale fluctuations become suppressed and short-wavelength production becomes important until the kinematic limit of $m_\chi = m_\phi$ is reached. Furthermore for this range of spectator masses, we must require $\trh \lesssim 7$~TeV to avoid overproduction. For larger $k$, as the long-wavelength production of $\chi$ requires very large masses (of order $\He$ and short-wavelength production requires relatively low masses ($\lesssim 100~\gev$) as can be seen by comparing Eqs.~(\ref{Eq:casek4}, \ref{Eq:casek6}) and Eqs.~(\ref{Eq:omega0k4}, \ref{Eq:omega0k6}).

On the other hand, we have seen that adding self-interactions of the form $\lambda_\chi \chi^p$ in the potential $V(\chi)$, 
significantly alters the allowed parameter space. If the self-interactions are sufficiently strong at the time oscillations begin (now when $\meff \simeq \He$) the energy density of the   spectator redshifts as $a^{-6p/(p+2)}$ for $p\le2k$ and $a^{-6p k/(k+2)(p-2)}$ for $p>2k$, in both cases faster than $a^{-3}$ for $p>2$. 
For large self-couplings which satisfy the isocurvature constraint, a wide range of masses is allowed $m_\chi = 10-10^{13}~\gev$
with reheating temperatures between $2000-2\times 10^{15}~\gev$ for $p=4$ as seen in the upper left panel of Fig.~\ref{rhochip}. For small lambda (still satisfying the isocurvature constraint),  the spectator mass must be less than roughly 1 TeV, and sub-GeV and even sub-eV masses are allowed. The analytic results for these cases are summarized in Eqs.~(\ref{Eq:k2p4}-\ref{Eq:k6p4}).

For $k=4$, the universe expands as if it were radiation-dominated, and hence the resultant relic density of the spectator field does not depend on $\trh$. Recall that in the absence of self-interactions, there were no consistent solutions. The presence of self-interactions now opens up the low mass range ($m_\chi \lesssim 10~\gev$) for $p = 4, 6$ and 8 as seen in Fig.~\ref{lk4}. These are now consistent with the short-wavelength production which still provides an upper limit of $\sim 120~\gev$.

Finally, we also considered the possibility for interactions
between the inflaton and the spectator field of the form,
$V(\phi,\chi)= \sigma_{n,m}\phi^n \chi^m$. The simplest case $k=m=n=2$ was considered in detail in \cite{Choi:2024bdn}. For suitably large couplings, $\sigma_{n,m}$, the spectator redshifts more quickly than $a^{-3}$ if the interaction terms dominate over the bare mass. For $n=m=2$, $\rho_\chi$ initially scales as $a^{-9/2}$. As in the case of self-interactions, the interactions with the inflaton lead to a significant effective mass for the spectator which moderates the size of the long-wavelength fluctuations. More generally, we found the complete set of solutions for the evolution of $\rho_\chi$
for all $k, m$, and $n$. These are given in Eq.~(\ref{lotsofposs}). In each case, there is a contribution to $\rho_\chi$ which scales as $a^{6n/(k+2)}$ reflecting the redshift of the effective mass of $\chi$. For $n > k$, the mass redshift faster than the Hubble parameter, and oscillations only begin when the bare mass term dominates. We described one interesting example of this case with $k=m=2$ and $n=4$. In this case, as expected, the spectator field remains frozen initially and its energy density decreases due to the evolution of the inflaton. The final relic density is largely independent of the spectator mass and we obtain an upper limit on $\trh$ which depends on the coupling $\sigma_{4,2}$.

In summary, we have explored the parameter space of spectator scalar field as a candidate for dark matter. Such a scalar is perhaps the simplest possible candidate for dark matter as it requires only stability. The model is highly constrained if there are no non-gravitational interactions, and there is a wide range of plausible masses and reheating temperatures depending on the 
nature and strength of either the spectator's self-interactions or interactions with the inflaton.

\acknowledgments
This project has received support from the European Union's Horizon 2020 research and innovation program under the Marie Sklodowska-Curie grant agreement No 860881-HIDDeN and the CNRS-IRP project UCMN. The authors would like to acknowledge Simon Clery, Mathieu Gross, and Jong-Hyun Yoon for useful discussions. The work of K.A.O.~was supported in part by DOE grant DE-SC0011842 at the University of Minnesota. The work of M.A.G.G.~was supported by the DGAPA-PAPIIT grant IA100525 at UNAM, and the CONAHCYT ``Ciencia de Frontera'' grant CF-2023-I-17. S.V. was supported in part by DOE grant DE-SC0022148 at the University of Florida.

\clearpage

\appendix

\section{Stochastic Approach}
\label{app:sa}
As discussed in Refs.~\cite{Starobinsky:1986fx, Starobinsky:1994bd, Kamenshchik:2021tjh}, the expectation value of the spectator field, $\chi$,
can be computed using a stochastic approach. This method treats the long-wavelength superhorizon modes of the spectator field $\chi(t, \mathbf{x})$ in a de Sitter background as a classical stochastic variable, $\varphi$, governed by the probability distribution $\rho(t, \varphi)$. The evolution of $\rho(t, \varphi)$ is described by the Fokker-Planck equation:
\begin{equation}
    \label{eqapp:fokkerplanck}
    \frac{\partial \rho}{\partial t} \; = \; \frac{H_I^3}{8\pi^2} \frac{\partial^2 \rho}{\partial \varphi^2} + \frac{1}{3H_I} \frac{\partial}{\partial \phi} \left(\frac{\partial V}{\partial \varphi} \rho(t, \varphi) \right) \, ,
\end{equation}
where $H_I$ is the Hubble parameter during inflation, and $V(\varphi)$ is the effective potential for the stochastic variable, $\varphi$.

For a potential with an effective mass term and a self-interaction term
\begin{equation}
    \label{appeq:effpotquartic}
    V(\varphi) \; = \; \frac{1}{2} m_{\chi, \rm eff}^2 \varphi^2 + \lambda_\chi M_P^4 \left( \frac{\varphi}{M_P} \right)^p \, ,
\end{equation}
the equilibrium solution to the Fokker-Planck equation at late times is
\begin{equation}
\rho_{\mathrm{eq}}(\varphi)=N^{-1} \exp \left(-\frac{8 \pi^2}{3 H_I^4} V(\varphi)\right) \, ,
\end{equation}
where $N$ is the normalization factor, determined by
\begin{equation}
    \int_{-\infty}^{\infty} \rho_{\mathrm{eq}}(\varphi) d \varphi=1 \, .
\end{equation}

We first consider the case $p = 4$. Using the effective potential~(\ref{appeq:effpotquartic}), the normalization constant $N$ is given by:
\begin{equation}
N \; = \; \int_{-\infty}^{\infty} \exp \left[-\frac{8 \pi^2}{3 H_I^4}\left(\frac{m_{\chi, \rm eff}^2 \varphi^2}{2} + \lambda_{\chi} \varphi^4 \right)\right] d \varphi \; = \; \frac{m_{\chi, \rm eff}}{2\sqrt{2\lambda_{\chi}}} \exp(z) \mathcal{K}_{\frac{1}{4}}(z) \, ,
\end{equation}
where $z \equiv \frac{\pi^2 m_{\rm eff}^4}{12 \lambda_{\chi} H_I^4}$ and $\mathcal{K}_{\frac{1}{4}}(z)$ is a modified Bessel function of the second kind. 

Next, we compute the expectation value of $\langle \chi^2 \rangle$. Using the equilibrium probability distribution $\rho_{\rm eq}(\varphi)$, the expectation value is given by:
\begin{equation}
    \begin{aligned}
    \langle \chi^2 \rangle \; = \; \langle \varphi^2 \rangle \; = \; \int_{-\infty}^{\infty} \varphi^2 \rho_{\rm eq}(\varphi) d\varphi \; = \;  \frac{1}{N}\int_{-\infty}^{\infty} \varphi^2 \exp\left[-\frac{8\pi^2}{3H^4} \left( \frac{m_{\chi, \rm eff}^2 \varphi^2}{2}+ \lambda_{\chi} \varphi^4 \right)\right] d\varphi   \\ 
    \; = \; \frac{1}{N}\frac{\pi m_{\chi, \rm{eff}}^3}{32 \lambda_{\chi}^{3/2}} \exp(z) \left[ \mathcal{I}_{\frac{1}{4}}(z) - \mathcal{I}_{-\frac{1}{4}}(z) + \mathcal{I}_{\frac{5}{4}}(z) - \mathcal{I}_{\frac{3}{4}}(z) \right]
+ \frac{1}{N} \frac{3H_I^4}{16 \pi m_{\chi, \rm{eff}} \lambda_{\chi}^{1/2}} \exp(z) \mathcal{I}_{\frac{1}{4}}(z) \, .
    \end{aligned}
\end{equation}
Here $\mathcal{I}_{\frac{1}{4}}(z)$ is the modified Bessel function of the first kind. Using the properties of the Bessel functions, we can rewrite the expectation value in the following form:
\begin{equation}
    \langle \chi^2 \rangle \; = \; \langle \varphi^2 \rangle \; = \; \frac{m_{\chi, \rm eff}^2}{8 \lambda_{\chi}} \left( \frac{\mathcal{K}_{\frac{3}{4}}(z)}{\mathcal{K}_{\frac{1}{4}}(z)} - 1 \right) \, .
\end{equation}
In the limit $\lambda_{\chi} H_I^4 \ll m_{\chi, \rm eff}^4$, the expectation value expands to
\begin{equation}
    \langle \chi^2 \rangle \; \simeq \; \frac{3 H_I^4}{8 \pi^2 m_{\chi, \rm eff}^2} - \frac{27 \lambda_{\chi} H_I^8}{16 \pi^4 m_{\chi, \rm eff}^6} \, .
\end{equation}
If $\lambda_{\chi} = 0$, the standard result $ \langle \chi^2 \rangle \; \simeq \; \frac{3 H_I^4}{8 \pi^2 m_{\chi, \rm eff}^2}$ is recovered. In the opposite limit $\lambda_{\chi} H_I^4 \gg m_{\chi, \rm eff}^4$, the expectation value becomes
\begin{equation}
    \label{appeq:expquartic}
    \langle \chi^2 \rangle \; \simeq \;  \sqrt{\frac{3}{2\pi^2}} \frac{\Gamma(3/4)}{\Gamma(1/4)} \frac{H_I^2}{2 \sqrt{\lambda_{\chi}}} \; \simeq \; 0.07 \frac{H_I^2}{\sqrt{\lambda_{\chi}}} \, .
\end{equation}

Next, we consider a more general case where the potential takes the form $V(\varphi) \simeq \lambda_{\chi}  M_P^4 \left( \frac{\varphi}{M_P} \right)^p$, assuming that the effective mass contribution in Eq.~(\ref{appeq:effpotquartic}) is subdominant. In this scenario, the normalization constant $N$ is given by
\begin{equation}
N \; = \; \int_{-\infty}^{\infty} \exp \left[-\frac{8 \pi^2}{3 H_I^4}\left( \lambda_{\chi} \varphi^p M_P^{4-p} \right)\right] d \varphi \; = \; 2 \Gamma(1/p) \left(\frac{3H_I^4 }{8\pi^2 \lambda_{\chi}M_P^{4-p}} \right)^{\frac{1}{p}} \, .
\end{equation}
The expectation value is then given by
\begin{equation}
    \begin{aligned}
    \langle \chi^2 \rangle \; = \; \langle \varphi^2 \rangle & \; = \; \int_{-\infty}^{\infty} \varphi^2 \rho_{\rm eq}(\varphi) d\varphi \; = \;  \frac{1}{N}\int_{-\infty}^{\infty} \varphi^2 \exp \left[-\frac{8 \pi^2}{3 H_I^4}\left(\lambda_{\chi} \varphi^p M_P^{4-p} \right)\right] d\varphi \\
    &\; = \; \frac{2 \Gamma(3/p)}{N} \left(\frac{3H_I^4 }{8\pi^2 \lambda_{\chi}M_P^{4-p}} \right)^{\frac{3}{p}} \; = \; \left(\frac{3 H_I^4}{8 \pi^2 \lambda_{\chi} M_P^{4-p}} \right)^{\frac{2}{p}} \frac{\Gamma(3/p)}{\Gamma(1/p)} \, .
    \end{aligned}\label{inivalp}
\end{equation}
For $p = 2$, this reduces to $ \langle \chi^2 \rangle \; \simeq \; \frac{3 H_I^4}{16 \pi^2 \lambda_{\chi} M_P^2}$, and for $p=4$, it reproduces Eq.~(\ref{appeq:expquartic}). For higher values of $p$, the results are
\begin{equation}
    \begin{aligned}
    \label{appeq:generalexpect6}
    &\langle \chi^2 \rangle \; \simeq \; 0.11 \left(\frac{H_I^4 M_P^2}{\lambda_{\chi}} \right)^{1/3} \,, \quad &(p=6)\, , \\
    &\langle \chi^2 \rangle  \; \simeq \; 0.14 \left(\frac{H_I^4 M_P^4}{\lambda_{\chi}} \right)^{1/4} \,, \quad &(p=8)\, , \\
    &\langle \chi^2 \rangle  \; \simeq \; 0.16 \left( \frac{H_I^4 M_P^6}{\lambda_{\chi}}\right)^{1/5} \,, \quad &(p=10)\, .
    \end{aligned}
\end{equation}

\section{Correlation Functions}
\label{app:corfuncts}
Following the computations in Refs.~\cite{Starobinsky:1994bd} and~\cite{Markkanen:2019kpv}, we consider the two-point correlation function for a local field function $f(\phi)$:
\begin{equation}
G_f(t_1, t_2; \vec{r}_1, \vec{r}_2) \equiv \langle f(\phi(t_1, \vec{r}_1)) f(\phi(t_2, \vec{r}_2)) \rangle \, .
\end{equation}
Due to de Sitter invariance, any correlator of a scalar observable $f(\phi)$ depends only on the de Sitter invariant quantity
\begin{equation}
y = \cosh H_I(t_1 - t_2) - \frac{H_I^2}{2} e^{H_I(t_1 + t_2)} |\vec{r}_1 - \vec{r}_2|^2 \, ,
\end{equation}
where $\vec{r}_1$ and $\vec{r}_2$ are comoving position vectors. For $|y| \gg 1$, both time-like and space-like separations can be expressed as
\begin{equation}
G_f(t_1, t_2; \vec{r}_1, \vec{r}_2) = G_f\left( H_I^{-1} \ln|2y - 1| \right) \, ,
\end{equation}
where the right-hand side represents the temporal correlation function
\begin{equation}
G_f(t) \equiv G_f(t; 0) = \langle f(\phi(0)) f(\phi(t)) \rangle \, ,
\end{equation}
corresponding to $\vec{r}_2 = \vec{r}_1$. This can be computed straightforwardly using the stochastic approach. 

We use the following form of the Fokker-Planck equation~(\ref{eqapp:fokkerplanck}):
\begin{equation}
\tilde{\rho}(t; \varphi) = e^{\frac{4\pi^2 V(\varphi)}{3H_I^4}} \rho(t; \varphi) \, ,
\end{equation}
which satisfies
\begin{equation}
\frac{\partial \tilde{P}(t; \phi)}{\partial t} = \frac{3H_I^3}{4\pi^2} \tilde{D}_\phi \tilde{P}(t; \phi) \, ,
\end{equation}
with
\begin{equation}
\tilde{D}_\phi = \frac{1}{2} \frac{\partial^2}{\partial \phi^2} - \frac{1}{2} \left( v'(\phi)^2 - v''(\phi) \right), \quad v(\phi) = \frac{4\pi^2}{3H_I^4} V(\phi) \, .
\end{equation}
This linear equation allows solutions via separation of variables, with independent solutions of the form $\tilde{P}_n(t; \phi) = e^{-\Lambda_n t} \psi_n(\phi)$, where $\psi_n(\phi)$ satisfies the time-independent Schrödinger-like eigenvalue equation
\begin{equation}
\tilde{D}_\phi \psi_n(\phi) \; = \; -\frac{4\pi^2 \Lambda_n}{H_I^3} \psi_n(\phi) \, .
\end{equation}
To solve this eigenvalue equation, we consider a general potential of the form
\begin{equation}
    V(\phi) \; = \; \lambda_{\chi}  M_P^4 \left( \frac{\phi}{M_P} \right)^p \, ,
\end{equation}
which leads to the Schrödinger-like equation
\begin{equation}
    \label{eqapp:fokkerplanck2}
    \frac{1}{2} \left\{ \frac{\partial^2}{\partial \phi^2} - \left( \frac{4\pi^2}{3H_I^4} \right)^2 \left( p^2 \lambda_{\chi}^2 M_P^{8-2p} \phi^{2p-2} \right) + \frac{4\pi^2}{3H_I^4} \left( p(p-1) \lambda_{\chi} M_P^{4-p} \phi^{p-2} \right) \right\} \psi_n(\phi) = -\frac{4\pi^2}{H_I^3} \Lambda_n \psi_n(\phi) \, .
\end{equation}

To simplify, we introduce dimensionless variables
\begin{equation}
\label{eqapp:eigensystem1}
\left\{ \frac{\partial^2}{\partial x^2} - U(x) \right\} \psi_n(x) = -8\pi^2 \frac{\Lambda_n(\alpha) M_P^{2-\frac{8}{p}}}{\left(p \lambda \right)^{2/p} H_I^{3-\frac{8}{p}}} \psi_n(x) \, ,
\end{equation}
where
\begin{equation}
    x \; \equiv \; \frac{\left(p \lambda \right)^{1/p}}{H_I^{4/p} M_P^{1- \frac{4}{p}}} \phi   \, ,
\end{equation}
and
\begin{equation}
    U(x) \; = \; \left( \frac{4\pi^2}{3} \right)^2 x^{2p-2} - \frac{4\pi^2}{3} p(p-1)x^{p-2} \, .
\end{equation}
We solve this eigensystem for $p = \{2, 4, 6, 8, 10 \}$ and summarize the lowest five eigenvalues in Table~\ref{tab:appeigen}. The results illustrate the dependence of the eigenvalues on the parameters $p$ and $\lambda_\chi$, highlighting the effects of self-interactions in the potential.

\begin{table}[ht!]
\centering
\begin{tabular}{l|l|l|l|l|l||}
\cline{2-6}
& $\Lambda_0$ & $\Lambda_1$ & $\Lambda_2$ & $\Lambda_3$ & $\Lambda_4$ \\ 
\hline
\multicolumn{1}{||l|}{$p=2$ }                 
    &$0$&$\frac{1}{3} \frac{\lambda_{\chi} M_P^2}{H_I}$&$\frac{2}{3}\frac{\lambda_{\chi} M_P^2}{H_I}$&$\frac{\lambda_{\chi} M_P^2}{H_I}$&$\frac{4}{3} \frac{\lambda_{\chi} M_P^2}{H_I}$
    \\ \hline
\multicolumn{1}{||l|}{$p=4$}   
    &$0$&$0.09\sqrt{\lambda_{\chi}} H_I$&$0.29\sqrt{\lambda_{\chi}} H_I$&$0.54\sqrt{\lambda_{\chi}} H_I$&$0.83\sqrt{\lambda_{\chi}} H_I$
    \\ \hline
\multicolumn{1}{||l|}{$p=6$}  
     &$0$&$0.06 \frac{H_I^{5/3} \lambda^{1/4}}{M_P^{2/3}}$&$0.21\frac{H_I^{5/3} \lambda^{1/4}}{M_P^{2/3}}$&$0.43\frac{H_I^{5/3} \lambda^{1/4}}{M_P^{2/3}}$&$0.70\frac{H_I^{5/3} \lambda^{1/4}}{M_P^{2/3}}$
     \\ \hline
\multicolumn{1}{||l|}{$p=8$}  
&$0$&$0.05 \frac{H_I^2 \lambda^{1/4}}{M_P}$&$0.18  \frac{H_I^2 \lambda^{1/4}}{M_P}$&$0.38  \frac{H_I^2 \lambda^{1/4}}{M_P}$&$0.64  \frac{H_I^2 \lambda^{1/4}}{M_P}$
\\ \hline
\multicolumn{1}{||l|}{$p=10$}  
&$0$&$0.04 \frac{H_I^{11/5} \lambda^{1/5}}{M_P^{6/5}}$&$0.16 \frac{H_I^{11/5} \lambda^{1/5}}{M_P^{6/5}}$&$0.35 \frac{H_I^{11/5} \lambda^{1/5}}{M_P^{6/5}}$&$0.60 \frac{H_I^{11/5} \lambda^{1/5}}{M_P^{6/5}}$
   \\ \hline
\end{tabular}
\caption{\em The lowest five eigenvalues of Eq.~(\ref{eqapp:eigensystem1}) for $p = \{2, 4, 6, 8, 10 \}$}
\label{tab:appeigen}
\end{table}

\section{Fragmentation of the Spectator Field}
\label{app:fragmentation}
In this appendix, we discuss the computation of the fragmentation of the spectator field, $\chi$. We follow a similar methodology to that used for the inflaton field in Section~\ref{sec:fragm}. The fragmentation of $\chi$ arises due to its self-interactions $\lambda_{\chi} \chi^p$, which lead to the excitation of spatially inhomogeneous fluctuations, $\delta \chi(t,\bx)$, continuously depleting the energy density of the homogeneous component, $\chi(t)$, analogous to the fragmentation of the inflaton condensate. At linear order, and neglecting metric perturbations, the growth of $\delta \chi$ is governed by the following equation:
\begin{equation}
\delta \ddot{\chi} + 3H \delta \dot{\chi} - \frac{\nabla^2 \delta\chi}{a^2} + p(p-1)\lambda_{\chi} M_P^2 \left( \frac{\chi(t)}{M_P} \right)^{p-2} \delta \chi \; = \; 0 \, .
\end{equation}
In general, when the spectator scalar field $\chi$ begins oscillating, it behaves as an underdamped anharmonic oscillator. Its dynamics can be parametrized in terms of an envelope function $\chi_0(t)$, which encodes the redshift due to cosmic expansion, and a quasi-periodic function $\mathcal{P}(t)$, which describes the short time-scale oscillations:
\begin{equation}
\label{app:chioscgen}
\chi(t) \; \simeq \;  \chi_0(t) \cdot \, \mathcal{P}(t) \, .
\end{equation}
The instantaneous effective mass of the spectator field is given by $m_{\chi}(t) = \sqrt{V_{\chi}''(\chi_0(t))}$. As the field oscillates, the effective mass and the equation of state parameter evolve, eventually leading to the fragmentation of the homogeneous condensate into localized bath of free particles.

We now consider the linear regime and integrate the Boltzmann equation for the spectator field fluctuations. The fragmentation interactions of the spectator field quanta can be obtained by expanding the coupling term $\lambda_{\chi} \chi^p$ and replacing $\chi(t) \rightarrow \chi(t) + \delta \chi(t, \mathbf{x})$. This leads to the following interaction Lagrangian term:\footnote{We note that there are additional terms, such as $\chi ^{p-3}\delta \chi^3$, but these lead to strongly-suppressed production rates due to phase-space volume and are therefore neglected.}
\begin{equation}
\label{app:interactionfragmchi}
\mathcal{L}_I \; = \; \frac{p(p-1)}{2} \lambda_{\chi} M_P^2 \left( \frac{\chi}{M_P}\right)^{p-2} \delta\chi^2 \, .
\end{equation}

Following Refs.~\cite{Nurmi:2015ema, Kainulainen:2016vzv, gkmo2, Garcia:2022vwm}, the Boltzmann equation in the presence of anharmonic oscillations of the inflaton background can be expressed as:
\begin{equation}
\begin{aligned}
\frac{\partial f_{\delta\chi}}{\partial t} - H |\mathbf{P}| \frac{\partial f_{\delta\chi}}{\partial |\mathbf{P}|} = \frac{1}{P^0} \sum_{n=1}^{\infty} \int \frac{d^3 \mathbf{K}_n}{(2\pi)^3 n_\chi} \frac{d^3 \mathbf{P}'}{(2\pi)^3 2P'^0} (2\pi)^4 \delta^{(4)} \left( K_n - P - P' \right) \left| \overline{\mathcal{M}_n} \right|^2 \\
\times \bigg[ f_\chi (K_n) \left(1 + f_{\delta\chi}(P)\right) \left(1 + f_{\delta\chi} (P')\right) - f_{\delta\chi} (P) f_{\delta\chi} (P') \left(1 + f_\chi (K_n)\right) \bigg] \, .
\end{aligned}
\end{equation}
Here, the physical four-momenta of the fluctuations are denoted by $P$ and $P'$, with $f_{\chi} = (2\pi)^3 n_{\chi}(t) \delta^{(3)}(\mathbf{k})$, where $n_{\chi}(t)$ is the instantaneous inflaton number density. The function $f_{\delta \chi}$ represents the Bose-Einstein distribution.\footnote{In general, we disregard Bose enhancement effects for the $\chi$-decay products.} The physical four-momentum of the spectator field condensate is $K_n = (E_n, \mathbf{0})$, where $E_n = n \, \omega_\chi$ denotes the energy of the $n^\text{th}$ oscillating mode. The transition amplitude $\mathcal{M}_n$ corresponds to the production of a pair of inflaton quanta $|f\rangle = |\delta\chi \, \delta\chi\rangle$ from the $n^\text{th}$ Fourier mode of the coherently oscillating spectator field. The transition probability can be written as:
\begin{equation}
\left| \langle f \left| i \int d^4 x \, \mathcal{L}_I \right| 0 \rangle \right|^2 = \text{Vol}_4 \sum_{n=-\infty}^{\infty} \left| \mathcal{M}_n \right|^2 (2\pi)^4 \delta^{(4)} (K_n - P - P') \, ,
\end{equation}
where $\text{Vol}_4$ represents the space-time volume. The matrix elements $\mathcal{M}_n$ are determined from the interaction Lagrangian and the corresponding mode decomposition of the oscillating spectator field. These elements govern the rate at which energy is transferred from the coherent oscillations of the spectator field into the excitation of fluctuations, ultimately leading to fragmentation. From the interaction term~(\ref{app:interactionfragmchi}), the matrix elements $\mathcal{M}_n$ are given by:
\begin{equation}
\mathcal{M}_n \; = \; p(p-1) \lambda_{\chi} M_P^2 \left(\frac{\chi_0}{M_P} \right)^{p-2} \left(\mathcal{P}^p \right)_n \quad \Rightarrow \quad \left| \overline{\mathcal{M}_n} \right|^2 \; = \; \frac{p^2(p-1)^2}{2} \lambda_{\chi}^2 M_P^4 \left(\frac{\chi_0}{M_P} \right)^{2p-4} \left| \left(\mathcal{P}^p \right)_n \right|^2 \, .
\end{equation}
Here, the mean amplitude squared $\left| \overline{\mathcal{M}_n} \right|^2$ includes a factor of 2 accounting for identical particles $\delta \chi$ in the final state. The coefficients $\left( \mathcal{P}^p \right)_n$ are the Fourier coefficients from the harmonic decomposition over one oscillation of the square of the quasi-periodic function $\mathcal{P}(t)$ defined in Eq.~(\ref{app:chioscgen}). These coefficients are derived from the expansion:
\begin{equation}
V_{\chi}(\chi) \; = \; V_{\chi}(\chi_0) \sum_{n=-\infty}^{+\infty} \left(\mathcal{P}^p \right)_n e^{-i n \omega_{\chi} t} \; = \; \rho_\chi \sum_{n=-\infty}^{+\infty} \left(\mathcal{P}^p \right)_n e^{-i n \omega_{\chi} t} \, ,
\end{equation}
where the oscillation frequency $\omega_{\chi}$ is given by \cite{gkmo2}
\begin{equation}
\omega_{\chi} \; = \; m_\chi \sqrt{\frac{\pi p}{2(p - 1)} } \, \frac{\Gamma\left( \frac{1}{2} + \frac{1}{p} \right)}{\Gamma\left( \frac{1}{p} \right)} \, .\label{freqchi}
\end{equation}

Following Ref.~\cite{gkmo2}, the fragmentation rate can be expressed as:
\begin{equation}
\Gamma_{\delta \chi } \; = \; \frac{1}{8\pi (1 + w_\chi) \rho_\chi} \sum_{n=1}^{\infty} \left| \overline{\mathcal{M}_n} \right|^2 E_n \, \beta_n (m_A, m_B) \, ,
\end{equation}
where 
\begin{equation}
\beta_n (m_A, m_B) \; \equiv \; \sqrt{ \left( 1 - \frac{(m_A + m_B)^2}{E_n^2} \right) \left( 1 - \frac{(m_A - m_B)^2}{E_n^2} \right) } \, .
\label{betachi}\end{equation}
Here, $m_A$ and $m_B$ represent the masses of the decay products (in this case, $\delta \chi$ quanta), and $E_n = n \omega_{\chi}$ is the energy associated with the $n^\text{th}$ oscillation mode. This formulation captures both the kinematic constraints and the energy transfer efficiency in the fragmentation process. For $m_A = m_B = m_{\chi}$, we can express the above equation as 
\begin{equation}
\label{eq:fragmentation1}
\Gamma_{\delta \chi} \; = \; \frac{p^2(p-1)^2 \lambda_{\chi}}{16 \pi (1 + w_\chi)} \left( \frac{\chi_0}{M_P}\right)^{p-4}  \omega_{\chi}\sum_{n=1}^{\infty} n \left|\left(\mathcal{P}^p \right)_n \right|^2  \, \sqrt{ 1 - \left( \frac{2 m_\chi}{n \omega_\chi} \right)^2} \, .
\end{equation}

Using $w_{\chi} = \frac{p - 2}{p + 2}$ and $m_{\chi} = \sqrt{\lambda_{\chi}} \sqrt{p(p - 1)} \left( \frac{\chi_0}{M_P} \right)^{\frac{p - 2}{2}} M_P$, we can rewrite the fragmentation rate as:
\begin{equation}
\Gamma_{\delta \chi} \; = \; \frac{(p+2)p^{3/2}(p-1)^{5/2} \lambda_{\chi}^{3/2} \omega_\chi}{32\pi m_\chi} \left( \frac{\chi_0}{M_P} \right)^{\frac{3p-10}{2}} M_P \sum_{n=1}^{\infty} n \left| \left( \mathcal{P}^p \right)_n \right|^2 \, \sqrt{ 1 - \left( \frac{2 m_\chi}{n \omega_\chi} \right)^2 } \, .
\end{equation}

For specific values of $p$, we find the following fragmentation rates:

\begin{align}
\label{eq:fragp4}
\Gamma_{\delta \chi} &\; \simeq \; 2.2 \times 10^{-3} \, \lambda_{\chi}^{3/2} \, \chi_{\rm end} \, , & \qquad (p=4) \, , \\
\label{eq:fragp6}
\Gamma_{\delta \chi} &\; \simeq \; 3.4 \times 10^{-2} \, \lambda_{\chi}^{3/2} \left( \frac{\chi_{\rm end}}{M_P} \right)^3 \chi_{\rm end} \, , & \qquad (p=6) \, , \\
\label{eq:fragp8}
\Gamma_{\delta \chi} &\; \simeq \; 0.16 \, \lambda_{\chi}^{3/2} \left( \frac{\chi_{\rm end}}{M_P} \right)^6 \chi_{\rm end} \, . & \qquad (p=8) \, .
\end{align}
In these expressions, we replaced $\chi_0$ with $\chi_{\rm end}$, which corresponds to the amplitude of the spectator field at the end of inflation. Note that in the above analysis, we have assumed that the envelope $\chi_0(t)$ remains approximately constant over one oscillation, which is a valid approximation for $p \leq 2k$. In this regime, the redshift due to the expansion of the Universe has a negligible effect on the oscillatory dynamics of the spectator field. However, for $p > 2k$, the redshift of $\chi_0$ dominates over the oscillations. Consequently, the effective frequency $\omega_\chi$ is significantly smaller than the one given in Eq.~\eqref{freqchi}. This reduction in frequency leads to a stronger kinematic suppression in the fragmentation rate due to the factor given in Eq.~\eqref{betachi}. As a result, the true upper bound on the self-coupling $\lambda_\chi$ is expected to be much higher than the estimates provided above for $p=6$ and $p=8$. This indicates that fragmentation effects are even less relevant for larger values of $p$ in this regime, and the spectator field can remain stable over longer periods without significantly altering the cosmological dynamics.

\end{document}